\documentclass[onecolumn,notitlepage,nofootinbib,superscriptaddress,floatfix]{revtex4-2}

\usepackage{float}
\makeatletter
\let\newfloat\newfloat@ltx
\makeatother

\usepackage{algorithm}
\usepackage{algorithmic}
\usepackage{array} 

\newcommand{\dimSym}{d_{\mathrm{sym}}}

\usepackage[utf8]{inputenc}       
\usepackage[english]{babel}

\usepackage{amsmath,amssymb,amsthm,mathtools}
\usepackage{physics}

\usepackage{graphicx}

\usepackage[shortlabels]{enumitem}

\usepackage[caption=false]{subfig} 
\usepackage{xcolor}
\usepackage{booktabs}
\usepackage{adjustbox}

\usepackage{array}

\makeatletter

\newcounter{subsubsubsection}[subsubsection]
\renewcommand{\thesubsubsubsection}{\thesubsubsection.\arabic{subsubsubsection}}

\newcommand{\subsubsubsection}[1]{%
  \refstepcounter{subsubsubsection}%
  \phantomsection%
  \addcontentsline{toc}{subsubsubsection}{%
    \protect\numberline{\thesubsubsubsection}#1%
  }%
  \vspace{1.0em}%
  \begin{center}
    \normalsize\normalfont \thesubsubsubsection\quad #1
  \end{center}
  \vspace{0.4em}%
}

\newcommand*\l@subsubsubsection[2]{%
  \ifnum\c@tocdepth>3\relax
    \vskip \z@ \@plus .2\p@
    {%
      \leftskip 4.8em\relax
      \rightskip \@tocrmarg
      \parfillskip -\rightskip
      \parindent 4.8em\relax
      \interlinepenalty\@M
      \leavevmode
      \@tempdima 1.6em\relax
      \advance\leftskip \@tempdima
      \null\nobreak\hskip -\leftskip
      \normalfont #1\nobreak\hfill\nobreak
      \hb@xt@\@pnumwidth{\hfil\normalfont #2}%
      \par
    }%
  \fi
}

\makeatother

\usepackage{tikz}
\usetikzlibrary{positioning,decorations.pathreplacing}
\usepackage{tikz-cd}
\usepackage{pgfplots}
\usepgfplotslibrary{groupplots,fillbetween}
\pgfplotsset{compat=1.18} 

\usepackage{csquotes}
\usepackage{enumitem}
\usepackage{IEEEtrantools}
\usepackage{relsize}
\usepackage{placeins}
\usepackage{comment}
\usepackage[normalem]{ulem}
\usepackage{systeme}
\usepackage{pgffor}

\definecolor{lb}{rgb}{.5,.1,-.7}

\usepackage[pdfpagelabels]{hyperref}
\hypersetup{
  colorlinks=true,
  linkcolor  = [rgb]{0.70,0.13,0.13},
  citecolor  = [rgb]{0.13,0.55,0.13},
  urlcolor   = [rgb]{0.25,0.41,0.88}
}

\theoremstyle{plain}
\newtheorem{theorem}{Theorem}[section]
\newtheorem{lemma}[theorem]{Lemma}
\newtheorem{proposition}[theorem]{Proposition}
\newtheorem{corollary}[theorem]{Corollary}
\theoremstyle{definition}
\newtheorem{definition}[theorem]{Definition}

\newtheorem{remark}[theorem]{Remark}

\newcommand{\fu}{Dahlem Center for Complex Quantum Systems, Freie Universit\"{a}t Berlin, 14195 Berlin, Germany}

\begin{document}
\title{Optimal learning of quantum channels in diamond distance}

\author{Antonio Anna Mele}
\affiliation{\fu}
\author{Lennart Bittel}
\affiliation{\fu}
\date{\today}

\begin{abstract}
Quantum process tomography---the task of estimating an unknown quantum channel---is a central problem in quantum information theory. A long-standing open question is how many uses of an unknown channel are required to learn it in \emph{diamond distance}, the standard metric for distinguishing quantum processes. While the analogous problem of quantum state tomography is by now well understood, for general channels it remained largely open beyond the unitary case. 
Here we establish the query complexity of channel tomography with optimal dependence on the dimension parameters, at any fixed constant accuracy in diamond distance. We design an algorithm showing that any channel with input and output dimensions $d_{\mathrm{in}},d_{\mathrm{out}}$ and Kraus rank at most $k$ can be learned to accuracy $\varepsilon$ using $O(d_{\mathrm{in}}d_{\mathrm{out}}k/\varepsilon^{2})$ channel uses. 
On the lower-bound side, we prove that $\Omega(d_{\mathrm{in}}d_{\mathrm{out}}k)$ channel uses are necessary and that, for channels with non-minimal Kraus rank, a separate $\Omega(1/\varepsilon^{2})$ contribution is unavoidable.
Since channels subsume states, unitaries, isometries, and measurements as special cases, our protocol provides a unified framework for the corresponding tomography tasks, yielding new guarantees for isometry and measurement tomography while recovering the known optimal scalings for state and unitary tomography. Notably, our algorithm follows the natural strategy of performing optimal tomography on the Choi state. The main technical contribution is to show that this suffices to control the induced diamond-distance error, thereby avoiding the worst-case dimension loss incurred by a naive conversion from Choi-state trace distance to channel diamond distance. Concretely, the protocol uses the channel only non-adaptively to prepare Choi-state copies, purifies them in parallel, and performs optimal pure-state tomography on the resulting purifications. Hence, we reduce channel tomography to pure-state tomography.
\end{abstract}

\maketitle
\vspace*{-0.7cm}
\section{Introduction}
In this work, we study the task of learning an accurate classical description of an unknown quantum channel from black-box access, using as few channel invocations as possible. This is the problem of \emph{quantum process tomography}~\cite{ChuangNielsen1997,PoyatosCiracZoller1997,Helsen_2022,MohseniRezakhaniLidar2008,EXP1,EXP2,EXP3,EXP4,EXP5,EXP6Blume_Kohout_2017,Scott_2008}, also referred to as quantum channel \emph{tomography} or \emph{learning}. Process tomography is a central primitive for calibrating, validating and debugging noisy quantum devices, and has been implemented across a range of platforms, including trapped ions, superconducting qubits and photonic architectures~\cite{MohseniRezakhaniLidar2008,EXP1,EXP2,EXP3,EXP4,EXP5,EXP6Blume_Kohout_2017}. 
In its most general form, a tomography protocol is given $N$ uses of an unknown channel, may probe it with arbitrary input states, may interleave these uses with adaptive quantum operations, and may finally perform a joint measurement on all output and auxiliary systems to produce a classical description of the channel. To make such learning guarantees precise, one must choose a metric on the space of quantum channels. Among the available choices, the \emph{diamond distance} is widely regarded as the standard one due to its operational meaning~\cite{Kitaev1997,GilchristLangfordNielsen2005,WatrousBook,Watrous2013CBnorm,aharonov1998quantumcircuitsmixedstates}: it captures the worst-case distinguishability between two quantum processes, optimized over arbitrary input states, auxiliary systems and measurements. Learning in this metric therefore asks for a classical surrogate of an unknown open-system dynamics that can replace the true process in any future experiment, up to a prescribed worst-case error. The central question is how many times the process must be used before such a surrogate can be guaranteed.

For comparison, the analogous problem of quantum state tomography---learning an unknown state in trace distance, the standard operational metric for states---is by now well understood~\cite{Hay98,OW16,Haah_2017,pelecanos2025mixedstatetomographyreduces}. In particular, the optimal dependence on dimension and accuracy for mixed-state tomography has been known for roughly a decade~\cite{OW16,Haah_2017} (with the full resolution in all relevant parameters, including fixed rank and failure probability, obtained only very recently~\cite{pelecanos2025mixedstatetomographyreduces,SSW25}). By contrast, the corresponding question for general quantum channels has remained open beyond special cases, most notably unitary channels~\cite{HaahKothariODonnellTang2023}.

In particular, for states on $\mathbb C^d$, the optimal number of copies required for tomography to fixed trace-distance accuracy scales with the number of degrees of freedom of the density matrix, namely $\Theta(d^2)$ for full-rank states. By analogy, one might expect the number of channel uses required for quantum process tomography to scale with the number of parameters describing a general channel. However, once one insists on diamond-distance guarantees, the best previously known tomography schemes for arbitrary $d$-dimensional channels required $\widetilde O(d^{6})$ channel uses~\cite{SurawyStepneyKahnKuengGuta2022,Oufkir_2023,barberarodriguez2025boostingprojectivemethodsquantum}, larger than the $\Theta(d^{4})$ number of independent parameters that are needed to describe a general channel. This left a substantial gap to the best known information-theoretic lower bounds, which scale as $\widetilde\Omega(d^{4})$ for constant accuracy~\cite{LBrosenthal2024quantumchanneltestingaveragecase}; see Table~\ref{tab:bounds-summary}. 
This raises a basic and conceptually clean question:
\begin{quote}
\emph{How many uses of an unknown quantum channel are needed to learn it to constant accuracy in diamond distance? More specifically, is this number governed by the number of independent parameters needed to describe a general channel?}
\end{quote}
In this work, we answer this question affirmatively.
We show that a number of channel uses scaling with the degrees of freedom of the unknown channel is both sufficient and necessary to achieve any fixed constant diamond-distance accuracy. In particular, channels with input and output dimensions $d_{\mathrm{in}}$ and $d_{\mathrm{out}}$ and Kraus rank at most $k$ admit a $\Theta(d_{\mathrm{in}}d_{\mathrm{out}}k)$-parameter description, and we give a channel-tomography algorithm achieving this query-complexity scaling and prove a matching lower bound, showing that no process-tomography procedure can asymptotically improve on this parameter-count dependence.
More generally, our main theorem can be stated as follows:
\begin{theorem}
\label{th:th1Main}
Let $\Lambda : \mathcal{L}(\mathbb{C}^{d_{\mathrm{in}}}) \to
\mathcal{L}(\mathbb{C}^{d_{\mathrm{out}}})$ be an unknown quantum channel with Kraus rank at most $k$.
Then, for any $\varepsilon\in(0,1)$ and any desired constant success probability, there exists a quantum process-tomography algorithm that uses
\begin{align}
  N = O\!\left(\frac{d_{\mathrm{in}}d_{\mathrm{out}}k}{\varepsilon^2}\right)
\end{align}
invocations of $\Lambda$ and outputs a classical description of a channel $\widehat{\Lambda}$ satisfying
$\|\widehat{\Lambda}-\Lambda\|_\diamond \le \varepsilon$, where $\|\cdot\|_\diamond$ denotes the diamond norm.
Conversely, any process-tomography algorithm achieving diamond-norm accuracy $\varepsilon$ must use
$N=\Omega(d_{\mathrm{in}}d_{\mathrm{out}}k)$ channel invocations. Moreover, if the allowed Kraus rank is non-minimal, namely if
$k>\lceil d_{\mathrm{in}}/d_{\mathrm{out}}\rceil$, then any such algorithm must also incur an
$\Omega(1/\varepsilon^2)$ lower-bound contribution.
\end{theorem}
Thus, Theorem~\ref{th:th1Main} establishes that the query complexity of channel tomography at constant accuracy in diamond distance is $\Theta(d_{\mathrm{in}}d_{\mathrm{out}}k)$. It also records a separate accuracy-dependent lower-bound: for channels with Kraus rank larger than the minimum allowed by the input and output dimensions, an $\Omega(\varepsilon^{-2})$ contribution to the query complexity is unavoidable. This contrasts with the minimal Kraus-rank case, which includes unitary channels, for which Heisenberg scaling $O(\varepsilon^{-1})$ is known to be achievable~\cite{HaahKothariODonnellTang2023}. We point out that, for readability, Theorem~\ref{th:th1Main} suppresses the explicit dependence on the failure probability in the upper bound and gives only the most relevant consequences of the lower-bound results proved later in the manuscript.

If no Kraus-rank assumption is imposed, then one may take
$k=d_{\mathrm{in}}d_{\mathrm{out}}$, and the upper bound becomes
$O(d_{\mathrm{in}}^{2}d_{\mathrm{out}}^{2}/\varepsilon^{2})$.
For channels acting on $\mathbb{C}^{d}$, i.e. when
$d_{\mathrm{in}}=d_{\mathrm{out}}=d$, this gives
$O(d^{4}/\varepsilon^{2})$.

\begin{figure}
  \centering
  \includegraphics[width=0.7\linewidth]{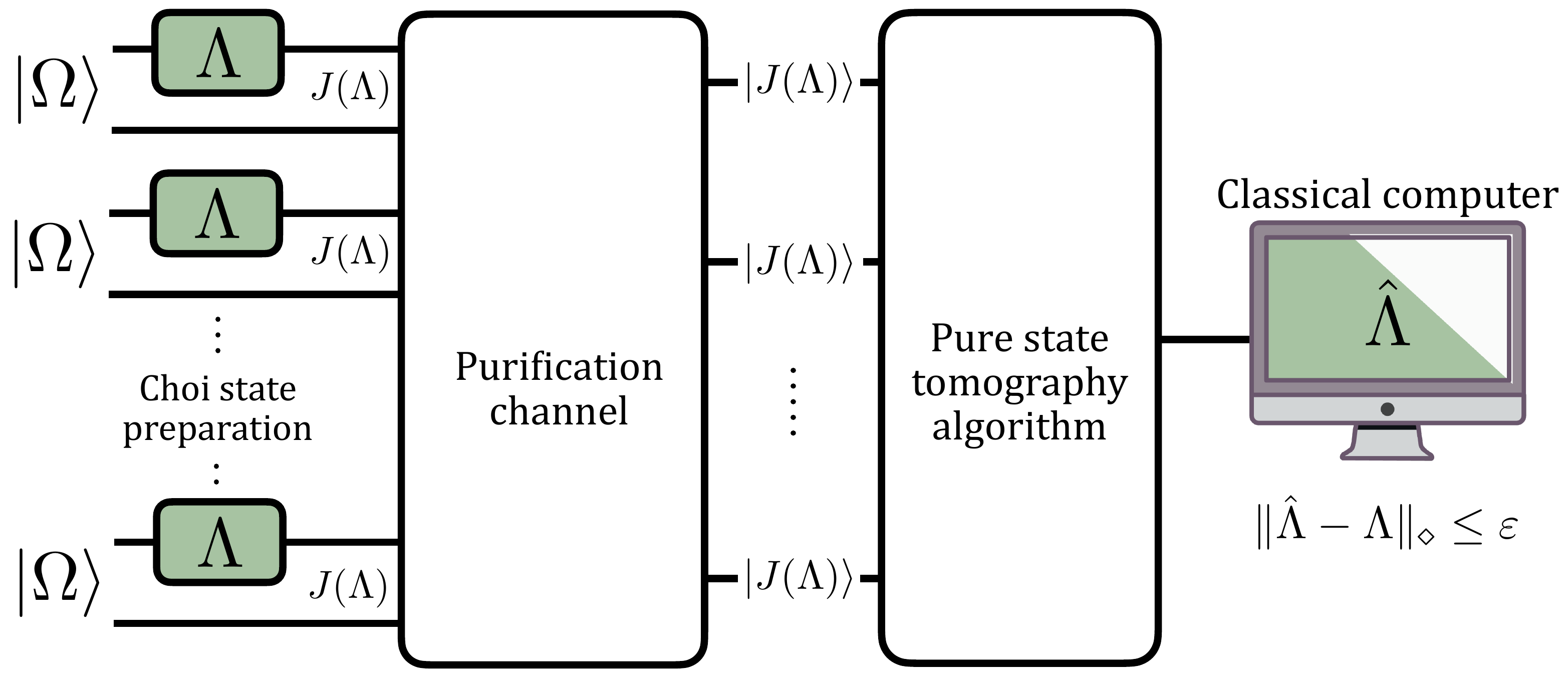}
  \caption{
    \textbf{Protocol for learning quantum channels.}
    Parallel invocations of the unknown channel $\Lambda$, applied to halves of
    maximally entangled states $\ket{\Omega}$, prepare copies of its normalised
    Choi state $\Phi_c = J(\Lambda)$.
    A purification channel then maps these Choi-state copies to copies of a
    random purification $\ket{\Phi_c}$, which are learned using a
    sample-optimal pure-state tomography scheme.
    Finally, a classical post-processing step, formulated as a semidefinite
    program projecting onto the set of CPTP maps, returns a channel estimate
    $\widehat{\Lambda}$ satisfying
    $\|\widehat{\Lambda}-\Lambda\|_\diamond \le \varepsilon$
    with high probability.
  }
  \label{fig:Fig1_CL}
\end{figure}

The algorithm achieving this scaling is conceptually simple (see Fig.~\ref{fig:Fig1_CL}).
We first prepare many copies of the Choi state of the unknown channel in parallel.
We then apply the recently introduced \emph{random purification channel}~\cite{tang2025conjugatequerieshelp} to these Choi states, producing multiple copies of a purification~\cite{tang2025conjugatequerieshelp,pelecanos2025mixedstatetomographyreduces,girardi2025randompurificationchannelsimple}.
Finally, we run a sample-optimal pure-state tomography procedure on the resulting purifications~\cite{Hay98,pelecanos2025mixedstatetomographyreduces,GutaKahnKuengTropp2020}, obtain an estimate of the purification, and reconstruct from it an estimate of the unknown channel, regularising the estimate by an SDP projection onto the set of CPTP maps.

In this sense, the protocol can be essentially viewed as applying a recently introduced sample-optimal mixed-state tomography strategy of Ref.~\cite{pelecanos2025mixedstatetomographyreduces} to the Choi state of the unknown channel. This is a conceptual strength of our result: we show that the most natural strategy, namely Choi-state tomography, in fact suffices for achieving such diamond-distance guarantees, even though it might a priori appear too crude. The apparent obstruction is that state tomography controls the Choi state in trace distance, whereas the operational metric for channels is the diamond distance, and a naive worst-case conversion from Choi-state trace distance to channel diamond distance incurs a dimension-dependent loss, which would destroy the desired scaling.

One of our main technical contributions is a sharp analysis showing that this worst-case dimension-dependent loss can be avoided for the estimator produced by the protocol. A key feature of the analysis is that we evaluate the estimator directly in diamond distance, using a convenient semidefinite-program (SDP) characterisation~\cite{Watrous2013CBnorm,SkrzypczykCavalcantiSDP}. A second key ingredient is that the residual state-estimation error after pure-state tomography can, without loss of generality, be modelled as Haar-random. This allows us to control the induced diamond-distance error of the reconstructed channel using concentration properties of random quantum states~\cite{AubrunSzarek2017AliceBob}. In particular, we prove that to guarantee diamond-distance error at most $\varepsilon$, it suffices to learn the purified Choi state, and hence the Choi state, to trace-distance accuracy $O(\varepsilon)$, with essentially no additional dimension-dependent overhead. 

Thus, our analysis shows that learning a quantum process in diamond distance can be reduced, without losing the optimal scaling in dimension, to pure-state tomography. This generalizes the mixed-to-pure-state tomography reduction of Ref.~\cite{pelecanos2025mixedstatetomographyreduces} from mixed states to quantum channels, a strict generalization since quantum states can be viewed as channels with one-dimensional input.

\begin{table}[h]
  \centering
  \begingroup
  \setlength{\tabcolsep}{10pt}
  \renewcommand{\arraystretch}{1.6}
  \begin{tabular}{@{} l c c c @{}}
    \toprule
    & \textbf{States}
    & \textbf{Channels [previous works]}
    & \textbf{Channels [this work]} \\
    \midrule
    \textbf{Upper bound}
      & $O\!\left(\dfrac{dk}{\varepsilon^{2}}\right)$
        ~\cite{OW16,Haah_2017,pelecanos2025mixedstatetomographyreduces}
      & $\widetilde O\!\left(
          \dfrac{d_{\mathrm{in}}^{3}d_{\mathrm{out}}^{3}}
          {\varepsilon^{2}}
        \right)$
        ~\cite{Oufkir_2023,SurawyStepneyKahnKuengGuta2022}
      & $O\!\left(
          \dfrac{d_{\mathrm{in}}d_{\mathrm{out}}k}
          {\varepsilon^{2}}
        \right)$ \\
    \addlinespace[4pt]
    \textbf{Lower bound}
      & $\Omega\!\left(\dfrac{dk}{\varepsilon^{2}}\right)$
        ~\cite{SSW25}
      & $\Omega\!\left(
          \dfrac{d_{\mathrm{in}}^{2}d_{\mathrm{out}}^{2}}
          {\log(d_{\mathrm{in}}d_{\mathrm{out}})}
        \right)$
        ~\cite{LBrosenthal2024quantumchanneltestingaveragecase}
      & $\Omega\!\left(d_{\mathrm{in}}d_{\mathrm{out}}k\right)$, 
        $\Omega\!\left(\dfrac{1}{\varepsilon^{2}}\right)$ \\
    \bottomrule
  \end{tabular}

\caption{
Query complexity for tomography of quantum states and quantum channels to accuracy~$\varepsilon$ in the corresponding operational metric.
For states, the metric is trace distance, $d$ is the Hilbert-space dimension, and $k$ is a rank upper bound; for channels, the metric is diamond distance, $d_{\mathrm{in}}$ and $d_{\mathrm{out}}$ are the input and output dimensions, and $k$ is a Kraus-rank upper bound.
The middle column gives the best prior bounds for general channel tomography (known only for unrestricted Kraus rank), while the right column gives our bounds.
At constant accuracy, our upper and lower bounds match in their dependence on $d_{\mathrm{in}}$, $d_{\mathrm{out}}$ and $k$, establishing dimension-optimal channel tomography in diamond distance.
The separate $\Omega(1/\varepsilon^{2})$ lower-bound contribution applies for non-minimal Kraus rank, namely $k>k_{\min}:=\lceil d_{\mathrm{in}}/d_{\mathrm{out}}\rceil$; for minimal Kraus rank this cannot hold uniformly, since the class includes unitaries, which admit Heisenberg scaling $\Theta(1/\varepsilon)$~\cite{HaahKothariODonnellTang2023}.
Our upper bound follows by reducing channel tomography to sample-optimal pure-state tomography on Choi-state purifications.
Since channels include states, isometries, and measurement devices as special cases, this gives a unified framework for several tomography tasks.
}
  \label{tab:bounds-summary}
  \endgroup
\end{table}

Two structural features of our scheme are worth highlighting. First, the protocol is entirely non-adaptive at the level of channel uses: all invocations of the unknown channel occur in a single parallel block, and all subsequent operations act only on the resulting Choi-state copies. In particular, this also shows that adaptivity offers no asymptotic advantage for achieving the dimension-optimal scaling at fixed constant diamond-distance accuracy. Second, coherence across Choi-state copies remains provably useful in this non-adaptive setting. Indeed, Ref.~\cite{Oufkir_2023} showed that any non-adaptive scheme restricted to incoherent measurements on individual channel outputs requires asymptotically more channel invocations to reach a given diamond-distance accuracy. In our protocol, this coherence is confined to the purification stage. Once purifications of the Choi state are available, the remaining tomography step can be carried out using any sample-optimal pure-state tomography procedure, including single-copy schemes~\cite{GutaKahnKuengTropp2020}.

We next briefly explain the converse direction. The lower bounds stated in Theorem~\ref{th:th1Main} are proved by two different reductions. The dimension-dependent lower bound $\Omega(d_{\mathrm{in}}d_{\mathrm{out}}k)$ follows from a packing argument: we construct many Kraus-rank-$k$ channels separated in diamond distance and reduce tomography to identifying which channel was given, following ideas used previously for unitary channels~\cite{HaahKothariODonnellTang2023,Bavaresco_2022}. The $\Omega(1/\varepsilon^{2})$ lower bound for channels with non-minimal Kraus rank follows from embedding the task of estimating the biases of multiple classical coins into channel tomography, so that the standard $\varepsilon^{-2}$ lower bound for coin-bias estimation carries over.

Beyond Theorem~\ref{th:th1Main}, our framework recovers and unifies several previously known query-optimal tomography results~\cite{Haah_2017,OW16,SSW25,pelecanos2025mixedstatetomographyreduces,HaahKothariODonnellTang2023}, and yields new guarantees for other basic tomography tasks in the strong metrics considered here~\cite{POVMnetzambrano2025fastquantummeasurementtomography,ISOMEyoshida2025quantumadvantagestorageretrieval}.

On the ``known results'' side, specialising to channels with $d_{\mathrm{in}}=1$ recovers the optimal $\Theta(kd/\varepsilon^2)$ scaling for trace-distance tomography of rank-$k$ states, matching the information-theoretic lower bounds~\cite{SSW25} and the upper bounds achieved by sample-optimal state-tomography schemes~\cite{Haah_2017,OW16,pelecanos2025mixedstatetomographyreduces}. For unitary channels, corresponding to $d_{\mathrm{in}}=d_{\mathrm{out}}=d$ and $k=1$, our Choi-state-based protocol recovers the optimal $\Theta(d^2)$ dimension dependence. Moreover, using our unitary-learning protocol as a black-box subroutine in the adaptive boosting scheme of Ref.~\cite{HaahKothariODonnellTang2023}, one can upgrade the $1/\varepsilon^2$ dependence to Heisenberg scaling $1/\varepsilon$, obtaining the optimal query complexity $\Theta(d^2/\varepsilon)$. Thus, our method provides a new route to query-optimal unitary tomography based on the natural Choi-state tomography approach, complementary to Ref.~\cite{HaahKothariODonnellTang2023}, which achieves the same optimal scaling through a protocol not based on Choi-state learning.

On the ``new consequences'' side, for Kraus-rank-one channels our result gives, to the best of our knowledge, the first algorithm for learning arbitrary isometries in diamond distance with query complexity $O(d_{\mathrm{in}}d_{\mathrm{out}}/\varepsilon^2)$. In sharp contrast with the unitary case, lower bounds rule out Heisenberg scaling $1/\varepsilon$ for general isometries~\cite{ISOMEyoshida2025quantumadvantagestorageretrieval} and show that this scaling is optimal up to logarithmic factors. 
Finally, we show that binary POVMs can be learned in operator norm with
$O(d^2/\varepsilon^2)$ uses of the unknown measurement device. We show that this scaling is optimal in the dimension at constant accuracy, and we also establish a separate
$\Omega(1/\varepsilon^2)$ lower bound. 

It is also important to clarify the practical scope of our result. The present result should primarily be seen as an information-theoretic achievability theorem rather than as a practically scalable tomography protocol. Although the query complexity is dimension-optimal, the protocol requires a coherent purification step acting jointly across many Choi-state copies, which is demanding in realistic multi-qubit regimes. In addition, the final CPTP regularisation is formulated as a semidefinite program whose size is polynomial in the Hilbert-space dimension, and hence generally not scalable in the number of qubits.
By contrast, the Kraus-rank-one case is substantially more feasible in practice. In this case the Choi state is already pure, so the purification step is unnecessary, and the task reduces directly to pure-state tomography on independent Choi-state copies~\cite{GutaKahnKuengTropp2020}. This also has a practically appealing feature compared with earlier optimal unitary-learning schemes~\cite{HaahKothariODonnellTang2023}: one repeatedly prepares and measures the same state rather than cycling through many different input settings.
In this sense, our protocol identifies the dimension-optimal query complexity of channel tomography in the strongest operational metric, while more practical incoherent schemes trade this query optimality for implementability~\cite{Oufkir_2023,barberarodriguez2025boostingprojectivemethodsquantum,POVMnetzambrano2025fastquantummeasurementtomography}.

\subsection*{Relation to concurrent works}
Before turning to the broader comparison with prior work, we clarify the relationship with two works that appeared after the first arXiv version of our manuscript~\cite{chen2025quantumchanneltomographyestimation,girardi2025randomstinespringsuperchannelconverting}.

First, Ref.~\cite{chen2025quantumchanneltomographyestimation} appeared shortly after our work and independently establishes the same upper-bound scaling for learning quantum channels in diamond distance in the constant-success-probability regime. The two approaches are technically different: Ref.~\cite{chen2025quantumchanneltomographyestimation} reduces channel tomography to isometry tomography by simulating access to a random Stinespring dilation, whereas our method proceeds via Choi-state tomography and a direct diamond-distance analysis of the resulting estimator. Our results additionally provide an explicit, non-trivial dependence on the failure probability. Ref.~\cite{chen2025quantumchanneltomographyestimation} further identifies additional channel families, beyond unitaries, for which scaling $O(1/\varepsilon)$ is achievable, namely minimal-Kraus-rank channels whose Stinespring isometry can be chosen unitary.

Second, Ref.~\cite{girardi2025randomstinespringsuperchannelconverting} studies a different but related algorithmic primitive: converting channel queries into queries to a random Stinespring isometry. Its upper-bound consequences rely explicitly on the isometry-tomography guarantees obtained in our work, which follow from Theorem~\ref{th:th1Main}. By contrast, its dimension-matching lower bound $\Omega(d_{\mathrm{in}}d_{\mathrm{out}}k)$ for constant diamond-norm accuracy was obtained independently. The first arXiv version of the present manuscript did not contain the lower-bound section; this section was added in a subsequent version, posted in coordination with Ref.~\cite{girardi2025randomstinespringsuperchannelconverting}, to clarify the relationship between the two independent lower-bound results.

We now place our results in the broader context of existing work on state, process, and measurement tomography.

\subsection{Related work}
We begin by situating our contributions within the broader literature on tomography and quantum learning theory~\cite{anshu2023survey}.
The picture for quantum \emph{state} tomography in trace distance is by now well understood and fairly complete~\cite{OW16,Haah_2017,pelecanos2025mixedstatetomographyreduces,KRT17,SSW25}.
For states of rank at most $r$ on $\mathbb C^d$, there exists a tomography protocol~\cite{pelecanos2025mixedstatetomographyreduces} that, given
$N=\Theta((rd+\log(1/\delta))/\varepsilon^{2})$ copies, outputs an estimate within trace distance $\varepsilon$ with probability at least $1-\delta$.
Matching lower bounds show that this scaling is optimal up to constant factors in all parameters~\cite{Haah_2017,Yuen_2023,SSW25}.
Thus, state tomography in trace distance is essentially a solved problem from the point of view of sample complexity.
The development of optimal schemes has a long history: early work established sample-optimal protocols for pure-state tomography based on coherent measurements across copies~\cite{Hay98,Har13}, and subsequent advances showed that the same asymptotic sample complexity can be achieved using only single-copy measurements~\cite{KRT17,Haah_2017,OW16}.
For mixed states of bounded rank, explicit algorithms achieving the optimal upper bound at any fixed constant failure probability were developed roughly a decade ago~\cite{OW16,Haah_2017}, while the fully matching lower bound was obtained only very recently~\cite{SSW25}. More recently, Ref.~\cite{pelecanos2025mixedstatetomographyreduces} introduced an alternative and conceptually elegant approach to sample-optimal mixed-state tomography, showing that it can be reduced to pure-state tomography via a random purification channel~\cite{tang2025conjugatequerieshelp}. This reduction also achieves the optimal dependence on the failure probability, and the same purification primitive is a key ingredient in our channel-tomography protocol. We refer the reader to Refs.~\cite{pelecanos2025mixedstatetomographyreduces,tang2025conjugatequerieshelp,girardi2025randompurificationchannelsimple} for further details on the random purification channel and its implementation via the Schur transform~\cite{bacon2005quantumschurtransformi,burchardt2025highdimensionalquantumschurtransforms}.

By contrast, the landscape for general quantum \emph{channel} (process) tomography has remained more fragmented. There is a large body of work on
process tomography on the experimental and heuristic
side~\cite{EXP1,EXP2,EXP3,EXP4,EXP5,EXP6Blume_Kohout_2017,Ahmed_2023,Knee_2018}, while much of the theory
with rigorous guarantees focuses on quantities that are technically convenient
yet weaker than diamond norm guarantees, such as trace or Hilbert--Schmidt distance of
the associated Choi state, or average gate fidelity~\cite{SurawyStepneyKahnKuengGuta2022,Kliesch_2021,Roth_2018,zhao2023learning,Knill_2008,Helsen_2022,Helsen_2023,Kiktenko_2020,Thinh_2019,Kliesch_2019,Flammia_2012,huang2023learningpredictarbitraryquantum,Kunjummen_2023,LOWERBrosenthal2024quantumchanneltestingaveragecase,fawzi2023quantumchannelcertificationincoherent,lang2025unifiedblockwisemeasurementdesign,Caro_2024,wadhwa2024agnosticprocesstomography}. Although these figures of merit are often adequate for specific applications,
they do not in general suffice to control the diamond distance on their own
without incurring polynomial overheads in the dimension, and may therefore be
too loose from a worst-case operational perspective~\cite{WatrousBook}.
Prior to our work, there was a substantial dimensional gap in the query-complexity of quantum channel tomography in diamond distance, as summarized in Table~\ref{tab:bounds-summary}.

On the lower-bound side, it was known that even with entangled inputs and fully coherent adaptive strategies, any estimator achieving constant diamond-distance error must use at least $N = \Omega\bigl(d_{\mathrm{in}}^{2} d_{\mathrm{out}}^{2} / \log(d_{\mathrm{in}} d_{\mathrm{out}})\bigr)$ queries of the unknown channel~\cite{LOWERBrosenthal2024quantumchanneltestingaveragecase,Oufkir_2023}.

On the upper-bound side, prior to our work the best general guarantees for learning quantum channels in diamond distance scaled as $N = O\!\left({d^{6}/\varepsilon^{2}}\right)$,
for a $d$-dimensional channel, leaving a quadratic gap to the information-theoretic lower bounds $\Omega(d^{4}/\log(d))$.
One route to the $O(d^{6}/\varepsilon^{2})$ scaling is a ``naive'' Choi-state reduction: prepare the Choi state of the unknown channel, apply an optimal mixed-state tomography algorithm with trace-distance guarantees, and then translate directly the resulting error bound to diamond distance, incurring a dimensional overhead. Without a more careful analysis (such as the one we develop, which exploits the finer structure of the tomography error distribution), this approach yields the query complexity $N = O\!\left(d_{\mathrm{in}}^{4} d_{\mathrm{out}}^{2}/\varepsilon^{2}\right).$
Refs.~\cite{SurawyStepneyKahnKuengGuta2022,Oufkir_2023,barberarodriguez2025boostingprojectivemethodsquantum} further showed that the scaling
$N = \Tilde O\!\left(d^{6}/\varepsilon^{2}\right)$ (more generally
$N = \Tilde  O\!\left(d_{\mathrm{in}}^{3} d_{\mathrm{out}}^{3}/\varepsilon^{2}\right)$)
can be achieved by more practical protocols based on incoherent, non-adaptive queries.
Our main theorem closes the dimensional gap between these upper bounds and the lower bound.
Ref.~\cite{Oufkir_2023} further analysed restricted models in which one is limited to non-adaptive, incoherent measurements on individual channel outputs, and showed that in this setting $N = \tilde{\Theta}\bigl(d_{\mathrm{in}}^{3} d_{\mathrm{out}}^{3} / \varepsilon^{2}\bigr)$ queries are both sufficient and necessary.

Significant progress has also been made on diamond-norm guarantees for
\emph{structured} families of channels~\cite{HaahKothariODonnellTang2023,grewal2025queryoptimalestimationunitarychannels,iyer2025mildlyinteractingfermionicunitariesefficiently,Huang_2024,vasconcelos2025learningshallowquantumcircuits,Fawzi_2025,Oszmaniec_2022,Trinh_2025,chen2025efficientselfconsistentlearninggate,Chen_2024,Bisio_2010,Yang_2020,zhao2023learning,Chen_2023}, such as Pauli-channels~\cite{Trinh_2025,chen2025efficientselfconsistentlearninggate,Chen_2024,Chen_2023} or structured unitary channels~\cite{HaahKothariODonnellTang2023,grewal2025queryoptimalestimationunitarychannels,iyer2025mildlyinteractingfermionicunitariesefficiently,Huang_2024,vasconcelos2025learningshallowquantumcircuits,Fawzi_2025,Oszmaniec_2022,Bisio_2010,Yang_2020,zhao2023learning}.
For general unitary channels, Ref.~\cite{HaahKothariODonnellTang2023} showed that one
can learn an unknown unitary on $\mathbb C^{d}$ in diamond distance with
$\Theta(d^{2})$ queries, and that combining such a learner with an adaptive
boosting procedure yields Heisenberg scaling $1/\varepsilon$ in the accuracy
parameter. Our approach recovers the same optimal $\Theta(d^{2})$ dependence
for unitary channels via Choi-state-tomography, and
the resulting unitary learner can be used as a drop-in replacement within the
boosting scheme of Ref.~\cite{HaahKothariODonnellTang2023} to achieve the optimal query complexity $\Theta(d^{2}/\varepsilon)$.

Our work is also connected to recent progress on the tomography of other
quantum primitives beyond general quantum
channels~\cite{li2024quantumnetworktomographylearning,ISOMEyoshida2025quantumadvantagestorageretrieval,POVMnetzambrano2025fastquantummeasurementtomography,Cattaneo_2023,Keith_2018,barberarodriguez2025boostingprojectivemethodsquantum}.
In particular, Ref.~\cite{ISOMEyoshida2025quantumadvantagestorageretrieval}
studied the problem of tomography of isometries and proved a lower bound
$N=\widetilde\Omega(d_{\mathrm{in}}d_{\mathrm{out}}/\varepsilon^{2})$
for generic isometric channels in diamond distance. This rules out
Heisenberg scaling in the accuracy parameter for general isometries, in sharp
contrast with the unitary case. Combined with this lower bound, our
$O(d_{\mathrm{in}}d_{\mathrm{out}}/\varepsilon^{2})$ upper bound gives
essentially optimal diamond-distance guarantees for learning arbitrary
isometries, up to logarithmic factors.
In the context of measurement (detector) tomography, Refs.~\cite{POVMnetzambrano2025fastquantummeasurementtomography,barberarodriguez2025boostingprojectivemethodsquantum} analyse practical schemes for learning POVMs using incoherent, non-adaptive measurements, and show that $\widetilde\Theta(d^{3}/\varepsilon^{2})$ queries are both sufficient and necessary in this restricted setting. Our results improve the dimensional scaling for binary POVMs to $O(d^{2}/\varepsilon^{2})$ in operator norm, using a coherent yet still non-adaptive protocol. This dependence on $d$ is optimal at constant accuracy, as shown by matching dimension lower bounds. We also prove a separate $\Omega(1/\varepsilon^{2})$ lower bound on the accuracy dependence for binary POVM tomography. 

Finally, there is an active line of research on the role of adaptivity and
coherence in quantum tomography and
learning~\cite{chen2023doesadaptivityhelpquantum,Trinh_2025,chen2021exponentialseparationslearningquantum,chen2024adaptivityhelpexponentiallyshadow,fawzi:hal-04107265,Salek_2022,girish2023poweradaptivityquantumquery,cervero2023weakschursamplinglogarithmic,hu2024sampleoptimalmemoryefficient,chen2024optimaltradeoffentanglementcopy}.
For state tomography with single-copy measurements, it was recently shown that adaptivity does not improve the asymptotic sample complexity, which for full-rank mixed states remains $\Theta(d^{3}/\varepsilon^{2})$ in trace distance~\cite{chen2023doesadaptivityhelpquantum}.
By contrast, coherent measurements across multiple copies can reduce the sample complexity of mixed-state tomography from $\Theta(d^{3}/\varepsilon^{2})$ to $\Theta(d^{2}/\varepsilon^{2})$~\cite{GutaKahnKuengTropp2020,OW16,pelecanos2025mixedstatetomographyreduces,Haah_2017,cervero2023weakschursamplinglogarithmic,hu2024sampleoptimalmemoryefficient,chen2024optimaltradeoffentanglementcopy}.
Importantly, the optimal coherent mixed-state tomography protocols can be implemented non-adaptively on the available copies.

In the channel setting, Ref.~\cite{Oufkir_2023} proved that any strategy based on non-adaptive, incoherent measurements on individual channel outputs incurs an unavoidable overhead in the query complexity, and in particular cannot beat $\Tilde{\Theta}\bigl(d_{\mathrm{in}}^{3} d_{\mathrm{out}}^{3}\bigr)$ queries.
Our results fit neatly into this picture: we show that optimal $\Theta\bigl(d_{\mathrm{in}}^{2} d_{\mathrm{out}}^{2}\bigr)$-scaling diamond-distance process tomography can be achieved by a fully non-adaptive yet coherent Choi-state-based protocol, and that, once the random purification channel is applied, the remaining work can be delegated to any sample-optimal single-copy pure-state tomography routine (that is, adaptivity is not needed, while coherence is needed only up to obtaining purified copies of the Choi state).

Other works have studied the learnability of quantum processes in different
models, such as PAC-style learning
settings~\cite{cheng2015learnabilityunknownquantummeasurements,raza2024onlinelearningquantumprocesses,Caro_2020,fanizza2025learningquantumprocessesinput},
classical shadow tomography~\cite{huang2023learningpredictarbitraryquantum,Helsen_2023,Levy_2024,Kunjummen_2023},
and quantum statistical query frameworks~\cite{Wadhwa_2025,Angrisani_2025}. These approaches
aim to predict the outcomes of many observables or approximate process
behavior without reconstructing the full channel in a strong operational
norm, and can therefore achieve substantially better sample complexity under
suitable structural assumptions, at the cost of weaker reconstruction
guarantees.

Having positioned our contributions within these lines of work, we now turn to
a more detailed presentation of our results and proof techniques.

\subsection{Our learning scheme and upper bound}
\label{subsec:main-result}

In this subsection, we summarise our channel-learning protocol and its main guarantee, and sketch the key proof ideas.
Full technical details are deferred to the preliminaries in Appendix~\ref{sec:appendixPRELIM} and to the complete analysis in Section~\ref{sec:main-diamond}.
Throughout, we fix an unknown quantum channel $\Lambda : \mathcal L(\mathcal H_{\mathrm{in}})\to \mathcal L(\mathcal H_{\mathrm{out}})$ with input and output dimensions $d_{\mathrm{in}} \coloneqq \dim(\mathcal H_{\mathrm{in}})$ and $d_{\mathrm{out}} \coloneqq \dim(\mathcal H_{\mathrm{out}})$, and Kraus rank at most $k$.
Our objective is to learn $\Lambda$ up to accuracy $\varepsilon$ in diamond distance, while using as few calls to the black-box channel as possible.

Our protocol is formulated in terms of the Choi state of the channel. Let $\Phi_c = J(\Lambda)$ denote the normalised Choi state of $\Lambda$, defined by
\begin{align}
  \Phi_c \coloneqq J(\Lambda)
  = (\Lambda \otimes \mathrm{id}_{\mathcal H_{\mathrm{in}}})(\Omega)
  \in \mathcal L(\mathcal H_{\mathrm{out}}\otimes\mathcal H_{\mathrm{in}}),
\end{align}
where $\Omega=\ketbra{\Omega}{\Omega}$ is the maximally entangled state on $\mathcal H_{\mathrm{in}} \otimes \mathcal H_{\mathrm{in}}$.
Since $\Lambda$ has Kraus rank at most $k$, the rank of $\Phi_c$ is at most $k$, and hence any purification of $\Phi_c$ lives in a Hilbert space of total dimension
\begin{align}
  d_{\mathrm{tot}} \coloneqq d_{\mathrm{in}} d_{\mathrm{out}}k.
\end{align}
The central idea is to reduce diamond-distance learning of $\Lambda$ to trace-distance pure-state tomography of such a purification of $\Phi_c$ of dimension $d_{\mathrm{tot}}$.

\begin{algorithm}
  \caption{Diamond-distance tomography for Kraus rank-\(k\) channels}
  \label{alg:diamond-tomography}
  \begin{algorithmic}[1]
    \REQUIRE Access to 
      \(\Lambda : \mathcal L(\mathcal H_{\mathrm{in}})
                  \to \mathcal L(\mathcal H_{\mathrm{out}})\)
      with Kraus rank \(\le k\); target accuracy \(\varepsilon \in (0,1)\);
      failure probability \(\delta \in (0,1)\).
    \STATE Choose \(N\) as prescribed by
           Theorem~\ref{th:main-diamond-tomography} (to obtain
           \(\varepsilon\)-accuracy in diamond distance).
    \STATE (Choi-state preparation) Using \(N\) parallel calls to \(\Lambda\),
           prepare \(N\) copies of the normalised Choi state
           \(\Phi_c = J(\Lambda)\) by feeding half of a maximally entangled
           state into each use of \(\Lambda\).
    \STATE (Purification) Apply the random purification
           channel~\cite{tang2025conjugatequerieshelp,pelecanos2025mixedstatetomographyreduces}
           of Lemma~\ref{le:purification-channel} to \(\Phi_c^{\otimes N}\)
           to obtain \(N\) copies of a random purification \(|\Phi_c\rangle\)
           of \(\Phi_c\).
    \STATE (Pure-state tomography) Run a sample-optimal pure-state tomography
           scheme \(\mathcal A\) on \(|\Phi_c\rangle^{\otimes N}\) to obtain
           an estimate \(|\hat\Phi_c\rangle\).
    \STATE (Choi reconstruction) Form the (normalised) Choi operator estimate
           \begin{align}
             \hat{\Phi}_c \coloneqq
             \tr_E\bigl(|\hat\Phi_c\rangle\langle\hat\Phi_c|\bigr),
           \end{align}
           and let \(\Lambda^{\mathrm{est}}\) be the completely positive map
           whose normalised Choi state is \(\hat{\Phi}_c\).
    \STATE (CPTP regularisation) Compute a CPTP map \(\hat\Lambda\) by
           solving (as in Lemma~\ref{le:cptp-regularisation}) the SDP
           \begin{align}
             \hat\Lambda \in
             \arg\min_{\Phi \in \mathsf{CPTP}}
             \bigl\|\Phi - \Lambda^{\mathrm{est}}\bigr\|_\diamond.
           \end{align}
    \ENSURE For \(N\) as above, the final estimator \(\hat\Lambda\) satisfies
           \(\tfrac12\|\hat\Lambda - \Lambda\|_\diamond \le \varepsilon\)
           with probability at least \(1 - \delta\).
  \end{algorithmic}
\end{algorithm}

Our protocol (Algorithm~\ref{alg:diamond-tomography}) has three main steps:
\begin{itemize}
  \item \textbf{Parallel, non-adaptive Choi-state preparation.}
  Given \(N\) uses of \(\Lambda\), we prepare the Choi-state copies \(\Phi_c^{\otimes N}\) by applying \(\Lambda^{\otimes N}\) in parallel to maximally entangled input states \(\Omega^{\otimes N}\) on \((\mathcal H_{\mathrm{in}}\otimes\mathcal H_{\mathrm{in}})^{\otimes N}\).
  All calls to the unknown channel occur in a single parallel block, and no adaptivity is required at the level of channel uses.

 \item \textbf{Purification via the random purification channel.}
  On the prepared Choi-state copies \(\Phi_c^{\otimes N}\), we apply the \emph{random purification channel} of Ref.~\cite{tang2025conjugatequerieshelp}
 \begin{align}
    \mathcal P^{(N)} :
    \mathcal D\bigl(
      (\mathcal H_{\mathrm{out}}\otimes\mathcal H_{\mathrm{in}})^{\otimes N}
    \bigr)
    \to
    \mathcal D\bigl(
      (\mathcal H_{\mathrm{out}}\otimes\mathcal H_{\mathrm{in}}\otimes\mathcal H_E)^{\otimes N}
    \bigr),
  \end{align}
  where \(\mathcal H_E \simeq \mathbb C^{k}\) is an environment register, so
  that
  \(
    \mathcal H_{\mathrm{out}}\otimes\mathcal H_{\mathrm{in}}\otimes\mathcal H_E
    \simeq \mathbb C^{d_{\mathrm{tot}}}
  \).
  This channel maps the mixed-state \(\Phi_c^{\otimes N}\) to \(N\) copies of a random
  purification
  \(
    |\Phi_c\rangle \in \mathbb C^{d_{\mathrm{tot}}}
      \simeq \mathcal H_{\mathrm{out}}\otimes\mathcal H_{\mathrm{in}}\otimes\mathcal H_E
  \),
  in the sense that
  \begin{align}
    \mathcal P^{(N)}\bigl(\Phi_c^{\otimes N}\bigr)
    = \mathbb E_{|\Phi_c'\rangle}
      \bigl[|\Phi_c'\rangle\langle\Phi_c'|^{\otimes N}\bigr],
  \end{align}
  where the expectation is over purifications \(|\Phi_c'\rangle\) drawn from
  the unitarily invariant (Haar) measure on the environment. 
  This is the only stage that really requires coherent (entangled) operations across different Choi-states
  copies, and it does not involve any further calls to the unknown channel. For Kraus-rank-one channels, the Choi state is already pure, so this purification step is unnecessary.

 \item \textbf{Pure-state tomography and CPTP regularisation.}
We perform pure-state tomography on \(|\Phi_c\rangle^{\otimes N}\) using a \emph{covariant}, sample-optimal pure-state tomography algorithm \(\mathcal A\) (e.g.,~\cite{Hay98,GutaKahnKuengTropp2020}) on \(\mathbb C^{d_{\mathrm{tot}}}\), obtaining a purified estimate \(|\hat\Phi_c\rangle\).
Here ``covariant'' means that rotating the unknown state by a unitary simply rotates the output estimate by the same unitary, without changing its error distribution; formally, covariance is defined in Definition~\ref{de:covariant}.
Tracing out the purifying register yields an intermediate CP map \(\Lambda^{\mathrm{est}}\) with Choi operator
\begin{align}
  J(\Lambda^{\mathrm{est}})
  = \mathrm{tr}_E\bigl(|\hat\Phi_c\rangle\langle\hat\Phi_c|\bigr).
\end{align}
As a final step, we \emph{regularise} this intermediate estimate by projecting it (via an SDP) onto the convex set of CPTP maps in diamond norm, namely
\begin{align}
  \hat\Lambda
  \in
  \arg\min_{\Phi \in \mathsf{CPTP}}
  \|\Phi - \Lambda^{\mathrm{est}}\|_\diamond.
\end{align}
\end{itemize}

The pure-state tomography algorithm \(\mathcal A\) can be any covariant
scheme on \(\mathbb C^{d_{\mathrm{tot}}}\) with trace distance accuracy guarantees.
Let \(\varepsilon_{\mathrm{pure}}
= \varepsilon_{\mathrm{pure}}(N,\delta,d_{\mathrm{tot}})\) denote its
worst-case trace-distance accuracy parameter, in the sense that when
\(\mathcal A\) is run on \(N\) copies of any pure state in
\(\mathbb C^{d_{\mathrm{tot}}}\), it outputs an estimated pure state
with trace distance accuracy at most \(\varepsilon_{\mathrm{pure}}\) with probability
at least \(1-\delta\).
The restriction to covariant schemes is without loss of generality for worst-case
pure-state tomography: by Lemma~\ref{le:haar-twirl-covariant}, any pure-state
tomography scheme can be converted into a covariant one by applying a
Haar-random unitary to the unknown state and undoing the same unitary on the
output estimate, without worsening its worst-case performance. This covariance
property is useful because it makes the error direction isotropic, which is the
key input for modelling the relevant error component as Haar-random.

Our main structural result,
Theorem~\ref{th:main-diamond-tomography}, relates the performance of such
a pure-state scheme to the resulting channel estimator \(\hat\Lambda\)
produced by Algorithm~\ref{alg:diamond-tomography}. It shows that
\begin{align}
  \frac12\|\hat\Lambda - \Lambda\|_\diamond
  \;\le\;
  \left(
    2 + \sqrt{\frac{1}{k d_{\mathrm{out}}}\log\frac{4}{\delta}}
 \, \right)
  \bigl[4 + S_\delta(d_{\mathrm{tot}})\bigr]\,
  \varepsilon_{\mathrm{pure}},
  \label{eq:main-summary-bound}
\end{align}
with probability at least \(1-\delta\), where
\(S_\delta(d_{\mathrm{tot}})\) is an explicit subleading function in the
total dimension and failure probability analysed in the proof.
In particular, the diamond-distance error of channel tomography essentially
scales linearly with the underlying trace-distance error of pure-state
tomography.

One convenient choice for \(\mathcal A\) is a projected least-squares (PLS) scheme based on single-copy measurements~\cite{GutaKahnKuengTropp2020},
or Hayashi’s covariant coherent scheme~\cite{Hay98}. Both are sample-optimal, in the sense that achieving trace-distance error
\(\varepsilon_{\mathrm{pure}}\) with failure probability at most \(\delta\) requires
\begin{align}
  N \;=\; \Theta\!\left(\frac{d_{\mathrm{tot}}+\log(\delta^{-1})}{\varepsilon_{\mathrm{pure}}^{2}}\right),
\end{align}
which is optimal in its dependence on \(d_{\mathrm{tot}}\), $\varepsilon_{\mathrm{pure}}$ and \(\delta\); see, e.g.,~\cite{pelecanos2025mixedstatetomographyreduces}.
Inverting this relation for \(\varepsilon_{\mathrm{pure}}\) and substituting into \eqref{eq:main-summary-bound},
together with \(d_{\mathrm{tot}} =d_{\mathrm{in}} d_{\mathrm{out}}  k\),
yields our main upper bound on the query complexity of diamond-distance channel learning.

 In the regime where the target failure probability is not
\emph{extremely} small compared to the total dimension, i.e.\ \(\delta
  > 4\exp(-d_{\mathrm{tot}})\),
we show that \(S_\delta(d_{\mathrm{tot}})\) remains uniformly bounded and
the prefactor in \eqref{eq:main-summary-bound} can be simplified, leading
to the following statement.

\begin{theorem}[Main theorem]
\label{th:main-informal}
Let \(\Lambda\) be a channel with input dimension \(d_{\mathrm{in}}\),
output dimension \(d_{\mathrm{out}}\), and Kraus rank at most \(k\).
Fix \(\varepsilon \in (0,1)\) and \(\delta \in (0,1)\) satisfying
\(\delta > 4 \exp(-d_{\mathrm{in}} d_{\mathrm{out}} k)\).
Then there exists a non-adaptive tomography protocol
(Algorithm~\ref{alg:diamond-tomography}) that uses
\begin{align}
  N
  = O\left(
    \frac{
      d_{\mathrm{in}} d_{\mathrm{out}}  k
      + d_{\mathrm{in}}\log(1/\delta)
    }{\varepsilon^2}
  \right)
\end{align}
queries to \(\Lambda\) and outputs a classical description of a quantum channel
\(\widehat\Lambda\) satisfying \(\|\hat\Lambda - \Lambda\|_\diamond \le \varepsilon\) with
probability at least \(1-\delta\).
\end{theorem}

For example, if \(\mathcal A\) is instantiated with Hayashi’s optimal
covariant pure-state scheme~\cite{Hay98}, the bound with leading constant
explicit becomes
$N = 256\, d_{\mathrm{in}} d_{\mathrm{out}} k/\varepsilon^{2}
  + O\bigl(d_{\mathrm{in}} \log(1/\delta)/\varepsilon^{2}\bigr)$.
  
If one prefers a statement with no restriction on the failure probability \(\delta\), a standard
boosting argument (see Remark~\ref{re:boosting-delta}) yields a clean bound,
valid for all \(\delta \in (0,1)\), namely
\(N = O\bigl(d_{\mathrm{in}} d_{\mathrm{out}} k \log(1/\delta)/\varepsilon^{2}\bigr)\).
In particular, for any fixed constant failure probability, this
reduces to exactly the scaling claimed in Theorem~\ref{th:th1Main}.

Before giving the proof sketch, it is useful to isolate the standard obstruction
that makes a naive Choi-state tomography approach appear insufficient for achieving the
desired dimension scaling in diamond-norm channel tomography. If \(\widehat J\)
is an estimate of the Choi state \(J(\Lambda)\) satisfying
\(\|\widehat J-J(\Lambda)\|_1\le \varepsilon\), then the standard worst-case
Choi-to-diamond conversion gives only
\begin{align}
  \|\widehat\Lambda-\Lambda\|_\diamond
  \le
  d_{\mathrm{in}}\|\widehat J-J(\Lambda)\|_1
  \le
  d_{\mathrm{in}}\varepsilon .
\end{align}
Thus, a Choi trace-norm error \(\varepsilon\) may translate into a diamond-norm
error larger by a factor \(d_{\mathrm{in}}\), producing an unwanted dimension
overhead in the query complexity. The proof sketch below explains why our estimator is not governed by this
worst-case conversion: the pure-state tomography error direction is effectively
Haar-random, and the resulting concentration cancels the \(d_{\mathrm{in}}\)
factor in the induced diamond-norm error. 


\begin{proof}[Proof sketch of the upper bound]
A formal statement and full proof are given in the appendix
(see Theorem~\ref{th:main-diamond-tomography}); here we only summarise the main mechanism.
Conditioned on achieving trace-distance error at most
\(\varepsilon_{\mathrm{pure}}\) in the pure-state tomography step, the estimated
Choi-state purification can be written as
\begin{align}
  |\hat\Phi_c\rangle
  =
  \sqrt{1-\varepsilon_{\mathrm{pure}}^{2}}\,|\Phi_c\rangle
  +
  \varepsilon_{\mathrm{pure}}\,|\psi_{\mathrm{err}}\rangle,
\end{align}
where, thanks to the covariance of the pure-state tomography algorithm
(Lemma~\ref{le:hayashi-haar-error}),
\( |\psi_{\mathrm{err}}\rangle \) is a Haar-random unit vector on the
subspace orthogonal to the Choi-state purification \( |\Phi_c\rangle \). Intuitively, covariance forces the
error direction to be isotropic; up to the subleading corrections, we may therefore treat \( |\psi_{\mathrm{err}}\rangle \) as
Haar-random in \(\mathbb C^{d_{\mathrm{tot}}}\).

We now translate this decomposition into a statement about channels. Let
\(\hat J=\tr_E(|\hat\Phi_c\rangle\langle\hat\Phi_c|)\) be the estimated
normalised Choi operator, and let
\(J(\Lambda)=\tr_E(|\Phi_c\rangle\langle\Phi_c|)\) be the true one. Expanding
\(\hat J-J(\Lambda)\) using the decomposition above gives a true-channel block,
an error block
\(\tr_E(|\psi_{\mathrm{err}}\rangle\langle\psi_{\mathrm{err}}|)\), and two
off-diagonal cross terms. We apply the triangle inequality to this decomposition.
The true-channel block has diamond norm \(1\), since it is the Choi operator of
a valid channel, and the off-diagonal terms are controlled by the diamond
Cauchy--Schwarz inequality (Lemma~\ref{le:diamond-CS}) in terms of the
true-channel block and the Haar-error block. Thus the diamond-norm error is
reduced to the diamond norm of the Haar-error block. More precisely, writing
\(\|X\|_\diamond\) for the diamond norm of the map with normalised Choi operator
\(X\), we obtain
\begin{align}
  \|\Lambda^{\mathrm{est}}-\Lambda\|_\diamond
  =
  \|\hat J-J(\Lambda)\|_\diamond
  \le
  O(\varepsilon_{\mathrm{pure}})
  \left(
    1+
    \left\|
      \tr_E\bigl(|\psi_{\mathrm{err}}\rangle\langle\psi_{\mathrm{err}}|\bigr)
    \right\|_\diamond
  \right).
\end{align}
It remains to show that this last diamond norm is \(O(1)\). Since
\(\tr_E(|\psi_{\mathrm{err}}\rangle\langle\psi_{\mathrm{err}}|)\) is a positive
Choi operator, the SDP characterisation of the diamond norm for positive Choi
operators (Lemma~\ref{le:diamond-positive-choi}) gives that the diamond norm
reduces to the operator norm of the input marginal:
\begin{align}
  \left\|
    \tr_E\bigl(|\psi_{\mathrm{err}}\rangle\langle\psi_{\mathrm{err}}|\bigr)
  \right\|_\diamond
  =
  d_{\mathrm{in}}
  \left\|
    \tr_{\mathrm{out},E}
    \bigl(|\psi_{\mathrm{err}}\rangle\langle\psi_{\mathrm{err}}|\bigr)
  \right\|_\infty
  =
  d_{\mathrm{in}}\cdot O\!\left(\frac1{d_{\mathrm{in}}}\right)
  =
  O(1),
\end{align}
where in the second step we used concentration of reduced states of Haar-random
vectors (see Lemma~\ref{le:random-induced-opnorm}), with high probability.
This is the essential cancellation: the \(d_{\mathrm{in}}\) factor from the
diamond-norm SDP characterisation is compensated by the
\(O(1/d_{\mathrm{in}})\) operator-norm scaling of a Haar-random reduced state.
Substituting this into the previous bound yields
\(\|\Lambda^{\mathrm{est}}-\Lambda\|_\diamond=O(\varepsilon_{\mathrm{pure}})\).
The final CPTP regularisation changes the error by at most a constant factor.
Finally, sample-optimal pure-state tomography gives
\(\varepsilon_{\mathrm{pure}}=O(\sqrt{d_{\mathrm{tot}}/N})\). Since
\(d_{\mathrm{tot}}=d_{\mathrm{in}}d_{\mathrm{out}}k\), choosing
\(N=O(d_{\mathrm{in}}d_{\mathrm{out}}k/\varepsilon^2)\) gives the claimed
query-complexity scaling.
\end{proof}
This proof sketch also clarifies why the result does not follow directly from mixed state tomography alone. The known ingredients we import are the random purification channel~\cite{tang2025conjugatequerieshelp,pelecanos2025mixedstatetomographyreduces,girardi2025randompurificationchannelsimple} and pure-state tomography~\cite{Hay98,pelecanos2025mixedstatetomographyreduces}, while our new technical contribution comes from directly controlling the diamond-distance of the specific Choi-state estimator produced by these subroutines.

\subsection{Applications: states, isometries, and POVMs}
\label{sec:applications}
Quantum channels provide a common framework that includes, as special cases,
quantum states (via $d_{\mathrm{in}} = 1$), isometries (via $k = 1$), in
particular unitaries (via $d_{\mathrm{in}} = d_{\mathrm{out}}$ and $k = 1$),
and POVMs (for instance, binary POVMs via $d_{\mathrm{out}} = 2$ and enforcing diagonal output states).
Accordingly, Theorem~\ref{th:main-informal} and its consequences
simultaneously capture and extend several tomography results for these
primitives. In this subsection we summarise the resulting guarantees, with
detailed formulations deferred to Appendix~\ref{sec:applicationsAPP}.
For any fixed constant success probability, specialising
Theorem~\ref{th:main-informal} to each of these settings implies the following.
\begin{itemize}
  \item \textbf{States.}
  When $d_{\mathrm{in}} = 1$, a channel is a state-preparation map and the
  diamond norm reduces to the trace distance. In this case we recover the
  optimal scaling $N = \Theta(d k / \varepsilon^{2})$ for trace-distance
  tomography of a rank-$k$ state in dimension $d$.

  \item \textbf{Isometries and unitaries.}
  For Kraus-rank-one channels $\Lambda(\rho) = V \rho V^\dagger$ with
  $V : \mathbb C^{d_{\mathrm{in}}} \to \mathbb C^{d_{\mathrm{out}}}$ an
  isometry, we obtain (to the best of our knowledge) the first explicit
  tomography scheme with query-complexity guarantees
  $N = O(d_{\mathrm{in}} d_{\mathrm{out}} / \varepsilon^{2})$ for learning
  arbitrary isometries in diamond distance. A recent work~\cite{ISOMEyoshida2025quantumadvantagestorageretrieval} proves a lower bound $\widetilde{\Omega}(d_{\mathrm{in}}d_{\mathrm{out}}/\varepsilon^2)$ for generic isometries, so our obtained scaling is tight up to logarithmic factors. For
  unitary channels ($d_{\mathrm{in}} = d_{\mathrm{out}} = d$), combining our
  Choi-state-based learner with the boosting scheme of
  Ref.~\cite{HaahKothariODonnellTang2023} further yields
  $N = \Theta(d^{2} / \varepsilon)$, achieving Heisenberg scaling in the
  accuracy parameter while preserving the optimal $d^{2}$ dimensional
  dependence. We also remark that in the isometric/unitary case the Choi state
  is already pure, so the purification step in our general construction is not
  needed; one can instead apply directly a sample-optimal pure-state tomography
  algorithm such as the projected least-squares scheme of
  Ref.~\cite{GutaKahnKuengTropp2020}, which uses only single-copy measurements.
  In particular, no coherent operations between Choi-state copies are required in this setting -- the Choi states can be prepared and measured sequentially.

  \item \textbf{Measurements.}
  If we restrict to channels with output dimension $d_{\mathrm{out}} = 2$ whose
  outputs are classical (i.e.\ diagonal in a fixed basis), we recover exactly
  the channels induced by binary POVMs: each such channel is determined by a
  single POVM element $E$ via the map
  $\rho \mapsto \sum_{b \in \{0,1\}} \tr(M_b \rho)\ketbra{b}{b}$,
  with $M_0 = E$ and $M_1 = \mathbb I - E$. In this setting, the diamond
  distance between two such channels coincides (up to a factor of $2$) with the
  operator-norm distance between the corresponding POVM elements, so diamond-norm tomography of the channel is equivalent to operator-norm learning of the
  POVM elements. For binary POVMs on $\mathbb C^{d}$, we obtain
  $N = O(d^{2} / \varepsilon^{2})$ samples for $\varepsilon$-accurate
  operator-norm learning. 
\end{itemize}

Thus, our results place previously separate guarantees for
states~\cite{pelecanos2025mixedstatetomographyreduces,Haah_2017,OW16},
isometries~\cite{ISOMEyoshida2025quantumadvantagestorageretrieval},
unitaries~\cite{HaahKothariODonnellTang2023}, and
measurements~\cite{POVMnetzambrano2025fastquantummeasurementtomography,barberarodriguez2025boostingprojectivemethodsquantum} within
a single, unified channel-based framework.

 \subsection{Lower bounds for channel tomography}
\label{subsec:lower-bounds-main}

We now turn to the lower bounds, deferring the fully explicit statements,
proofs, and additional discussion to Appendix~\ref{sec:lower-bound}. Recall
that the Kraus rank of a channel must satisfy
\(k_{\min}\le k\le d_{\mathrm{in}}d_{\mathrm{out}}\), where
\(k_{\min}\coloneqq \lceil d_{\mathrm{in}}/d_{\mathrm{out}}\rceil\).

\begin{theorem}[Query lower bounds for channel tomography]
\label{thm:lower-bounds-main}
Any tomography protocol that, for every channel
\(\Lambda:\mathcal L(\mathbb C^{d_{\mathrm{in}}})\to
\mathcal L(\mathbb C^{d_{\mathrm{out}}})\) of Kraus rank at most \(k\), uses \(N\) queries to \(\Lambda\) and outputs an
estimator \(\widehat\Lambda\) satisfying
\begin{align}
    \|\widehat\Lambda-\Lambda\|_\diamond\le \varepsilon
\end{align}
with probability at least \(1-\delta\), must obey
\begin{align}
  N
  =
  \Omega\!\left(
    \frac{d_{\mathrm{in}}d_{\mathrm{out}}k}{\varepsilon^{\beta}}
    +
    \frac{m}{\varepsilon^{2}}\log\!\frac{m}{\delta}
  \right),
  \label{eq:combined-delta-lb-mainn0}
\end{align}
where \(\beta=\beta(d_{\mathrm{in}},d_{\mathrm{out}},k)\le1\) and
\(m=m(d_{\mathrm{in}},d_{\mathrm{out}},k)\in[0,d_{\mathrm{in}}]\)
are given explicitly in Theorem~\ref{prop:combined-lb-delta-general-k}.
In particular, \(m=0\) at
minimal Kraus rank \(k=k_{\min}\), \(m>0\) whenever \(k>k_{\min}\), and
\(m=d_{\mathrm{in}}\) whenever \(k\ge 2d_{\mathrm{in}}\).
\end{theorem}
In particular, the first term gives the dimension-dependent lower bound
$\Omega(d_{\mathrm{in}}d_{\mathrm{out}}k)$ at fixed constant accuracy.
The second term gives a $\Omega\left(\frac{1}{\varepsilon^{2}}\log\frac{1}{\delta}\right)$ contribution
whenever the Kraus rank is non-minimal.
At the minimal Kraus rank $k=k_{\min}$, this $\Omega(\varepsilon^{-2})$-contribution vanishes (i.e.\ $m=0$), which is
consistent with the fact that the class of Kraus-rank-$k_{\min}$ channels contains unitary channels, for
which Heisenberg-scaling $N=O(1/\varepsilon)$ is achievable in diamond distance; see, e.g.,
Ref.~\cite{HaahKothariODonnellTang2023}. Moreover, when $k \ge 2 d_{\mathrm{in}}$ we have $m=d_{\mathrm{in}}$
and \eqref{eq:combined-delta-lb-mainn0} reduces to
\begin{align}
  N
  \;=\;
  \Omega\!\left(
    \frac{d_{\mathrm{in}} d_{\mathrm{out}} k}{\varepsilon^{\beta}}
    \;+\;
    \frac{d_{\mathrm{in}}}{\varepsilon^{2}}\log\!\frac{d_{\mathrm{in}}}{\delta}
  \right).
  \label{eq:combined-delta-lb-main-simplified}
\end{align}

Theorem~\ref{thm:lower-bounds-main} matches the dimension dependence of our upper bound, and in the regime
$\delta > 4\exp(-d_{\mathrm{in}}d_{\mathrm{out}}k)$ and $k \ge 2 d_{\mathrm{in}}$ it also matches its
$\delta$-dependence. Indeed, in this regime our upper bound (Theorem~\ref{th:th1Mainapp}) can be written as
\begin{align}
  N \;=\; O\!\left(
    \frac{d_{\mathrm{in}} d_{\mathrm{out}} k}{\varepsilon^{2}}
    \;+\;
    \frac{d_{\mathrm{in}}}{\varepsilon^{2}}\log\!\frac{d_{\mathrm{in}}}{\delta}
  \right).
\end{align}
However, the exponent $\beta$ governing the $\varepsilon$-dependence in the first term of
Eq.~\eqref{eq:combined-delta-lb-main-simplified} is not tight in general. As the proof sketch below indicates, this exponent is tied to the particular proof strategy based on packing construction and channel-identification bound used. We expect that refined lower-bound constructions may improve the exponent $\beta$, plausibly matching our upper bound in broader parameter regimes; we leave this to future work. 

For fixed constant accuracy and constant success probability,
Theorem~\ref{thm:lower-bounds-main} implies the matching lower bound
\(N=\Omega(d_{\mathrm{in}}d_{\mathrm{out}}k)\). We moreover show in
Appendix~\ref{sec:lower-bound} that this dimension lower bound remains valid
even if the protocol is granted purified access to a fixed Stinespring dilation
of \(\Lambda\), with the environment register retained, and no definite causal
order is imposed among the \(N\) uses.

We now sketch the proof. Full details are given in
Appendix~\ref{sec:lower-bound}.
\begin{proof}[Proof sketch]
The two terms in \eqref{eq:combined-delta-lb-mainn0} come from two independent
reductions. 

We first prove the dimension-dependent term by reducing tomography to channel
identification. Let \(\{\Lambda_x\}_{x=1}^M\) be a family of Kraus-rank-\(k\)
channels satisfying
\(\|\Lambda_x-\Lambda_y\|_\diamond>2\varepsilon\) for all \(x\neq y\), i.e., a packing-net of channels.
Any tomography protocol that outputs an \(\varepsilon\)-accurate estimate can
then be used to identify which channel \(\Lambda_x\) was given. Thus tomography
is at least as hard as identifying an element of a large diamond-distance
packing.
For this identification task, we prove that any \(N\)-query protocol trying to
identify one of the \(M\) channels chosen uniformly at random has a success probability that satisfies
\begin{align}
  p_{\mathrm{succ}}
  \coloneqq
  \frac1M\sum_{x=1}^M
  \Pr[\text{the protocol outputs }x\mid \Lambda_x]
  \le
  \frac{1}{M}
  \binom{N+D-1}{D-1},
  \qquad
  D:=d_{\mathrm{in}}d_{\mathrm{out}}k .
\end{align}
Intuitively, the binomial factor is the effective symmetric-subspace dimension
accessible from \(N\) uses of a Kraus-rank-\(k\) channel, even allowing fully general quantum strategies. Hence constant success probability forces
\(M\lesssim \binom{N+D-1}{D-1}\). Using the standard estimate
\(\binom{N+D-1}{D-1}\lesssim (1+N/(D-1))^{D-1}\), this implies
\(N\gtrsim (D-1)M^{1/(D-1)}\).

It remains to construct a sufficiently large packing, which will fix $M$. We do this by packing
Stinespring isometries and quotienting out the environment-unitary gauge
freedom, since environment-unitarily equivalent Stinespring isometries define
the same channel. The Stinespring-isometry manifold has real dimension
\(2d_{\mathrm{in}} d_{\mathrm{out}} k-d_{\mathrm{in}}^2\), and quotienting by
environment unitaries removes \(k^2\) dimensions. Thus the effective dimension
is $2d_{\mathrm{in}} d_{\mathrm{out}} k-d_{\mathrm{in}}^2-k^2.$
Accounting for the relation between Stinespring distance and diamond distance
gives a packing of size
\(M\gtrsim (1/\sqrt{\varepsilon})^{
2d_{\mathrm{in}} d_{\mathrm{out}} k-d_{\mathrm{in}}^2-k^2}\).
Combining this with the identification bound yields
\begin{align}
  N
  =
  \Omega\!\left(\frac{D}{\varepsilon^\beta}\right),
  \qquad
  \beta
  =
  \frac{
    2d_{\mathrm{in}} d_{\mathrm{out}} k-d_{\mathrm{in}}^2-k^2
  }{
    2(D-1)
  },
\end{align}
where \(D=d_{\mathrm{in}}d_{\mathrm{out}}k\). This gives the first term in
\eqref{eq:combined-delta-lb-mainn0}; in particular, at fixed constant accuracy
it gives \(N=\Omega(d_{\mathrm{in}}d_{\mathrm{out}}k)\).
This \(\varepsilon\)-dependent exponent should not be interpreted as sharp. It
is a by-product of the particular packing route used here: the argument first
separates Stinespring isometries and then transfers this separation to channels
in diamond distance. This is well suited to capturing the dimension of the
channel family at constant accuracy, but not necessarily the optimal local
scaling in \(\varepsilon\).

The second term has a different, statistical origin. We restrict to a subfamily
of Kraus-rank-\(k\) channels that encodes the biases of \(m\) independent
classical coins. Learning such a channel to diamond-distance accuracy
\(\varepsilon\) allows one to estimate these coin biases to accuracy
\(O(\varepsilon)\). Standard coin-bias lower bounds therefore imply
\(N=\Omega((m/\varepsilon^2)\log(m/\delta))\). Here \(m\) is the number of
independent biases that can be embedded while respecting the Kraus-rank
constraint: \(m=0\) at minimal Kraus rank, \(m>0\) as soon as the Kraus rank is
non-minimal, and \(m=d_{\mathrm{in}}\) when \(k\ge 2d_{\mathrm{in}}\), since in
that regime there is enough Kraus-rank slack to encode one independent coin
bias for each input basis state.

Since a tomography protocol must succeed uniformly over all channels of Kraus
rank at most \(k\), both reductions apply. Combining them gives
\eqref{eq:combined-delta-lb-mainn0}.
\end{proof} 

While for general Kraus-rank-$k$ channels our proof method captures the optimal dependence on
$(d_{\mathrm{in}},d_{\mathrm{out}},k)$, it does not for the target
accuracy $\varepsilon$. However, for important special cases---notably pure-state tomography in trace
distance and unitary-channel tomography in diamond norm--- we can show the same framework combined with the corresponding
packing bounds recovers the optimal scaling in \emph{both} the dimension parameters and $\varepsilon$.

Finally, we use the same approach to obtain a dimension-matching lower bound for learning binary measurement channels in diamond distance (equivalently for learning the associated POVM element in operator norm), showing that our \(\Theta(d^2)\) query-complexity scaling is necessary and sufficient at constant accuracy. Similarly as above, we also prove a separate accuracy- and failure probability-dependent lower bound of \(\Omega(d\varepsilon^{-2}\log(d/\delta))\) for learning binary measurements.

\subsection{Discussion and open problems}
In this work we introduced a tomography scheme for learning unknown quantum
channels with rigorous recovery guarantees in diamond distance. Our main result
shows that any channel with input dimension \(d_{\mathrm{in}}\), output
dimension \(d_{\mathrm{out}}\), and Kraus rank at most \(k\) can be learned to
diamond-distance accuracy \(\varepsilon\) using
\(N=O(d_{\mathrm{in}}d_{\mathrm{out}}k/\varepsilon^{2})\) queries. We also prove
two complementary lower bounds: a matching
\(\Omega(d_{\mathrm{in}}d_{\mathrm{out}}k)\) lower bound at constant accuracy,
and a separate \(\Omega(1/\varepsilon^{2})\) lower-bound contribution for
channels with non-minimal Kraus rank. Thus, as summarized in
Table~\ref{tab:bounds-summary}, our results close the \emph{dimensional}
query-complexity gap for quantum process tomography in diamond distance.

Operationally, this identifies the experimental query cost of reconstructing
open-system dynamics in the strongest sense: after \(N\) uses of the unknown
process, the learner outputs a classical surrogate that can replace the true
channel in any future experiment, up to error
\(\varepsilon\). Our query-complexity bound shows that, at constant accuracy, this cost is
set by the intrinsic number of channel degrees of freedom. Thus the result
provides an information-theoretic reference point for how costly it is to learn
general noisy quantum dynamics, independently of any particular experimental
platform or reconstruction heuristic.

Our analysis also has implications on the role of adaptivity and coherence in quantum
channel learning~\cite{Oufkir_2023,LOWERBrosenthal2024quantumchanneltestingaveragecase,chen2023doesadaptivityhelpquantum,chen2021exponentialseparationslearningquantum}.
We show that adaptivity does not improve the asymptotic query complexity for process tomography at fixed constant diamond-distance accuracy: all uses of the unknown channel in our
protocol occur in a single parallel block. At the same time, coherence across
channel uses is provably helpful in this non-adaptive setting: Ref.~\cite{Oufkir_2023} shows that any fully
incoherent, non-adaptive strategy necessarily incurs asymptotically larger
query complexity. Our scheme sits precisely at this intermediate point, being
non-adaptive but coherent: it prepares many copies of the Choi state, applies a
coherent random purification map~\cite{tang2025conjugatequerieshelp,pelecanos2025mixedstatetomographyreduces},
and finally reduces the problem to sample-optimal pure-state tomography on
independent copies of a fixed purification.

Conceptually, our results sharpen a common folklore statement. While the
Choi--Jamio{\l}kowski isomorphism guarantees that a quantum channel is uniquely
determined by its Choi state, this equivalence does not automatically translate
into an \emph{efficient} learning strategy. Indeed, for several channel-learning
tasks (including tasks beyond tomography), it has been argued that approaches
based on preparing Choi states and then applying a direct analogue of the
corresponding state-estimation procedure can be suboptimal compared to
channel-tailored approaches~\cite{SurawyStepneyKahnKuengGuta2022,Kliesch_2021,Roth_2018,zhao2023learning,Knill_2008,Helsen_2022,Helsen_2023,wadhwa2024agnosticprocesstomography,Kiktenko_2020,Thinh_2019,Kliesch_2019,Flammia_2012,huang2023learningpredictarbitraryquantum,Kunjummen_2023,LOWERBrosenthal2024quantumchanneltestingaveragecase,fawzi2023quantumchannelcertificationincoherent,lang2025unifiedblockwisemeasurementdesign,Caro_2024}.
Our results show that this concern does not apply to diamond-distance channel
tomography: Choi-state estimation \emph{does} suffice to achieve
dimension-optimal learning guarantees in diamond distance.

Finally, by specialising our protocol to the corresponding settings, we recover
the optimal $\Theta(d k/\varepsilon^{2})$ sample-complexity scaling for
trace-distance tomography of rank-$k$ states and the optimal
$\Theta(d^{2}/\varepsilon)$ query-complexity scaling for diamond-distance
learning of unitary channels. Beyond these cases, our general bounds also yield
new guarantees: for Kraus-rank-one channels, we obtain (to the best of
our knowledge) the first diamond-norm tomography scheme for arbitrary
isometries with query complexity
$O(d_{\mathrm{in}} d_{\mathrm{out}}/\varepsilon^{2})$, with the additional
practical experimental advantage that it does not require the random purification channel and
therefore reduces to single-copy measurements on the Choi
state; and for tomography of quantum measurements, we derive the
dimension-optimal upper bound $O(d^{2}/\varepsilon^{2})$ for diamond-norm learning
of binary measurement channels.

Our results suggest several natural directions for further work. A first
challenge is to close the remaining gap between the upper and lower bounds for
learning channels in diamond distance. The bounds match in their dependence on
\((d_{\mathrm{in}},d_{\mathrm{out}},k)\), and in a broad parameter regime also
in their dependence on the failure probability \(\delta\), but they do not yet
match in the target accuracy \(\varepsilon\). For fixed input and output dimension, the optimal $\varepsilon$-scaling depends sensitively on the admissible Kraus rank: once $k$ exceeds the minimal value $k_{\min}$, our reduction already enforces an $\Omega(1/\varepsilon^{2})$ contribution, whereas at the minimal Kraus rank the class contains unitary channels, for which Heisenberg scaling $\Theta(1/\varepsilon)$ is achievable and a uniform $\Omega(1/\varepsilon^{2})$ lower bound cannot hold in general. Characterising the optimal
joint dependence on
\((\varepsilon,d_{\mathrm{in}},d_{\mathrm{out}},k)\), specifically determining
the sharp \(\varepsilon\)-exponent \(\beta\) in
Eq.\eqref{eq:combined-delta-lb-main-simplified}, remains an appealing open
problem. We expect that sharpening the lower bound in this direction would
establish the full optimality of our protocol in broad Kraus-rank regimes.


A second direction is to better understand the role of coherence in channel
tomography. In particular, it would be interesting to determine the optimal
query complexity under explicit restrictions on coherent processing across
channel uses---for example, when one is limited to a number \(t\) of adaptive and coherent queries to the unknown channel interleaved with arbitrary quantum operations, including collective
measurements across the $t$-uses. For state tomography,
such trade-offs between the number of copies and the available quantum memory
(coherent versus incoherent measurements) have been analysed in
Refs.~\cite{chen2024optimaltradeoffentanglementcopy,hu2024sampleoptimalmemoryefficient};
developing an analogous theory for diamond-distance channel learning remains
largely open. One natural possibility is to study blockwise versions of our
protocol, in which the coherent part is applied only within blocks of at most
\(t\) Choi-state copies, while different blocks are processed independently and
combined only through classical post-processing, in the spirit of
limited-entanglement measurement schemes for state
tomography~\cite{pelecanos2025mixedstatetomographyreduces,PSW25}.



\subsection*{Acknowledgements}
We are grateful to Aadil Oufkir and Ewin Tang for helpful early discussions on this problem. 
In particular, we thank Aadil Oufkir for pointing us to relevant lower-bound results~\cite{LOWERBrosenthal2024quantumchanneltestingaveragecase}, and Ewin Tang for suggesting to model the error state in the
Choi-state learning algorithm as Haar-random, in analogy with her work~\cite{HaahKothariODonnellTang2023}.

We also thank Amira Abbas, Matthias Caro, Mina Doosti, Jens Eisert, Marco Fanizza,
Vishnu Iyer, Lorenzo Leone, Laura Lewis, Francesco Anna Mele, Ryan O'Donnell,
and Chirag Wadhwa for stimulating discussions at the 
1st AIMS Workshop and School on ``The Theory of Quantum Learning Algorithms'',
hosted at the African Institute for Mathematical Sciences (AIMS), where this project was conceived.
We are especially grateful to Marco Fanizza for conversations that sparked our excitement about this problem.

A.A.M.\ and L.B.\ acknowledge funding from the BMFTR (PhoQuant, QPIC, Hybrid++, QR.N, PASQUOPS, QSolid),
the BMWK (EniQmA), the DFG (CRC 183), the EU Quantum Flagship (PasQuans2, Millenion),
the Munich Quantum Valley, Berlin Quantum, and the European Research Council (ERC, AdG DebuQC).
A.A.M.\ further acknowledges support from a 2025 Google PhD Fellowship.

\bibliography{ref}

@misc{STINdistkretschmann2007continuitytheoremstinespringsdilation,
      title={A Continuity Theorem for Stinespring's Dilation}, 
      author={Dennis Kretschmann and Dirk Schlingemann and Reinhard F. Werner},
      year={2007},
      eprint={0710.2495},
      archivePrefix={arXiv},
      primaryClass={quant-ph},
      url={https://arxiv.org/abs/0710.2495}, 
}

@article{Chiribella_2013,
   title={Quantum computations without definite causal structure},
   volume={88},
   ISSN={1094-1622},
   url={http://dx.doi.org/10.1103/PhysRevA.88.022318},
   DOI={10.1103/physreva.88.022318},
   number={2},
   journal={Physical Review A},
   publisher={American Physical Society (APS)},
   author={Chiribella, Giulio and D’Ariano, Giacomo Mauro and Perinotti, Paolo and Valiron, Benoit},
   year={2013},
   month=aug }

@article{Bavaresco_2022,
   title={Unitary channel discrimination beyond group structures: Advantages of sequential and indefinite-causal-order strategies},
   volume={63},
   ISSN={1089-7658},
   url={http://dx.doi.org/10.1063/5.0075919},
   DOI={10.1063/5.0075919},
   number={4},
   journal={Journal of Mathematical Physics},
   publisher={AIP Publishing},
   author={Bavaresco, Jessica and Murao, Mio and Quintino, Marco Túlio},
   year={2022},
   month=apr }

@article{Hoffreumon_2021,
   title={The Multi-round Process Matrix},
   volume={5},
   ISSN={2521-327X},
   url={http://dx.doi.org/10.22331/q-2021-01-20-384},
   DOI={10.22331/q-2021-01-20-384},
   journal={Quantum},
   publisher={Verein zur Forderung des Open Access Publizierens in den Quantenwissenschaften},
   author={Hoffreumon, Timothée and Oreshkov, Ognyan},
   year={2021},
   month=jan, pages={384} }

@article{Ara_jo_2015,
  title   = {Witnessing causal nonseparability},
  author  = {Ara{\'u}jo, Mateus and Branciard, Cyril and Costa, Fabio and Feix, Adrien and Giarmatzi, Christina and Brukner, {\v{C}}aslav},
  journal = {New Journal of Physics},
  volume  = {17},
  number  = {10},
  pages   = {102001},
  year    = {2015},
  month   = oct,
  doi     = {10.1088/1367-2630/17/10/102001},
  url     = {https://doi.org/10.1088/1367-2630/17/10/102001}
}

@misc{barberarodriguez2025boostingprojectivemethodsquantum,
      title={Boosting projective methods for quantum process and detector tomography}, 
      author={Júlia Barberà-Rodríguez and Leonardo Zambrano and Antonio Acín and Donato Farina},
      year={2025},
      eprint={2406.11646},
      archivePrefix={arXiv},
      primaryClass={quant-ph},
      url={https://arxiv.org/abs/2406.11646}, 
}

@misc{REALhinrichs2016entropynumbersembeddingsschatten,
      title={Entropy numbers of embeddings of Schatten classes}, 
      author={Aicke Hinrichs and Joscha Prochno and Jan Vybiral},
      year={2016},
      eprint={1612.08105},
      archivePrefix={arXiv},
      primaryClass={math.FA},
      url={https://arxiv.org/abs/1612.08105}, 
}

@inproceedings{Tsybakov2008IntroductionTN,
  title={Introduction to Nonparametric Estimation},
  author={A. Tsybakov},
  booktitle={Springer Series in Statistics},
  year={2008},
  url={https://api.semanticscholar.org/CorpusID:42933599}
}

@misc{girardi2025randomstinespringsuperchannelconverting,
      title={Random Stinespring superchannel: converting channel queries into dilation isometry queries}, 
      author={Filippo Girardi and Francesco Anna Mele and Haimeng Zhao and Marco Fanizza and Ludovico Lami},
      year={2025},
      eprint={2512.20599},
      archivePrefix={arXiv},
      primaryClass={quant-ph},
      url={https://arxiv.org/abs/2512.20599}, 
}

@misc{chen2025quantumchanneltomographyestimation,
      title={Quantum channel tomography and estimation by local test}, 
      author={Kean Chen and Nengkun Yu and Zhicheng Zhang},
      year={2025},
      eprint={2512.13614},
      archivePrefix={arXiv},
      primaryClass={quant-ph},
      url={https://arxiv.org/abs/2512.13614}, 
}

@article{Oreshkov_2012,
  title   = {Quantum correlations with no causal order},
  author  = {Oreshkov, Ognyan and Costa, Fabio and Brukner, {\v{C}}aslav},
  journal = {Nature Communications},
  volume  = {3},
  pages   = {1092},
  year    = {2012},
  month   = oct,
  doi     = {10.1038/ncomms2076},
  url     = {https://doi.org/10.1038/ncomms2076}
}

@book{Vershynin_2018, place={Cambridge}, series={Cambridge Series in Statistical and Probabilistic Mathematics}, title={High-Dimensional Probability: An Introduction with Applications in Data Science}, publisher={Cambridge University Press}, author={Vershynin, Roman}, year={2018}, collection={Cambridge Series in Statistical and Probabilistic Mathematics}}

@misc{NETHOMszarek1997metricentropyhomogeneousspaces,
      title={Metric Entropy of Homogeneous Spaces}, 
      author={Stanislaw J. Szarek},
      year={1997},
      eprint={math/9701213},
      archivePrefix={arXiv},
      primaryClass={math.MG},
      url={https://arxiv.org/abs/math/9701213}, 
}

@article{Chiribella_2009,
   title={Theoretical framework for quantum networks},
   volume={80},
   ISSN={1094-1622},
   url={http://dx.doi.org/10.1103/PhysRevA.80.022339},
   DOI={10.1103/physreva.80.022339},
   number={2},
   journal={Physical Review A},
   publisher={American Physical Society (APS)},
   author={Chiribella, Giulio and D’Ariano, Giacomo Mauro and Perinotti, Paolo},
   year={2009},
   month=aug }

@article{Chiribella_2008,
   title={Quantum Circuit Architecture},
   volume={101},
   ISSN={1079-7114},
   url={http://dx.doi.org/10.1103/PhysRevLett.101.060401},
   DOI={10.1103/physrevlett.101.060401},
   number={6},
   journal={Physical Review Letters},
   publisher={American Physical Society (APS)},
   author={Chiribella, G. and D’Ariano, G. M. and Perinotti, P.},
   year={2008},
   month=aug }

@article{EXP1,
  author  = {O'Brien, Jeremy L. and Pryde, Geoff J. and Gilchrist, Alexei and James, Daniel F. V. and Langford, Nathan K. and Ralph, Timothy C. and White, Andrew G.},
  title   = {Quantum Process Tomography of a Controlled-NOT Gate},
  journal = {Physical Review Letters},
  volume  = {93},
  number  = {8},
  pages   = {080502},
  year    = {2004},
  doi     = {10.1103/PhysRevLett.93.080502},
  url     = {https://doi.org/10.1103/PhysRevLett.93.080502}
}

@article{Yuen_2023,
   title={An Improved Sample Complexity Lower Bound for (Fidelity) Quantum State Tomography},
   volume={7},
   ISSN={2521-327X},
   url={http://dx.doi.org/10.22331/q-2023-01-03-890},
   DOI={10.22331/q-2023-01-03-890},
   journal={Quantum},
   publisher={Verein zur Forderung des Open Access Publizierens in den Quantenwissenschaften},
   author={Yuen, Henry},
   year={2023},
   month=jan, pages={890} }

@article{Fawzi_2025,
   title={Lower Bounds on Learning Pauli Channels With Individual Measurements},
   volume={71},
   ISSN={1557-9654},
   url={http://dx.doi.org/10.1109/TIT.2025.3527902},
   DOI={10.1109/tit.2025.3527902},
   number={4},
   journal={IEEE Transactions on Information Theory},
   publisher={Institute of Electrical and Electronics Engineers (IEEE)},
   author={Fawzi, Omar and Oufkir, Aadil and França, Daniel Stilck},
   year={2025},
   month=apr, pages={2642–2661} }

@misc{girardi2025randompurificationchannelsimple,
      title={Random purification channel made simple}, 
      author={Filippo Girardi and Francesco Anna Mele and Ludovico Lami},
      year={2025},
      eprint={2511.23451},
      archivePrefix={arXiv},
      primaryClass={quant-ph},
      url={https://arxiv.org/abs/2511.23451}, 
}

@misc{vasconcelos2025learningshallowquantumcircuits,
      title={Learning shallow quantum circuits with many-qubit gates}, 
      author={Francisca Vasconcelos and Hsin-Yuan Huang},
      year={2025},
      eprint={2410.16693},
      archivePrefix={arXiv},
      primaryClass={quant-ph},
      url={https://arxiv.org/abs/2410.16693}, 
}

@inproceedings{Huang_2024, series={STOC ’24},
   title={Learning Shallow Quantum Circuits},
   url={http://dx.doi.org/10.1145/3618260.3649722},
   DOI={10.1145/3618260.3649722},
   booktitle={Proceedings of the 56th Annual ACM Symposium on Theory of Computing},
   publisher={ACM},
   author={Huang, Hsin-Yuan and Liu, Yunchao and Broughton, Michael and Kim, Isaac and Anshu, Anurag and Landau, Zeph and McClean, Jarrod R.},
   year={2024},
   month=jun, pages={1343–1351},
   collection={STOC ’24} }

@misc{iyer2025mildlyinteractingfermionicunitariesefficiently,
      title={Mildly-Interacting Fermionic Unitaries are Efficiently Learnable}, 
      author={Vishnu Iyer},
      year={2025},
      eprint={2504.11318},
      archivePrefix={arXiv},
      primaryClass={quant-ph},
      url={https://arxiv.org/abs/2504.11318}, 
}

@misc{grewal2025queryoptimalestimationunitarychannels,
      title={Query-Optimal Estimation of Unitary Channels via Pauli Dimensionality}, 
      author={Sabee Grewal and Daniel Liang},
      year={2025},
      eprint={2510.00168},
      archivePrefix={arXiv},
      primaryClass={quant-ph},
      url={https://arxiv.org/abs/2510.00168}, 
}

@misc{chen2023doesadaptivityhelpquantum,
      title={When Does Adaptivity Help for Quantum State Learning?}, 
      author={Sitan Chen and Brice Huang and Jerry Li and Allen Liu and Mark Sellke},
      year={2023},
      eprint={2206.05265},
      archivePrefix={arXiv},
      primaryClass={quant-ph},
      url={https://arxiv.org/abs/2206.05265}, 
}

@article{EXP2,
  author  = {Riebe, M. and Kim, K. and Schindler, P. and Monz, T. and Schmidt, P. O. and K{\"o}rber, T. K. and H{\"a}nsel, W. and H{\"a}ffner, H. and Roos, C. F. and Blatt, R.},
  title   = {Process Tomography of Ion Trap Quantum Gates},
  journal = {Physical Review Letters},
  volume  = {97},
  number  = {22},
  pages   = {220407},
  year    = {2006},
  doi     = {10.1103/PhysRevLett.97.220407},
  url     = {https://doi.org/10.1103/PhysRevLett.97.220407}
}

@article{EXP3,
  author  = {Bialczak, R. C. and Ansmann, M. and Hofheinz, M. and Lucero, E. and Neeley, M. and O'Connell, A. D. and Sank, D. and Wang, H. and Wenner, J. and Steffen, M. and Cleland, A. N. and Martinis, John M.},
  title   = {Quantum Process Tomography of a Universal Entangling Gate Implemented with Josephson Phase Qubits},
  journal = {Nature Physics},
  volume  = {6},
  pages   = {409--413},
  year    = {2010},
  doi     = {10.1038/nphys1639},
  url     = {https://doi.org/10.1038/nphys1639}
}

@article{EXP4,
  author  = {Ballance, C. J. and Harty, T. P. and Linke, N. M. and Sepiol, M. A. and Lucas, D. M.},
  title   = {High-Fidelity Quantum Logic Gates Using Trapped-Ion Hyperfine Qubits},
  journal = {Physical Review Letters},
  volume  = {117},
  number  = {6},
  pages   = {060504},
  year    = {2016},
  doi     = {10.1103/PhysRevLett.117.060504},
  url     = {https://doi.org/10.1103/PhysRevLett.117.060504}
}

@article{EXP5,
   title={Quantum process tomography of a high-dimensional quantum communication channel},
   volume={3},
   ISSN={2521-327X},
   url={http://dx.doi.org/10.22331/q-2019-05-06-138},
   DOI={10.22331/q-2019-05-06-138},
   journal={Quantum},
   publisher={Verein zur Forderung des Open Access Publizierens in den Quantenwissenschaften},
   author={Bouchard, Frédéric and Hufnagel, Felix and Koutný, Dominik and Abbas, Aazad and Sit, Alicia and Heshami, Khabat and Fickler, Robert and Karimi, Ebrahim},
   year={2019},
   month=may, pages={138} }

@article{Hay98,
   title={Asymptotic estimation theory for a finite-dimensional pure state model},
   volume={31},
   ISSN={1361-6447},
   url={http://dx.doi.org/10.1088/0305-4470/31/20/006},
   DOI={10.1088/0305-4470/31/20/006},
   number={20},
   journal={Journal of Physics A: Mathematical and General},
   publisher={IOP Publishing},
   author={Hayashi, Masahito},
   year={1998},
   month=may, pages={4633–4655} }

@article{HayashiHashimotoHoribe2005,
  author        = {Akihisa Hayashi and Takanori Hashimoto and Masahito Horibe},
  title         = {Reexamination of optimal quantum state estimation of pure states},
  journal       = {Physical Review A},
  volume        = {72},
  pages         = {032325},
  year          = {2005},
  doi           = {10.1103/PhysRevA.72.032325},
  url           = {https://doi.org/10.1103/PhysRevA.72.032325},
  eprint        = {quant-ph/0410207},
  archivePrefix = {arXiv},
  primaryClass  = {quant-ph}
}

@book{Holevo1982,
  author    = {Alexander S. Holevo},
  title     = {Probabilistic and Statistical Aspects of Quantum Theory},
  publisher = {North-Holland},
  address   = {Amsterdam},
  year      = {1982},
  series    = {North-Holland Series in Statistics and Probability},
  isbn      = {0444863338}
}

@article{Popescu_2006,
   title={Entanglement and the foundations of statistical mechanics},
   volume={2},
   ISSN={1745-2481},
   url={http://dx.doi.org/10.1038/nphys444},
   DOI={10.1038/nphys444},
   number={11},
   journal={Nature Physics},
   publisher={Springer Science and Business Media LLC},
   author={Popescu, Sandu and Short, Anthony J. and Winter, Andreas},
   year={2006},
   month=oct, pages={754–758} }

@article{Hayden_2004,
   title={Randomizing Quantum States: Constructions and Applications},
   volume={250},
   ISSN={1432-0916},
   url={http://dx.doi.org/10.1007/s00220-004-1087-6},
   DOI={10.1007/s00220-004-1087-6},
   number={2},
   journal={Communications in Mathematical Physics},
   publisher={Springer Science and Business Media LLC},
   author={Hayden, Patrick and Leung, Debbie and Shor, Peter W. and Winter, Andreas},
   year={2004},
   month=jul, pages={371–391} }

@book{JohnsonKotzBalakrishnan2000,
  author    = {Norman L. Johnson and Samuel Kotz and N. Balakrishnan},
  title     = {Continuous Multivariate Distributions. Volume 1: Models and Applications},
  edition   = {2},
  year      = {2000},
  publisher = {John Wiley \& Sons},
  series    = {Wiley Series in Probability and Statistics},
  isbn      = {0-471-18387-3}
}

@misc{DirichletWiki,
  title        = {Dirichlet distribution --- {W}ikipedia{,} The Free Encyclopedia},
  howpublished = {\url{https://en.wikipedia.org/wiki/Dirichlet_distribution}},
  note         = {Accessed: 2025-12-03}
}

@book{AubrunSzarek2017AliceBob,
  author    = {Aubrun, Guillaume and Szarek, Stanis{\l}aw J.},
  title     = {Alice and Bob Meet Banach: The Interface of Asymptotic Geometric Analysis and Quantum Information Theory},
  series    = {Mathematical Surveys and Monographs},
  volume    = {223},
  publisher = {American Mathematical Society},
  address   = {Providence, RI},
  year      = {2017},
}

@misc{tang2025conjugatequerieshelp,
      title={Conjugate queries can help}, 
      author={Ewin Tang and John Wright and Mark Zhandry},
      year={2025},
      eprint={2510.07622},
      archivePrefix={arXiv},
      primaryClass={quant-ph},
      url={https://arxiv.org/abs/2510.07622}, 
}

@book{Helstrom1976,
  author    = {Carl W. Helstrom},
  title     = {Quantum Detection and Estimation Theory},
  series    = {Mathematics in Science and Engineering},
  volume    = {123},
  publisher = {Academic Press},
  address   = {New York},
  year      = {1976},
  isbn      = {0123400503}
}

@misc{OW16,
      title={Efficient quantum tomography}, 
      author={Ryan O'Donnell and John Wright},
      year={2015},
      eprint={1508.01907},
      archivePrefix={arXiv},
      primaryClass={quant-ph},
      url={https://arxiv.org/abs/1508.01907}, 
}

@article{KRT17,
  author  = {Richard Kueng and Holger Rauhut and Ulrich Terstiege},
  title   = {Low rank matrix recovery from rank one measurements},
  journal = {Applied and Computational Harmonic Analysis},
  volume  = {42},
  number  = {1},
  pages   = {88--116},
  year    = {2017},
  doi     = {10.1016/j.acha.2015.07.007},
}

@misc{Har13,
  author       = {Aram W. Harrow},
  title        = {The Church of the Symmetric Subspace},
  howpublished = {arXiv preprint},
  year         = {2013},
  eprint       = {1308.6595},
  archivePrefix = {arXiv},
}

@misc{PSW25,
  author       = {Angelos Pelecanos and Jack Spilecki and John Wright},
  title        = {The debiased Keyl's algorithm: a new unbiased estimator for full state tomography},
  howpublished = {arXiv preprint},
  year         = {2025},
  eprint       = {2510.07788},
  archivePrefix = {arXiv},
}

@misc{SSW25,
  author       = {Thilo Scharnhorst and Jack Spilecki and John Wright},
  title        = {Optimal lower bounds for quantum state tomography},
  howpublished = {arXiv preprint},
  year         = {2025},
  eprint       = {2510.07699},
  archivePrefix = {arXiv},
}

@book{SkrzypczykCavalcantiSDP,
  author    = {Paul Skrzypczyk and Daniel Cavalcanti},
  title     = {Semidefinite Programming in Quantum Information Science},
  series    = {IOP Series in Quantum Technology},
  publisher = {IOP Publishing},
  address   = {Bristol},
  year      = {2023},
  isbn      = {978-0-7503-3341-2},
  doi       = {10.1088/978-0-7503-3343-6},
}

@article{Watrous2013CBnorm,
  author  = {John Watrous},
  title   = {Simpler Semidefinite Programs for Completely Bounded Norms},
  journal = {Chicago Journal of Theoretical Computer Science},
  volume  = {2013},
  number  = {8},
  year    = {2013},
  doi     = {10.4086/cjtcs.2013.008},
  eprint  = {1207.5726},
  archivePrefix = {arXiv},
  primaryClass  = {quant-ph},
}

@article{ChuangNielsen1997,
   title={Prescription for experimental determination of the dynamics of a quantum black box},
   volume={44},
   ISSN={1362-3044},
   url={http://dx.doi.org/10.1080/09500349708231894},
   DOI={10.1080/09500349708231894},
   number={11–12},
   journal={Journal of Modern Optics},
   publisher={Informa UK Limited},
   author={Chuang, Isaac L. and Nielsen, M. A.},
   year={1997},
   month=nov, pages={2455–2467} }

@article{PoyatosCiracZoller1997,
  title        = {Complete characterization of a quantum process: The two-bit quantum gate},
  author       = {Poyatos, J. F. and Cirac, J. I. and Zoller, P.},
  journal      = {Physical Review Letters},
  volume       = {78},
  number       = {2},
  pages        = {390--393},
  year         = {1997},
  doi          = {10.1103/PhysRevLett.78.390}
}

@article{MohseniRezakhaniLidar2008,
  title        = {Quantum-process tomography: Resource analysis of different strategies},
  author       = {Mohseni, M. and Rezakhani, A. T. and Lidar, D. A.},
  journal      = {Physical Review A},
  volume       = {77},
  number       = {3},
  pages        = {032322},
  year         = {2008},
  doi          = {10.1103/PhysRevA.77.032322}
}

@article{Kitaev1997,
  title={Quantum computations: algorithms and error correction},
  author={A Yu Kitaev},
  journal={Russian Mathematical Surveys},
  year={1997},
  volume={52},
  pages={1191 - 1249},
  url={https://api.semanticscholar.org/CorpusID:250816585}
}

@article{GilchristLangfordNielsen2005,
   title={Distance measures to compare real and ideal quantum processes},
   volume={71},
   ISSN={1094-1622},
   url={http://dx.doi.org/10.1103/PhysRevA.71.062310},
   DOI={10.1103/physreva.71.062310},
   number={6},
   journal={Physical Review A},
   publisher={American Physical Society (APS)},
   author={Gilchrist, Alexei and Langford, Nathan K. and Nielsen, Michael A.},
   year={2005},
   month=jun }

@book{WatrousBook, place={Cambridge}, title={The Theory of Quantum Information}, publisher={Cambridge University Press}, author={Watrous, John}, year={2018}}

@misc{ISOMEyoshida2025quantumadvantagestorageretrieval,
      title={Quantum Advantage in Storage and Retrieval of Isometry Channels}, 
      author={Satoshi Yoshida and Jisho Miyazaki and Mio Murao},
      year={2025},
      eprint={2507.10784},
      archivePrefix={arXiv},
      primaryClass={quant-ph},
      url={https://arxiv.org/abs/2507.10784}, 
}

@misc{fawzi2023quantumchannelcertificationincoherent,
      title={Quantum Channel Certification with Incoherent Strategies}, 
      author={Omar Fawzi and Nicolas Flammarion and Aurélien Garivier and Aadil Oufkir},
      year={2023},
      eprint={2303.01188},
      archivePrefix={arXiv},
      primaryClass={quant-ph},
      url={https://arxiv.org/abs/2303.01188}, 
}

@article{Chen_2024,
   title={Tight Bounds on Pauli Channel Learning without Entanglement},
   volume={132},
   ISSN={1079-7114},
   url={http://dx.doi.org/10.1103/PhysRevLett.132.180805},
   DOI={10.1103/physrevlett.132.180805},
   number={18},
   journal={Physical Review Letters},
   publisher={American Physical Society (APS)},
   author={Chen, Senrui and Oh, Changhun and Zhou, Sisi and Huang, Hsin-Yuan and Jiang, Liang},
   year={2024},
   month=may }

@misc{chen2025efficientselfconsistentlearninggate,
      title={Efficient self-consistent learning of gate set Pauli noise}, 
      author={Senrui Chen and Zhihan Zhang and Liang Jiang and Steven T. Flammia},
      year={2025},
      eprint={2410.03906},
      archivePrefix={arXiv},
      primaryClass={quant-ph},
      url={https://arxiv.org/abs/2410.03906}, 
}

@misc{girish2023poweradaptivityquantumquery,
      title={The Power of Adaptivity in Quantum Query Algorithms}, 
      author={Uma Girish and Makrand Sinha and Avishay Tal and Kewen Wu},
      year={2023},
      eprint={2311.16057},
      archivePrefix={arXiv},
      primaryClass={quant-ph},
      url={https://arxiv.org/abs/2311.16057}, 
}

@article{Salek_2022,
   title={Usefulness of adaptive strategies in asymptotic quantum channel discrimination},
   volume={105},
   ISSN={2469-9934},
   url={http://dx.doi.org/10.1103/PhysRevA.105.022419},
   DOI={10.1103/physreva.105.022419},
   number={2},
   journal={Physical Review A},
   publisher={American Physical Society (APS)},
   author={Salek, Farzin and Hayashi, Masahito and Winter, Andreas},
   year={2022},
   month=feb }

@misc{chen2024optimaltradeoffentanglementcopy,
      title={An optimal tradeoff between entanglement and copy complexity for state tomography}, 
      author={Sitan Chen and Jerry Li and Allen Liu},
      year={2024},
      eprint={2402.16353},
      archivePrefix={arXiv},
      primaryClass={quant-ph},
      url={https://arxiv.org/abs/2402.16353}, 
}

@misc{hu2024sampleoptimalmemoryefficient,
      title={Sample Optimal and Memory Efficient Quantum State Tomography}, 
      author={Yanglin Hu and Enrique Cervero-Martín and Elias Theil and Laura Mančinska and Marco Tomamichel},
      year={2024},
      eprint={2410.16220},
      archivePrefix={arXiv},
      primaryClass={quant-ph},
      url={https://arxiv.org/abs/2410.16220}, 
}

@misc{cervero2023weakschursamplinglogarithmic,
      title={Weak Schur sampling with logarithmic quantum memory}, 
      author={Enrique Cervero and Laura Mančinska},
      year={2023},
      eprint={2309.11947},
      archivePrefix={arXiv},
      primaryClass={quant-ph},
      url={https://arxiv.org/abs/2309.11947}, 
}

@misc{fanizza2025learningquantumprocessesinput,
      title={Learning quantum processes without input control}, 
      author={Marco Fanizza and Yihui Quek and Matteo Rosati},
      year={2025},
      eprint={2211.05005},
      archivePrefix={arXiv},
      primaryClass={quant-ph},
      url={https://arxiv.org/abs/2211.05005}, 
}

@misc{cheng2015learnabilityunknownquantummeasurements,
      title={The Learnability of Unknown Quantum Measurements}, 
      author={Hao-Chung Cheng and Min-Hsiu Hsieh and Ping-Cheng Yeh},
      year={2015},
      eprint={1501.00559},
      archivePrefix={arXiv},
      primaryClass={quant-ph},
      url={https://arxiv.org/abs/1501.00559}, 
}

@misc{aharonov1998quantumcircuitsmixedstates,
      title={Quantum Circuits with Mixed States}, 
      author={Dorit Aharonov and Alexei Kitaev and Noam Nisan},
      year={1998},
      eprint={quant-ph/9806029},
      archivePrefix={arXiv},
      primaryClass={quant-ph},
      url={https://arxiv.org/abs/quant-ph/9806029}, 
}

@article{fawzi:hal-04107265,
  TITLE = {{On Adaptivity in Quantum Testing}},
  AUTHOR = {Fawzi, Omar and Flammarion, Nicolas and Garivier, Aur{\'e}lien and Oufkir, Aadil},
  URL = {https://hal.science/hal-04107265},
  JOURNAL = {{Transactions on Machine Learning Research Journal}},
  PUBLISHER = {{[Amherst Massachusetts]: OpenReview.net, 2022}},
  PAGES = {1-33},
  YEAR = {2023},
  MONTH = Sep,
  HAL_ID = {hal-04107265},
  HAL_VERSION = {v1},
}

@misc{chen2024adaptivityhelpexponentiallyshadow,
      title={Adaptivity can help exponentially for shadow tomography}, 
      author={Sitan Chen and Weiyuan Gong and Zhihan Zhang},
      year={2024},
      eprint={2412.19022},
      archivePrefix={arXiv},
      primaryClass={quant-ph},
      url={https://arxiv.org/abs/2412.19022}, 
}

@article{Keith_2018,
   title={Joint quantum-state and measurement tomography with incomplete measurements},
   volume={98},
   ISSN={2469-9934},
   url={http://dx.doi.org/10.1103/PhysRevA.98.042318},
   DOI={10.1103/physreva.98.042318},
   number={4},
   journal={Physical Review A},
   publisher={American Physical Society (APS)},
   author={Keith, Adam C. and Baldwin, Charles H. and Glancy, Scott and Knill, E.},
   year={2018},
   month=oct }

@article{Cattaneo_2023,
   title={Self-consistent quantum measurement tomography based on semidefinite programming},
   volume={5},
   ISSN={2643-1564},
   url={http://dx.doi.org/10.1103/PhysRevResearch.5.033154},
   DOI={10.1103/physrevresearch.5.033154},
   number={3},
   journal={Physical Review Research},
   publisher={American Physical Society (APS)},
   author={Cattaneo, Marco and Rossi, Matteo A. C. and Korhonen, Keijo and Borrelli, Elsi-Mari and García-Pérez, Guillermo and Zimborás, Zoltán and Cavalcanti, Daniel},
   year={2023},
   month=sep }

@misc{li2024quantumnetworktomographylearning,
      title={Quantum Network Tomography via Learning Isometries on Stiefel Manifold}, 
      author={Ze-Tong Li and Xin-Lin He and Cong-Cong Zheng and Yu-Qian Dong and Tian Luan and Xu-Tao Yu and Zai-Chen Zhang},
      year={2024},
      eprint={2404.06988},
      archivePrefix={arXiv},
      primaryClass={quant-ph},
      url={https://arxiv.org/abs/2404.06988}, 
}

@article{Chen_2023,
   title={The learnability of Pauli noise},
   volume={14},
   ISSN={2041-1723},
   url={http://dx.doi.org/10.1038/s41467-022-35759-4},
   DOI={10.1038/s41467-022-35759-4},
   number={1},
   journal={Nature Communications},
   publisher={Springer Science and Business Media LLC},
   author={Chen, Senrui and Liu, Yunchao and Otten, Matthew and Seif, Alireza and Fefferman, Bill and Jiang, Liang},
   year={2023},
   month=jan }

@misc{lang2025unifiedblockwisemeasurementdesign,
      title={A Unified Blockwise Measurement Design for Learning Quantum Channels and Lindbladians via Low-Rank Matrix Sensing}, 
      author={Quanjun Lang and Jianfeng Lu},
      year={2025},
      eprint={2501.14080},
      archivePrefix={arXiv},
      primaryClass={quant-ph},
      url={https://arxiv.org/abs/2501.14080}, 
}

@article{Yang_2020,
   title={Optimal Universal Programming of Unitary Gates},
   volume={125},
   ISSN={1079-7114},
   url={http://dx.doi.org/10.1103/PhysRevLett.125.210501},
   DOI={10.1103/physrevlett.125.210501},
   number={21},
   journal={Physical Review Letters},
   publisher={American Physical Society (APS)},
   author={Yang, Yuxiang and Renner, Renato and Chiribella, Giulio},
   year={2020},
   month=nov }

@article{Kunjummen_2023,
   title={Shadow process tomography of quantum channels},
   volume={107},
   ISSN={2469-9934},
   url={http://dx.doi.org/10.1103/PhysRevA.107.042403},
   DOI={10.1103/physreva.107.042403},
   number={4},
   journal={Physical Review A},
   publisher={American Physical Society (APS)},
   author={Kunjummen, Jonathan and Tran, Minh C. and Carney, Daniel and Taylor, Jacob M.},
   year={2023},
   month=apr }

@article{Bisio_2010,
   title={Optimal quantum learning of a unitary transformation},
   volume={81},
   ISSN={1094-1622},
   url={http://dx.doi.org/10.1103/PhysRevA.81.032324},
   DOI={10.1103/physreva.81.032324},
   number={3},
   journal={Physical Review A},
   publisher={American Physical Society (APS)},
   author={Bisio, Alessandro and Chiribella, Giulio and D’Ariano, Giacomo Mauro and Facchini, Stefano and Perinotti, Paolo},
   year={2010},
   month=mar }

@article{Oszmaniec_2022,
   title={Fermion Sampling: A Robust Quantum Computational Advantage Scheme Using Fermionic Linear Optics and Magic Input States},
   volume={3},
   ISSN={2691-3399},
   url={http://dx.doi.org/10.1103/PRXQuantum.3.020328},
   DOI={10.1103/prxquantum.3.020328},
   number={2},
   journal={PRX Quantum},
   publisher={American Physical Society (APS)},
   author={Oszmaniec, Michał and Dangniam, Ninnat and Morales, Mauro E.S. and Zimborás, Zoltán},
   year={2022},
   month=may }

@misc{huang2023learningpredictarbitraryquantum,
      title={Learning to predict arbitrary quantum processes}, 
      author={Hsin-Yuan Huang and Sitan Chen and John Preskill},
      year={2023},
      eprint={2210.14894},
      archivePrefix={arXiv},
      primaryClass={quant-ph},
      url={https://arxiv.org/abs/2210.14894}, 
}

@article{Caro_2020,
   title={Pseudo-dimension of quantum circuits},
   volume={2},
   ISSN={2524-4914},
   url={http://dx.doi.org/10.1007/s42484-020-00027-5},
   DOI={10.1007/s42484-020-00027-5},
   number={2},
   journal={Quantum Machine Intelligence},
   publisher={Springer Science and Business Media LLC},
   author={Caro, Matthias C. and Datta, Ishaun},
   year={2020},
   month=nov }

@misc{raza2024onlinelearningquantumprocesses,
      title={Online learning of quantum processes}, 
      author={Asad Raza and Matthias C. Caro and Jens Eisert and Sumeet Khatri},
      year={2024},
      eprint={2406.04250},
      archivePrefix={arXiv},
      primaryClass={quant-ph},
      url={https://arxiv.org/abs/2406.04250}, 
}

@article{Caro_2024,
   title={Learning Quantum Processes and Hamiltonians via the Pauli Transfer Matrix},
   volume={5},
   ISSN={2643-6817},
   url={http://dx.doi.org/10.1145/3670418},
   DOI={10.1145/3670418},
   number={2},
   journal={ACM Transactions on Quantum Computing},
   publisher={Association for Computing Machinery (ACM)},
   author={Caro, Matthias C.},
   year={2024},
   month=jun, pages={1–53} }

@article{Wadhwa_2025,
   title={Learning Quantum Processes with Quantum Statistical Queries},
   volume={9},
   ISSN={2521-327X},
   url={http://dx.doi.org/10.22331/q-2025-05-12-1739},
   DOI={10.22331/q-2025-05-12-1739},
   journal={Quantum},
   publisher={Verein zur Forderung des Open Access Publizierens in den Quantenwissenschaften},
   author={Wadhwa, Chirag and Doosti, Mina},
   year={2025},
   month=may, pages={1739} }

@article{Flammia_2012,
   title={Quantum tomography via compressed sensing: error bounds, sample complexity and efficient estimators},
   volume={14},
   ISSN={1367-2630},
   url={http://dx.doi.org/10.1088/1367-2630/14/9/095022},
   DOI={10.1088/1367-2630/14/9/095022},
   number={9},
   journal={New Journal of Physics},
   publisher={IOP Publishing},
   author={Flammia, Steven T and Gross, David and Liu, Yi-Kai and Eisert, Jens},
   year={2012},
   month=sep, pages={095022} }

@misc{chen2021exponentialseparationslearningquantum,
      title={Exponential separations between learning with and without quantum memory}, 
      author={Sitan Chen and Jordan Cotler and Hsin-Yuan Huang and Jerry Li},
      year={2021},
      eprint={2111.05881},
      archivePrefix={arXiv},
      primaryClass={quant-ph},
      url={https://arxiv.org/abs/2111.05881}, 
}

@article{Trinh_2025,
   title={Adaptivity is not helpful for Pauli channel learning},
   volume={9},
   ISSN={2521-327X},
   url={http://dx.doi.org/10.22331/q-2025-09-24-1864},
   DOI={10.22331/q-2025-09-24-1864},
   journal={Quantum},
   publisher={Verein zur Forderung des Open Access Publizierens in den Quantenwissenschaften},
   author={Trinh, Xuan Du and Yu, Nengkun},
   year={2025},
   month=sep, pages={1864} }

@article{Kliesch_2019,
   title={Guaranteed recovery of quantum processes from few measurements},
   volume={3},
   ISSN={2521-327X},
   url={http://dx.doi.org/10.22331/q-2019-08-12-171},
   DOI={10.22331/q-2019-08-12-171},
   journal={Quantum},
   publisher={Verein zur Forderung des Open Access Publizierens in den Quantenwissenschaften},
   author={Kliesch, Martin and Kueng, Richard and Eisert, Jens and Gross, David},
   year={2019},
   month=aug, pages={171} }

@article{Thinh_2019,
   title={Practical and reliable error bars for quantum process tomography},
   volume={99},
   ISSN={2469-9934},
   url={http://dx.doi.org/10.1103/PhysRevA.99.052311},
   DOI={10.1103/physreva.99.052311},
   number={5},
   journal={Physical Review A},
   publisher={American Physical Society (APS)},
   author={Thinh, Le Phuc and Faist, Philippe and Helsen, Jonas and Elkouss, David and Wehner, Stephanie},
   year={2019},
   month=may }

@article{Kiktenko_2020,
   title={Estimating the precision for quantum process tomography},
   volume={59},
   ISSN={0091-3286},
   url={http://dx.doi.org/10.1117/1.OE.59.6.061614},
   DOI={10.1117/1.oe.59.6.061614},
   number={06},
   journal={Optical Engineering},
   publisher={SPIE-Intl Soc Optical Eng},
   author={Kiktenko, Evgeniy O. and Kublikova, Daria N. and Fedorov, Aleksey K.},
   year={2020},
   month=jan, pages={1} }

@article{Knee_2018,
   title={Quantum process tomography via completely positive and trace-preserving projection},
   volume={98},
   ISSN={2469-9934},
   url={http://dx.doi.org/10.1103/PhysRevA.98.062336},
   DOI={10.1103/physreva.98.062336},
   number={6},
   journal={Physical Review A},
   publisher={American Physical Society (APS)},
   author={Knee, George C. and Bolduc, Eliot and Leach, Jonathan and Gauger, Erik M.},
   year={2018},
   month=dec }

@article{Ahmed_2023,
   title={Gradient-Descent Quantum Process Tomography by Learning Kraus Operators},
   volume={130},
   ISSN={1079-7114},
   url={http://dx.doi.org/10.1103/PhysRevLett.130.150402},
   DOI={10.1103/physrevlett.130.150402},
   number={15},
   journal={Physical Review Letters},
   publisher={American Physical Society (APS)},
   author={Ahmed, Shahnawaz and Quijandría, Fernando and Kockum, Anton Frisk},
   year={2023},
   month=apr }

@article{Helsen_2023,
   title={Shadow estimation of gate-set properties from random sequences},
   volume={14},
   ISSN={2041-1723},
   url={http://dx.doi.org/10.1038/s41467-023-39382-9},
   DOI={10.1038/s41467-023-39382-9},
   number={1},
   journal={Nature Communications},
   publisher={Springer Science and Business Media LLC},
   author={Helsen, J. and Ioannou, M. and Kitzinger, J. and Onorati, E. and Werner, A. H. and Eisert, J. and Roth, I.},
   year={2023},
   month=aug }

@article{Helsen_2022,
   title={General Framework for Randomized Benchmarking},
   volume={3},
   ISSN={2691-3399},
   url={http://dx.doi.org/10.1103/PRXQuantum.3.020357},
   DOI={10.1103/prxquantum.3.020357},
   number={2},
   journal={PRX Quantum},
   publisher={American Physical Society (APS)},
   author={Helsen, J. and Roth, I. and Onorati, E. and Werner, A.H. and Eisert, J.},
   year={2022},
   month=jun }

@article{Knill_2008,
   title={Randomized benchmarking of quantum gates},
   volume={77},
   ISSN={1094-1622},
   url={http://dx.doi.org/10.1103/PhysRevA.77.012307},
   DOI={10.1103/physreva.77.012307},
   number={1},
   journal={Physical Review A},
   publisher={American Physical Society (APS)},
   author={Knill, E. and Leibfried, D. and Reichle, R. and Britton, J. and Blakestad, R. B. and Jost, J. D. and Langer, C. and Ozeri, R. and Seidelin, S. and Wineland, D. J.},
   year={2008},
   month=jan }

@article{EXP6Blume_Kohout_2017,
   title={Demonstration of qubit operations below a rigorous fault tolerance threshold with gate set tomography},
   volume={8},
   ISSN={2041-1723},
   url={http://dx.doi.org/10.1038/ncomms14485},
   DOI={10.1038/ncomms14485},
   number={1},
   journal={Nature Communications},
   publisher={Springer Science and Business Media LLC},
   author={Blume-Kohout, Robin and Gamble, John King and Nielsen, Erik and Rudinger, Kenneth and Mizrahi, Jonathan and Fortier, Kevin and Maunz, Peter},
   year={2017},
   month=feb }

@article{Kliesch_2021,
   title={Theory of Quantum System Certification},
   volume={2},
   ISSN={2691-3399},
   url={http://dx.doi.org/10.1103/PRXQuantum.2.010201},
   DOI={10.1103/prxquantum.2.010201},
   number={1},
   journal={PRX Quantum},
   publisher={American Physical Society (APS)},
   author={Kliesch, Martin and Roth, Ingo},
   year={2021},
   month=jan }

@article{Roth_2018,
   title={Recovering Quantum Gates from Few Average Gate Fidelities},
   volume={121},
   ISSN={1079-7114},
   url={http://dx.doi.org/10.1103/PhysRevLett.121.170502},
   DOI={10.1103/physrevlett.121.170502},
   number={17},
   journal={Physical Review Letters},
   publisher={American Physical Society (APS)},
   author={Roth, I. and Kueng, R. and Kimmel, S. and Liu, Y.-K. and Gross, D. and Eisert, J. and Kliesch, M.},
   year={2018},
   month=oct }

@misc{bacon2005quantumschurtransformi,
      title={The Quantum Schur Transform: I. Efficient Qudit Circuits}, 
      author={Dave Bacon and Isaac L. Chuang and Aram W. Harrow},
      year={2005},
      eprint={quant-ph/0601001},
      archivePrefix={arXiv},
      primaryClass={quant-ph},
      url={https://arxiv.org/abs/quant-ph/0601001}, 
}

@misc{burchardt2025highdimensionalquantumschurtransforms,
      title={High-dimensional quantum Schur transforms}, 
      author={Adam Burchardt and Jiani Fei and Dmitry Grinko and Martin Larocca and Maris Ozols and Sydney Timmerman and Vladyslav Visnevskyi},
      year={2025},
      eprint={2509.22640},
      archivePrefix={arXiv},
      primaryClass={quant-ph},
      url={https://arxiv.org/abs/2509.22640}, 
}

@inproceedings{Oufkir_2023,
   title={Sample-Optimal Quantum Process Tomography with non-adaptive Incoherent Measurements},
   url={http://dx.doi.org/10.1109/ISIT54713.2023.10206538},
   DOI={10.1109/isit54713.2023.10206538},
   booktitle={2023 IEEE International Symposium on Information Theory (ISIT)},
   publisher={IEEE},
   author={Oufkir, Aadil},
   year={2023},
   month=jun, pages={1919–1924} }

@article{SurawyStepneyKahnKuengGuta2022,
   title={Projected Least-Squares Quantum Process Tomography},
   volume={6},
   ISSN={2521-327X},
   url={http://dx.doi.org/10.22331/q-2022-10-20-844},
   DOI={10.22331/q-2022-10-20-844},
   journal={Quantum},
   publisher={Verein zur Forderung des Open Access Publizierens in den Quantenwissenschaften},
   author={Surawy-Stepney, Trystan and Kahn, Jonas and Kueng, Richard and Guta, Madalin},
   year={2022},
   month=oct, pages={844} }

@inproceedings{HaahKothariODonnellTang2023,
   title={Query-optimal estimation of unitary channels in diamond distance},
   url={http://dx.doi.org/10.1109/FOCS57990.2023.00028},
   DOI={10.1109/focs57990.2023.00028},
   booktitle={2023 IEEE 64th Annual Symposium on Foundations of Computer Science (FOCS)},
   publisher={IEEE},
   author={Haah, Jeongwan and Kothari, Robin and O’Donnell, Ryan and Tang, Ewin},
   year={2023},
   month=nov, pages={363–390} }

@misc{pelecanos2025mixedstatetomographyreduces,
      title={Mixed state tomography reduces to pure state tomography}, 
      author={Angelos Pelecanos and Jack Spilecki and Ewin Tang and John Wright},
      year={2025},
      eprint={2511.15806},
      archivePrefix={arXiv},
      primaryClass={quant-ph},
      url={https://arxiv.org/abs/2511.15806}, 
}

@misc{LOWERBrosenthal2024quantumchanneltestingaveragecase,
      title={Quantum Channel Testing in Average-Case Distance}, 
      author={Gregory Rosenthal and Hugo Aaronson and Sathyawageeswar Subramanian and Animesh Datta and Tom Gur},
      year={2024},
      eprint={2409.12566},
      archivePrefix={arXiv},
      primaryClass={quant-ph},
      url={https://arxiv.org/abs/2409.12566}, 
}

@misc{POVMnetzambrano2025fastquantummeasurementtomography,
      title={Fast quantum measurement tomography with dimension-optimal error bounds}, 
      author={Leonardo Zambrano and Sergi Ramos-Calderer and Richard Kueng},
      year={2025},
      eprint={2507.04500},
      archivePrefix={arXiv},
      primaryClass={quant-ph},
      url={https://arxiv.org/abs/2507.04500}, 
}

@misc{GutaKahnKuengTropp2020,
      title={Fast state tomography with optimal error bounds}, 
      author={Madalin Guta and Jonas Kahn and Richard Kueng and Joel A. Tropp},
      year={2018},
      eprint={1809.11162},
      archivePrefix={arXiv},
      primaryClass={quant-ph},
      url={https://arxiv.org/abs/1809.11162}, 
}

@misc{LBrosenthal2024quantumchanneltestingaveragecase,
      title={Quantum Channel Testing in Average-Case Distance}, 
      author={Gregory Rosenthal and Hugo Aaronson and Sathyawageeswar Subramanian and Animesh Datta and Tom Gur},
      year={2024},
      eprint={2409.12566},
      archivePrefix={arXiv},
      primaryClass={quant-ph},
      url={https://arxiv.org/abs/2409.12566}, 
}

@article{Scott_2008,
   title={Optimizing quantum process tomography with unitary 2-designs},
   volume={41},
   ISSN={1751-8121},
   url={http://dx.doi.org/10.1088/1751-8113/41/5/055308},
   DOI={10.1088/1751-8113/41/5/055308},
   number={5},
   journal={Journal of Physics A: Mathematical and Theoretical},
   publisher={IOP Publishing},
   author={Scott, A J},
   year={2008},
   month=jan, pages={055308} }

@article{anshu2023survey,
	author = {Anshu, Anurag and Arunachalam, Srinivasan},
	date = {2024/01/01},
	doi = {10.1038/s42254-023-00662-4},
	id = {Anshu2024},
	isbn = {2522-5820},
	journal = {Nature Reviews Physics},
	number = {1},
	pages = {59--69},
	title = {A survey on the complexity of learning quantum states},
	url = {https://doi.org/10.1038/s42254-023-00662-4},
	volume = {6},
	year = {2024}}

@article{Levy_2024,
   title={Classical shadows for quantum process tomography on near-term quantum computers},
   volume={6},
   ISSN={2643-1564},
   url={http://dx.doi.org/10.1103/PhysRevResearch.6.013029},
   DOI={10.1103/physrevresearch.6.013029},
   number={1},
   journal={Physical Review Research},
   publisher={American Physical Society (APS)},
   author={Levy, Ryan and Luo, Di and Clark, Bryan K.},
   year={2024},
   month=jan }

@misc{wadhwa2024agnosticprocesstomography,
      title={Agnostic Process Tomography}, 
      author={Chirag Wadhwa and Laura Lewis and Elham Kashefi and Mina Doosti},
      year={2024},
      eprint={2410.11957},
      archivePrefix={arXiv},
      primaryClass={quant-ph},
      url={https://arxiv.org/abs/2410.11957}, 
}

@article{Angrisani_2025,
   title={Learning unitaries with quantum statistical queries},
   volume={9},
   ISSN={2521-327X},
   url={http://dx.doi.org/10.22331/q-2025-07-30-1817},
   DOI={10.22331/q-2025-07-30-1817},
   journal={Quantum},
   publisher={Verein zur Forderung des Open Access Publizierens in den Quantenwissenschaften},
   author={Angrisani, Armando},
   year={2025},
   month=jul, pages={1817} }

@article{zhao2023learning,
   title={Learning Quantum States and Unitaries of Bounded Gate Complexity},
   volume={5},
   ISSN={2691-3399},
   url={http://dx.doi.org/10.1103/PRXQuantum.5.040306},
   DOI={10.1103/prxquantum.5.040306},
   number={4},
   journal={PRX Quantum},
   publisher={American Physical Society (APS)},
   author={Zhao, Haimeng and Lewis, Laura and Kannan, Ishaan and Quek, Yihui and Huang, Hsin-Yuan and Caro, Matthias C.},
   year={2024},
   month=oct }

@article{Haah_2017,
   title={Sample-optimal tomography of quantum states},
   ISSN={1557-9654},
   url={http://dx.doi.org/10.1109/TIT.2017.2719044},
   DOI={10.1109/tit.2017.2719044},
   journal={IEEE Transactions on Information Theory},
   publisher={Institute of Electrical and Electronics Engineers (IEEE)},
   author={Haah, Jeongwan and Harrow, Aram W. and Ji, Zhengfeng and Wu, Xiaodi and Yu, Nengkun},
   year={2017},
   pages={1–1} }

\clearpage
\onecolumngrid
\begin{center}
\vspace*{\baselineskip}
{\Large\textbf{Supplemental Material}}
\end{center}

\tableofcontents
\vspace*{2\baselineskip}

\section{Preliminaries and technical tools}
\label{sec:appendixPRELIM}
In this section we fix notation and recall the basic notions, some of the previous results crucial for our work and some technical lemmas that will be used throughout.

\subsection{Notation}
\label{sub:notation}
All Hilbert spaces are finite dimensional. We write
$\mathcal H$, $\mathcal H_{\mathrm{in}}$, $\mathcal H_{\mathrm{out}}$, etc.\ for
Hilbert spaces, and $\mathcal L(\mathcal H)$ and $\mathcal D(\mathcal H)$ for
linear and density operators on $\mathcal H$, respectively. The identity on
$\mathcal H$ is denoted by $\mathbb I_{\mathcal H}$, or simply $\mathbb I$ when
there is no ambiguity. For $X_{AB} \in \mathcal L(\mathcal H_A \otimes \mathcal H_B)$
we write $\tr_B(X_{AB})$ for the partial trace over $B$.

We denote the input and output dimensions of a channel by
$d_{\mathrm{in}} \coloneqq \dim(\mathcal H_{\mathrm{in}})$ and
$d_{\mathrm{out}} \coloneqq \dim(\mathcal H_{\mathrm{out}})$.
Fix an orthonormal basis $\{\ket{j}\}_{j=1}^{d_{\mathrm{in}}}$ of
$\mathcal H_{\mathrm{in}}$. The (normalized) maximally entangled vector on
$\mathcal H_{\mathrm{in}} \otimes \mathcal H_{\mathrm{in}}$ is
$\ket{\Omega} \coloneqq d_{\mathrm{in}}^{-1/2} \sum_{j=1}^{d_{\mathrm{in}}}
\ket{j} \otimes \ket{j}$, and we write
$\Omega \coloneqq \ketbra{\Omega}{\Omega}$ for the corresponding maximally
entangled state.

For an operator $X \in \mathcal L(\mathcal H)$, the Schatten $p$-norm is defined,
for $1 \le p < \infty$, by $\|X\|_p \coloneqq \bigl(\tr(|X|^p)\bigr)^{1/p}$,
where $|X| \coloneqq \sqrt{X^\dagger X}$. Equivalently, $\|X\|_p$ coincides with
the $\ell_p$-norm of the vector of singular values of $X$; in particular,
$\|X\|_1$ is the trace norm. The quantity $\|X\|_\infty$ denotes the operator
norm and coincides with the largest singular value of $X$. For an operator
$X \in \mathcal L(\mathcal H)$ we write $X \ge 0$ if $X$ is positive semidefinite.
A linear map $V : \mathcal H_{\mathrm{in}} \to \mathcal H_{\mathrm{out}}$ between Hilbert spaces is called an
\emph{isometry} if $V^\dagger V = \mathbb I_{\mathcal H_{\mathrm{in}}}$. In particular, if
$\dim(\mathcal H_{\mathrm{in}}) = \dim(\mathcal H_{\mathrm{out}})$, an isometry is unitary.

We write $\mathcal N_d(m,V)$ for the $d$-dimensional real Gaussian distribution
with mean vector $m \in \mathbb R^d$ and covariance matrix
$V \in \mathbb R^{d \times d}$ (symmetric and positive semidefinite).
Thus a random vector $X \in \mathbb R^d$ satisfies
$X \sim \mathcal N_d(m,V)$ if its probability density function is
\begin{align}
  p_X(x)
  = \frac{1}{\sqrt{(2\pi)^d \det V}}\,
    \exp\Bigl(-\tfrac12 (x-m)^\top V^{-1}(x-m)\Bigr),
  \qquad x \in \mathbb R^d.
\end{align}
In the one-dimensional case we write $\mathcal N_1(m,\sigma^2)$ for the
normal distribution with mean $m \in \mathbb R$ and variance $\sigma^2 > 0$;
in particular, the standard normal distribution is $\mathcal N_1(0,1)$.

We also use the Beta distribution.
For parameters $\alpha,\beta>0$, a random variable $X$ is said to have a
$\mathrm{Beta}(\alpha,\beta)$ distribution, written
$X \sim \mathrm{Beta}(\alpha,\beta)$, if it takes values in $[0,1]$ and its
probability density function is
\begin{align}
  p_X(x)
  \coloneqq \frac{x^{\alpha-1}(1-x)^{\beta-1}}{B(\alpha,\beta)},
  \qquad x \in [0,1],
\end{align}
where $B(\alpha,\beta)$ denotes the Beta function, defined by $B(\alpha,\beta)
  \coloneqq \int_0^1 t^{\alpha-1}(1-t)^{\beta-1}\,\mathrm dt.$

We use standard asymptotic notation $O(\cdot)$, $\Omega(\cdot)$ and $\Theta(\cdot)$
with respect to the relevant problem parameters. Given nonnegative functions
$f,g : \mathbb N^m \to [0,\infty)$, we write $f = O(g)$ if there exist constants
$C > 0$ and $n_0 \in \mathbb N$ such that
$f(\mathbf n) \le C\,g(\mathbf n)$ for all $\mathbf n \in \mathbb N^m$ with
$\mathbf n \ge n_0$ coordinatewise. We write $f = \Omega(g)$ if $g = O(f)$, and
$f = \Theta(g)$ if both $f = O(g)$ and $f = \Omega(g)$ hold. Finally, we write
$f = \tilde \Omega(g)$ when the relation $f = \Omega(g)$ holds up to
polylogarithmic factors in the relevant parameters (in our setting, this amounts
to an additional logarithmic factor in the denominator).

\subsection{Preliminaries on quantum information theory}
In this subsection we briefly recall basic notions from quantum information theory that will be used throughout, including quantum states and their purifications, POVMs, quantum channels, the Choi representation, and the diamond norm together with its SDP characterisation. For a more comprehensive introduction to these topics, we refer the reader to standard textbooks, e.g., Ref.~\cite{WatrousBook}.

\subsubsection{Quantum states, purifications and POVMs}
A quantum state on a Hilbert space $\mathcal H$ is a density operator $\rho \in \mathcal D(\mathcal H)$, i.e., a positive semidefinite operator $\rho\ge 0$ with $\tr(\rho) = 1$. The rank of $\rho$, denoted $\mathrm{rank}(\rho)$, is its matrix rank; states with $\mathrm{rank}(\rho) = 1$ are called \emph{pure} and can be written as $\rho = \ketbra{\psi}{\psi}$ for some unit vector $\ket{\psi} \in \mathcal H$. The trace distance between two quantum states $\rho,\sigma \in \mathcal D(\mathcal H)$ is defined as $d_{\mathrm{tr}}(\rho,\sigma) \coloneqq \tfrac12\|\rho - \sigma\|_1$. It is an operationally meaningful metric: the optimal success probability for distinguishing $\rho$ from $\sigma$ in a single-shot binary discrimination task is $p_{\mathrm{succ}} = \tfrac12\bigl(1 + d_{\mathrm{tr}}(\rho,\sigma)\bigr)$. Moreover, if $\rho = \ketbra{\psi}{\psi}$ and $\sigma = \ketbra{\phi}{\phi}$ are pure, then $d_{\mathrm{tr}}(\rho,\sigma) = \sqrt{1 - |\braket{\psi}{\phi}|^2}$.

Given a state $\rho \in \mathcal D(\mathcal H)$, a \emph{purification} of $\rho$ is a pure state $\ket{\Psi} \in \mathcal H \otimes \mathcal H_E$ on a larger Hilbert space such that $\tr_E(\ketbra{\Psi}{\Psi}) = \rho$, where $\mathcal H_E$ is an auxiliary environment space. There exists a purification of $\rho$ on an environment of minimal dimension $\mathrm{rank}(\rho)$; for example, if $\rho = \sum_{i=1}^r \lambda_i \ketbra{i}{i}$ is a spectral decomposition with $r = \mathrm{rank}(\rho)$, then $\ket{\Psi_0} \coloneqq \sum_{i=1}^r \sqrt{\lambda_i}\,\ket{i} \otimes \ket{i} \in \mathcal H \otimes \mathbb C^r$ is a purification of $\rho$.

All purifications of a given state with the same environment Hilbert space are equivalent up to a unitary on the environment. More precisely, if $\mathcal H_E$ is chosen with minimal dimension $\dim(\mathcal H_E) = \mathrm{rank}(\rho)$ and $\ket{\Psi_0}, \ket{\Psi'} \in \mathcal H \otimes \mathcal H_E$ are two purifications of $\rho$, then there exists a unitary $U_E$ on $\mathcal H_E$ such that $\ket{\Psi'} = (\mathbb I_{\mathcal H} \otimes U_E)\ket{\Psi_0}$.

A positive operator-valued measure (POVM) with outcome set $\mathcal X$ on $\mathcal H$ is a family $\{M_x\}_{x \in \mathcal X} \subset \mathcal L(\mathcal H)$ of positive semidefinite operators satisfying $\sum_{x \in \mathcal X} M_x = \mathbb I_{\mathcal H}$. Measuring a state $\rho \in \mathcal D(\mathcal H)$ with $\{M_x\}$ yields outcome $x$ with probability $\tr(M_x \rho)$. A POVM with two outcomes $\mathcal X = \{0,1\}$ is called \emph{binary}; it is fully specified by a single operator $M$ with $0 \le M \le \mathbb I_{\mathcal H}$, in which case we write $\{M, \mathbb I_{\mathcal H} - M\}$.

\subsubsection{Quantum channels and Choi representation}
\label{subsec:qchannels}
A (quantum) channel is a completely positive trace-preserving (CPTP) map
$\Lambda : \mathcal L(\mathcal H_{\mathrm{in}}) \to \mathcal L(\mathcal H_{\mathrm{out}})$.
Equivalently, $\Lambda$ admits a Kraus decomposition
$\Lambda(\rho) = \sum_{\alpha=1}^{k} K_\alpha \rho K_\alpha^\dagger$ for operators
$K_\alpha : \mathcal H_{\mathrm{in}} \to \mathcal H_{\mathrm{out}}$ satisfying
$\sum_{\alpha=1}^{k} K_\alpha^\dagger K_\alpha = \mathbb I_{\mathcal H_{\mathrm{in}}}$;
the smallest such $k$ is the Kraus rank of $\Lambda$. Equivalently (Stinespring dilation), there exists an auxiliary Hilbert space $\mathcal H_E$ and an isometry
$V:\mathcal H_{\mathrm{in}}\to \mathcal H_{\mathrm{out}}\otimes \mathcal H_E$ with $V^\dagger V=\mathbb I_{\mathcal H_{\mathrm{in}}}$
such that $\Lambda(\rho)=\Tr_E[V\rho V^\dagger]$ for all states $\rho$.
Given any Kraus family $\{K_\alpha\}_{\alpha=1}^k$ and an orthonormal basis $\{\ket{\alpha}\}_{\alpha=1}^k$ of $\mathcal H_E\simeq\mathbb C^k$,
one may take $V \coloneqq \sum_{\alpha=1}^k K_\alpha\otimes \ket{\alpha}$, in which case
$V^\dagger V=\sum_{\alpha=1}^k K_\alpha^\dagger K_\alpha=\mathbb I_{\mathcal H_{\mathrm{in}}}$ and indeed
$\Tr_E[V\rho V^\dagger]=\sum_{\alpha=1}^k K_\alpha \rho K_\alpha^\dagger$.
Conversely, for any isometry $V$ and any choice of basis $\{\ket{\alpha}\}$ of $\mathcal H_E$, the operators
$K_\alpha \coloneqq (\mathbb I_{\mathcal H_{\mathrm{out}}}\otimes\bra{\alpha})V$ form a Kraus representation of $\Lambda$.
The dilation is not unique: replacing $V$ by $(\mathbb I_{\mathcal H_{\mathrm{out}}}\otimes U)V$ for any unitary $U$ on $\mathcal H_E$
does not change the induced channel, and one can always choose $\dim(\mathcal H_E)=k$ equal to the minimal Kraus rank (equivalently, the Choi rank).

A linear map $\Phi : \mathcal L(\mathcal H_{\mathrm{in}}) \to \mathcal L(\mathcal H_{\mathrm{out}})$
is said to be \emph{Hermiticity preserving} if it maps Hermitian operators to Hermitian operators.
In particular, every positive (and hence every completely positive) map is Hermiticity preserving.

We use the standard Choi--Jamiołkowski representation. For a linear map
$\Phi : \mathcal L(\mathcal H_{\mathrm{in}}) \to \mathcal L(\mathcal H_{\mathrm{out}})$,
its Choi operator is $J(\Phi) \coloneqq (\Phi \otimes \mathrm{id})(\Omega)\in \mathcal L(\mathcal H_{\mathrm{out}} \otimes \mathcal H_{\mathrm{in}})$,
where $\mathrm{id}$ denotes the identity map on $\mathcal L(\mathcal H_{\mathrm{in}})$ and
$\Omega = \ket{\Omega}\bra{\Omega}$ with $\ket{\Omega} = d_{\mathrm{in}}^{-1/2}\sum_{j=1}^{d_{\mathrm{in}}} \ket{j} \otimes \ket{j}$.
Then $\Phi$ is completely positive if and only if $J(\Phi)\ge 0$, and trace preserving if and only if
$\tr_{\mathrm{out}}(J(\Phi))=\mathbb I_{\mathcal H_{\mathrm{in}}}/d_{\mathrm{in}}$.
For a channel $\Lambda$ we write $\Phi_\Lambda \coloneqq J(\Lambda)$ and refer to $\Phi_\Lambda$ as the Choi state of $\Lambda$; its rank
$k \coloneqq \mathrm{rank}(\Phi_\Lambda)$ is the Choi rank and coincides with the minimal Kraus rank.
We adopt the standard vectorization map
$\mathrm{vec} : \mathcal L(\mathcal H_{\mathrm{in}},\mathcal H_{\mathrm{out}}) \to \mathcal H_{\mathrm{out}} \otimes \mathcal H_{\mathrm{in}}$,
defined on matrix units by $\mathrm{vec}(\ket{i}\bra{j}) \coloneqq \ket{i} \otimes \ket{j}$ and extended linearly, and denote its inverse by
$\mathrm{vec}^{-1}$.
With our convention for $\Omega$, any Kraus representation $\{K_\alpha\}$ of $\Lambda$ satisfies
$J(\Lambda)=d_{\mathrm{in}}^{-1}\sum_{\alpha} \mathrm{vec}(K_\alpha)\mathrm{vec}(K_\alpha)^\dagger$.
Conversely, let $\Phi_\Lambda=\sum_{\alpha=1}^{k} \lambda_\alpha \ket{v_\alpha}\bra{v_\alpha}$ be a spectral decomposition of the Choi state
with $\lambda_\alpha>0$ and $\{\ket{v_\alpha}\}$ orthonormal in $\mathcal H_{\mathrm{out}} \otimes \mathcal H_{\mathrm{in}}$.
Defining $K_\alpha \coloneqq \sqrt{d_{\mathrm{in}}\lambda_\alpha}\,\mathrm{vec}^{-1}(\ket{v_\alpha})$ gives a Kraus family such that
$\Lambda(\rho)=\sum_{\alpha=1}^{k} K_\alpha \rho K_\alpha^\dagger$ and $\sum_{\alpha} K_\alpha^\dagger K_\alpha = \mathbb I_{\mathcal H_{\mathrm{in}}}$.
In particular, the Choi rank $\mathrm{rank}(\Phi_\Lambda)$ equals the minimal number of Kraus operators required to represent $\Lambda$.

Channels with Kraus rank $k=1$ are isometries, i.e., maps of the form $\Lambda(\rho)=V\rho V^\dagger$ with $V^\dagger V=\mathbb I_{\mathcal H_{\mathrm{in}}}$. When $d_{\mathrm{in}}=d_{\mathrm{out}}$, such channels are unitary channels.

\begin{lemma}[Dimension constraint from Kraus rank]
\label{le:dimen-constraint-kraus}
Let $\Lambda : \mathcal L(\mathcal H_{\mathrm{in}}) \to \mathcal L(\mathcal H_{\mathrm{out}})$ be a quantum channel (CPTP map) with
$d_{\mathrm{in}} \coloneqq \dim(\mathcal H_{\mathrm{in}})$ and $d_{\mathrm{out}} \coloneqq \dim(\mathcal H_{\mathrm{out}})$.
Assume that $\Lambda$ has minimal Kraus rank $k$ (equivalently, its Choi operator $J(\Lambda)$ has rank $k$). Then
$\lceil d_{\mathrm{in}}/d_{\mathrm{out}}\rceil \le k \le d_{\mathrm{in}}d_{\mathrm{out}}$.
\end{lemma}

\begin{proof}
For the upper bound, note that $J(\Lambda)$ acts on $\mathcal H_{\mathrm{out}}\otimes \mathcal H_{\mathrm{in}}\simeq \mathbb C^{d_{\mathrm{out}}d_{\mathrm{in}}}$, hence
$k=\mathrm{rank}(J(\Lambda))\le d_{\mathrm{out}}d_{\mathrm{in}}$.
For the lower bound, use a minimal Stinespring dilation: since $\Lambda$ has minimal Kraus rank $k$, there exists a Hilbert space $\mathcal H_E$ with
$\dim(\mathcal H_E)=k$ and an isometry $V:\mathcal H_{\mathrm{in}}\to \mathcal H_{\mathrm{out}}\otimes \mathcal H_E$ such that
$\Lambda(\rho)=\Tr_E[V\rho V^\dagger]$.
Let $\{e_j\}_{j=1}^{d_{\mathrm{in}}}$ be an orthonormal basis of $\mathcal H_{\mathrm{in}}$.
Because $V^\dagger V=\mathbb I_{\mathcal H_{\mathrm{in}}}$, the vectors $\{V e_j\}_{j=1}^{d_{\mathrm{in}}}$ are orthonormal in
$\mathcal H_{\mathrm{out}}\otimes \mathcal H_E$.
In particular, $\mathcal H_{\mathrm{out}}\otimes \mathcal H_E$ contains an orthonormal set of size $d_{\mathrm{in}}$, so its dimension must be at least
$d_{\mathrm{in}}$, i.e., $d_{\mathrm{out}}k=\dim(\mathcal H_{\mathrm{out}}\otimes \mathcal H_E)\ge d_{\mathrm{in}}$.
Thus $k\ge d_{\mathrm{in}}/d_{\mathrm{out}}$, and since $k$ is an integer we conclude $k\ge \lceil d_{\mathrm{in}}/d_{\mathrm{out}}\rceil$.
\end{proof}

\subsubsection{Diamond norm and its SDP formulation}
Given a linear map
$\Phi : \mathcal L(\mathcal H_{\mathrm{in}}) \to \mathcal L(\mathcal H_{\mathrm{out}})$,
its \emph{diamond norm} (also known as the completely bounded trace norm) is defined as
\begin{align}
  \|\Phi\|_\diamond
  \coloneqq \sup_{d' \ge 1}\,
  \sup_{\rho \in \mathcal D(\mathcal H_{\mathrm{in}} \otimes \mathbb C^{d'})}
  \bigl\|(\Phi \otimes \mathrm{id}_{\mathbb C^{d'}})(\rho)\bigr\|_1 .
  \label{eq:def-diamond}
\end{align}
It is standard that the supremum may be restricted, without loss of generality, to
$d' = d_{\mathrm{in}}$ and to pure states
$\rho = \ket{\psi}\bra{\psi}$ on
$\mathcal H_{\mathrm{in}} \otimes \mathbb C^{d_{\mathrm{in}}}$; see, e.g.,
Ref.~\cite[Sec.~3.3]{WatrousBook}. For a Hermitian operator $X$, the trace norm admits the variational characterisation
\begin{align}
  \|X\|_1
  = \max_{\substack{H = H^{\dagger}\\ -\,\mathbb I \le H \le \mathbb I}}
    \tr(HX),
  \label{eq:trace-norm-dual}
\end{align}
where the maximum is taken over Hermitian $H$ satisfying $-\,\mathbb I \le H \le \mathbb I$. For two quantum channels
$\Lambda,\Gamma : \mathcal L(\mathcal H_{\mathrm{in}}) \to
\mathcal L(\mathcal H_{\mathrm{out}})$,
we define their \emph{diamond distance} by
\begin{align}
  d_\diamond(\Lambda,\Gamma)
  \coloneqq \tfrac12\,\|\Lambda - \Gamma\|_\diamond,
\end{align}
which always obeys $0 \le d_\diamond(\Lambda,\Gamma) \le 1$.
With this convention, $d_\diamond(\Lambda,\Gamma)$ plays the exact analogue of the trace
distance for states: the optimal success probability for distinguishing $\Lambda$ from
$\Gamma$ using a single channel use with arbitrary entanglement assistance is $p_{\mathrm{succ}}
  = \tfrac12\bigl(1 + d_\diamond(\Lambda,\Gamma)\bigr),$ (see, e.g., Ref.~\cite{WatrousBook}). A particularly convenient way to express the diamond norm is via the (normalized) Choi operator
\begin{align}
  J(\Phi)
  \coloneqq (\Phi \otimes \mathrm{id})(\Omega)
  \in \mathcal L(\mathcal H_{\mathrm{out}} \otimes \mathcal H_{\mathrm{in}}),
\end{align}
where $\Omega = \ketbra{\Omega}$ is the maximally entangled state
introduced above. In this representation, the diamond norm admits a natural
semidefinite-program (SDP) formulation directly in terms of $J(\Phi)$.
For a more detailed discussion of SDP characterisations of the diamond norm we
refer the reader to, e.g., Refs.~\cite{WatrousBook,Watrous2013CBnorm,SkrzypczykCavalcantiSDP}.

\begin{lemma}[SDP formulation of the diamond norm; cf.\ Eq.~(7.23) in Ref.~\cite{SkrzypczykCavalcantiSDP}]
\label{le:diamond-sdp}
Let
$\Phi : \mathcal L(\mathcal H_{\mathrm{in}}) \to \mathcal L(\mathcal H_{\mathrm{out}})$
be a Hermiticity-preserving linear map, and let $J(\Phi)$ be its (normalized) Choi
operator as above. Then its diamond norm admits the semidefinite-program representation
\begin{align}
  \|\Phi\|_\diamond
  = d_{\mathrm{in}}\,
    \max_{\substack{Y = Y^\dagger\\ \sigma \in \mathcal D(\mathcal H_{\mathrm{in}})}}
      &\tr\bigl(J(\Phi)\,Y\bigr)
  \label{eq:diamond-sdp-primal-J}\\
  \text{subject to}\quad
  &Y \in \mathcal L(\mathcal H_{\mathrm{out}} \otimes \mathcal H_{\mathrm{in}}),\nonumber\\
  &-\,\mathbb I_{\mathrm{out}} \otimes \sigma
    \;\le\; Y \;\le\;
    \mathbb I_{\mathrm{out}} \otimes \sigma .\nonumber
\end{align}
\end{lemma}

\begin{proof}
For self-consistency, we briefly recall the standard derivation.
Without loss of generality, we may restrict in the definition of
$\|\Phi\|_\diamond$ to a $d_{\mathrm{in}}$-dimensional auxiliary register and to pure input
states, and use the variational characterisation of the trace norm:
\begin{align}
  \|\Phi\|_\diamond
  &= \sup_{\ket{\psi}\in\mathcal H_{\mathrm{in}}\otimes\mathbb C^{d_{\mathrm{in}}}}
     \bigl\|(\Phi\otimes\mathrm{id})(\ketbra{\psi}{\psi})\bigr\|_1 = \sup_{\ket{\psi}}\;
     \max_{\substack{H=H^\dagger\\ -\mathbb I\le H\le\mathbb I}}
     \tr\bigl(H\,(\Phi\otimes\mathrm{id})(\ketbra{\psi}{\psi})\bigr).
\end{align}

Fix an orthonormal basis $\{\ket{j}\}$ of $\mathcal H_{\mathrm{in}}$ and set
$\ket{\Omega}
 = d_{\mathrm{in}}^{-1/2}\sum_{j=1}^{d_{\mathrm{in}}}\ket{j}\otimes\ket{j}$
on $\mathcal H_{\mathrm{in}}\otimes\mathcal H_R$ with
$\mathcal H_R\simeq\mathcal H_{\mathrm{in}}$.

Every unit vector \(|\psi\rangle\in H_{\rm in}\otimes H_R\) can be written as
\[
  |\psi\rangle
  =
  \sqrt{d_{\rm in}}\,(\mathbb I_{\rm in}\otimes A)|\Omega\rangle
\]
for some \(A\in\mathbb C^{d_{\rm in}\times d_{\rm in}}\) satisfying
\(\operatorname{tr}(A^\dagger A)=1\). Hence
\[
  |\psi\rangle\langle\psi|
  =
  d_{\rm in}(I_{\rm in}\otimes A)\Omega(\mathbb I_{\rm in}\otimes A^\dagger).
\]
Using that $A$ acts only on the reference system and that
$J(\Phi)=(\Phi\otimes\mathrm{id})(\Omega)$, we obtain
\begin{align}
  (\Phi\otimes\mathrm{id})(\ketbra{\psi}{\psi})
  &=d_{\mathrm{in}} (\mathbb I_{\mathrm{out}}\otimes A)\,J(\Phi)\,(\mathbb I_{\mathrm{out}}\otimes A^\dagger),\\
  \tr\bigl(H\,(\Phi\otimes\mathrm{id})(\ketbra{\psi}{\psi})\bigr)
  &= d_{\mathrm{in}}\tr\Bigl(
       J(\Phi)\,
       (\mathbb I_{\mathrm{out}}\otimes A^\dagger)\,
       H\,
       (\mathbb I_{\mathrm{out}}\otimes A)
     \Bigr).
\end{align}
Define $\sigma \coloneqq A^\dagger A  \in\mathcal D(\mathcal H_R)$ and
\begin{align}
  Y \coloneqq  
  (\mathbb I_{\mathrm{out}}\otimes A^\dagger)\,
  H\,(\mathbb I_{\mathrm{out}}\otimes A).
\end{align}
Then $Y=Y^\dagger$ and
\begin{align}
  \tr\bigl(H\,(\Phi\otimes\mathrm{id})(\ketbra{\psi}{\psi})\bigr)
  = d_{\mathrm{in}}\,\tr\bigl(J(\Phi)\,Y\bigr).
\end{align}
Moreover, from $-\mathbb I\le H\le\mathbb I$ it follows that
$-\mathbb I_{\mathrm{out}}\otimes\sigma \le Y \le \mathbb I_{\mathrm{out}}\otimes\sigma$.
Thus any feasible pair $(\ket{\psi},H)$ induces a feasible pair $(\sigma,Y)$
for \eqref{eq:diamond-sdp-primal-J} with objective value
$d_{\mathrm{in}}\tr(J(\Phi)Y)$, so the right-hand side of
\eqref{eq:diamond-sdp-primal-J} is at least $\|\Phi\|_\diamond$.

Conversely, given any feasible $(\sigma,Y)$ for \eqref{eq:diamond-sdp-primal-J}
one may choose $A$ with $A^\dagger A=\sigma$, define
$|\psi\rangle
=
\sqrt{d_{\rm in}}\,(\mathbb I_{\rm in}\otimes A)|\Omega\rangle$, and recover some
$H$ with $-\mathbb I\le H\le\mathbb I$ such that
$Y = 
(\mathbb I_{\mathrm{out}}\otimes A^\dagger)\,H\,(\mathbb I_{\mathrm{out}}\otimes A)$.
This shows that the two optimisations are equivalent and proves
\eqref{eq:diamond-sdp-primal-J}.
\end{proof}


In the proof of our main theorem, we will also need the case in which the Choi operator entering the SDP in
Lemma~\ref{le:diamond-sdp} is positive semidefinite, $J(\Phi) \ge 0$
(in fact, it will be a quantum state).
In this situation, the optimisation problem~\eqref{eq:diamond-sdp-primal-J} admits a
particularly simple form.

\begin{lemma}[Diamond norm for positive Choi operator]
\label{le:diamond-positive-choi}
Let $\Phi : \mathcal L(\mathcal H_{\mathrm{in}}) \to \mathcal L(\mathcal H_{\mathrm{out}})$
be Hermiticity preserving, and assume that its normalized Choi operator
$J(\Phi)$ is positive semidefinite. Then
\begin{align}
  \|\Phi\|_\diamond
  &= d_{\mathrm{in}}\,
     \max_{\sigma \in \mathcal D(\mathcal H_{\mathrm{in}})}
       \tr\bigl(J(\Phi)\,(\mathbb I_{\mathrm{out}} \otimes \sigma)\bigr)
     \label{eq:diamond-positive-choi-1}\\
  &= d_{\mathrm{in}}\,
     \bigl\|\tr_{\mathrm{out}} J(\Phi)\bigr\|_\infty .
     \label{eq:diamond-positive-choi-2}
\end{align}
\end{lemma}

\begin{proof}
By Lemma~\ref{le:diamond-sdp} we have
\begin{align}
  \|\Phi\|_\diamond
  = d_{\mathrm{in}}\,
    \max_{\substack{Y = Y^\dagger\\ \sigma \in \mathcal D(\mathcal H_{\mathrm{in}})}}
      \tr\bigl(J(\Phi)\,Y\bigr)
  \quad\text{subject to}\quad
  -\,\mathbb I_{\mathrm{out}} \otimes \sigma \le Y \le \mathbb I_{\mathrm{out}} \otimes \sigma.
\end{align}
For any $\sigma \in \mathcal D(\mathcal H_{\mathrm{in}})$, we have
$\tr\bigl(J(\Phi)\,Y\bigr) \le \tr\bigl(J(\Phi) (\mathbb I_{\mathrm{out}} \otimes \sigma)\bigr)$, since $J(\Phi) \ge 0$.
Thus, for fixed $\sigma$, the maximum over $Y$ is achieved at
$Y = \mathbb I_{\mathrm{out}} \otimes \sigma$, which yields
\begin{align}
  \|\Phi\|_\diamond
  = d_{\mathrm{in}}\,
    \max_{\sigma \in \mathcal D(\mathcal H_{\mathrm{in}})}
      \tr\bigl(J(\Phi)\,(\mathbb I_{\mathrm{out}} \otimes \sigma)\bigr),
\end{align}
proving Eq.\eqref{eq:diamond-positive-choi-1}. For Eq.\eqref{eq:diamond-positive-choi-2}, note that for any
$\sigma \in \mathcal D(\mathcal H_{\mathrm{in}})$ we have
$\tr\bigl(J(\Phi)\,(\mathbb I_{\mathrm{out}} \otimes \sigma)\bigr)
= \tr\bigl(\sigma\,\tr_{\mathrm{out}} J(\Phi)\bigr)$.
Maximising the right-hand side over all density operators $\sigma$ gives
$\max_{\sigma} \tr\bigl(\sigma\,\tr_{\mathrm{out}} J(\Phi)\bigr)
= \|\tr_{\mathrm{out}} J(\Phi)\|_\infty$, which establishes
Eq.\eqref{eq:diamond-positive-choi-2}.
\end{proof}

We will also use in our main theorem the following technical lemma controlling ``off-diagonal''
Choi blocks in terms of the corresponding diagonal ones.
Here, for any Hermitian operator
$X \in \mathcal L(\mathcal H_{\mathrm{out}}\otimes\mathcal H_{\mathrm{in}})$
we write $\|X\|_\diamond$ for the diamond norm of the Hermiticity-preserving
map $\Phi_X$ with normalized Choi operator $J(\Phi_X)=X$.

\begin{lemma}[Diamond-norm Cauchy--Schwarz for purifications]
\label{le:diamond-CS}
Let
\(
  \ket{\Phi_1},\ket{\Phi_2}
  \in \mathcal H_{\mathrm{out}}\otimes\mathcal H_{\mathrm{in}}\otimes\mathcal H_E
\)
be pure states and define
\begin{align}
  J_{ij}
  \coloneqq \tr_E\bigl(\ketbra{\Phi_i}{\Phi_j}\bigr)
  \in \mathcal L(\mathcal H_{\mathrm{out}}\otimes\mathcal H_{\mathrm{in}}),
  \qquad i,j\in\{1,2\}.
\end{align}
Then
\begin{align}
  \bigl\|J_{12}+J_{21}\bigr\|_\diamond
  \;\le\;
  2\,\sqrt{\|J_{11}\|_\diamond\,\|J_{22}\|_\diamond}.
  \label{eq:diamond-CS}
\end{align}
\end{lemma}

\begin{proof}
Let $\Xi$ be the Hermiticity-preserving map whose normalized Choi operator is
$J(\Xi) \coloneqq J_{12} + J_{21}$.
By Lemma~\ref{le:diamond-sdp},
\begin{align}
  \|\Xi\|_\diamond
  = d_{\mathrm{in}}\,
    \max_{\substack{Y = Y^\dagger\\ \sigma \in \mathcal D(\mathcal H_{\mathrm{in}})}}
    \tr\bigl((J_{12}+J_{21})\,Y\bigr)
  \quad\text{subject to}\quad
  -\,\mathbb I_{\mathrm{out}}\otimes\sigma
  \;\le\; Y \;\le\;
  \mathbb I_{\mathrm{out}}\otimes\sigma.
\end{align}
Take an optimal pair $(\bar Y,\bar\sigma)$ and set
$A \coloneqq \mathbb I_{\mathrm{out}}\otimes\bar\sigma \ge 0$.
Then $\bar Y = \bar Y^\dagger$ and the constraints become
\begin{align}
  -A \;\le\; \bar Y \;\le\; A.
\end{align}
By Lemma~\ref{le:interval-contraction}, there exists a Hermitian operator
$K$ with $\|K\|_\infty \le 1$ such that
\begin{align}
  \bar Y = A^{1/2} K A^{1/2}.
\end{align}
Define
\begin{align}
  \bar W \coloneqq \bar Y \otimes \mathbb I_E
  = (A^{1/2}\otimes\mathbb I_E)\,(K\otimes\mathbb I_E)\,(A^{1/2}\otimes\mathbb I_E).
\end{align}
By the definition of the partial trace,
\begin{align}
  \tr\bigl(J_{12}\,\bar Y\bigr)
  &= \tr\Bigl(\tr_E\bigl(\ketbra{\Phi_1}{\Phi_2}\bigr)\,\bar Y\Bigr)
   = \tr\Bigl(\ketbra{\Phi_1}{\Phi_2}\,(\bar Y\otimes\mathbb I_E)\Bigr)
   = \bra{\Phi_2}\bar W\ket{\Phi_1}.
\end{align}
Moreover, since $J_{21} = J_{12}^\dagger$ and $\bar Y$ is Hermitian,
\begin{align}
  \tr(J_{21}\bar Y)
  = \tr(J_{12}^\dagger \bar Y)
  = \tr\bigl((J_{12}\bar Y)^\dagger\bigr)
  = \tr(J_{12}\bar Y)^*.
\end{align}
Hence
\begin{align}
  \|\Xi\|_\diamond
  &= d_{\mathrm{in}}\,
     \tr\bigl((J_{12}+J_{21})\bar Y\bigr) \nonumber\\
  &= d_{\mathrm{in}}\bigl(\tr(J_{12}\bar Y) + \tr(J_{21}\bar Y)\bigr) \nonumber\\
  &= 2d_{\mathrm{in}}\,\Re\,\tr(J_{12}\bar Y) \nonumber\\
  &\le 2d_{\mathrm{in}}\bigl|\tr(J_{12}\bar Y)\bigr|
   = 2d_{\mathrm{in}}\bigl|\bra{\Phi_2}\bar W\ket{\Phi_1}\bigr|.
\end{align}
Now define
\begin{align}
  \ket{\Psi_i}
  \coloneqq (A^{1/2}\otimes\mathbb I_E)\ket{\Phi_i},
  \qquad i=1,2.
\end{align}
Then
\begin{align}
  \bra{\Phi_2}\bar W\ket{\Phi_1}
  &= \bra{\Phi_2}(A^{1/2}\otimes\mathbb I_E)\,(K\otimes\mathbb I_E)\,
     (A^{1/2}\otimes\mathbb I_E)\ket{\Phi_1} = \bra{\Psi_2}(K\otimes\mathbb I_E)\ket{\Psi_1}.
\end{align}
By the Cauchy--Schwarz inequality and $\|K\|_\infty \le 1$,
\begin{align}
  \bigl|\bra{\Psi_2}(K\otimes\mathbb I_E)\ket{\Psi_1}\bigr|
  &\le \|K\otimes\mathbb I_E\|_\infty\,\|\Psi_1\|_2\,\|\Psi_2\|_2 \le \|\Psi_1\|_2\,\|\Psi_2\|_2.
\end{align}
Thus
\begin{align}
  \|\Xi\|_\diamond
  \le 2d_{\mathrm{in}}\,\|\Psi_1\|_2\,\|\Psi_2\|_2.
\end{align}
We now compute the norms $\|\Psi_i\|$.
Using the definition of $A$ and the partial trace,
\begin{align}
  \|\Psi_i\|_2^2
  = \bra{\Phi_i}(A\otimes\mathbb I_E)\ket{\Phi_i}
  = \tr\bigl(\ketbra{\Phi_i}{\Phi_i}\,(A\otimes\mathbb I_E)\bigr)
  = \tr\bigl(\tr_E(\ketbra{\Phi_i}{\Phi_i})\,A\bigr)
  = \tr\bigl(J_{ii}(\mathbb I_{\mathrm{out}}\otimes\bar\sigma)\bigr)
  \eqqcolon \alpha_i,
\end{align}
for $i=1,2$. Hence
\begin{align}
  \|\Xi\|_\diamond
  \le 2d_{\mathrm{in}}\,\sqrt{\alpha_1\alpha_2}.
\end{align}

Finally, for each $i=1,2$, consider the completely positive map whose normalized
Choi operator is $J_{ii} \ge 0$. By Lemma~\ref{le:diamond-positive-choi},
\begin{align}
  \|J_{ii}\|_\diamond
  = d_{\mathrm{in}}\max_{\sigma\in\mathcal D(\mathcal H_{\mathrm{in}})}
      \tr\bigl(J_{ii}(\mathbb I_{\mathrm{out}}\otimes\sigma)\bigr)
  \;\ge\;
  d_{\mathrm{in}}\tr\bigl(J_{ii}(\mathbb I_{\mathrm{out}}\otimes\bar\sigma)\bigr)
  = d_{\mathrm{in}}\alpha_i,
\end{align}
so $\alpha_i \le \|J_{ii}\|_\diamond/d_{\mathrm{in}}$ and therefore
\begin{align}
  \sqrt{\alpha_1\alpha_2}
  \le \frac{1}{d_{\mathrm{in}}}\sqrt{\|J_{11}\|_\diamond\,\|J_{22}\|_\diamond}.
\end{align}
Combining this with the bound on $\|\Xi\|_\diamond$ gives
\begin{align}
  \|\Xi\|_\diamond
  \le 2\,\sqrt{\|J_{11}\|_\diamond\,\|J_{22}\|_\diamond},
\end{align}
which is exactly \eqref{eq:diamond-CS}.
\end{proof}

\begin{lemma}
\label{le:interval-contraction}
Let $A \ge 0$ and $Y = Y^\dagger$ be operators on a finite-dimensional Hilbert space.
Assume that
\begin{align}
\label{eq:interval-assumption}
  -A \;\le\; Y \;\le\; A .
\end{align}
Then there exists a Hermitian operator $K$ with $\|K\|_\infty \le 1$ such that
$Y = A^{1/2} K A^{1/2}$.
\end{lemma}

\begin{proof}
Let $\ket{v} \in \ker(A)$. From Eq.\eqref{eq:interval-assumption}, we have $\bra{v}Y\ket{v}=0$. In particular, $\bra{v}(A+Y)\ket{v} = 0$. Since $A+Y \ge 0$, the condition
$\bra{v}(A+Y)\ket{v} = 0$ implies $(A+Y)\ket{v} = 0$, and using $A\ket{v} = 0$ we conclude
$Y\ket{v} = 0$. Thus $\ker(A) \subseteq \ker(Y)$, which is equivalent to
$\mathrm{supp}(Y)\subseteq \mathrm{supp}(A)$.

Restrict $A$ and $Y$ to $\mathrm{supp}(A)$ and denote the restrictions by
$A_{\mathrm{supp}}$ and $Y_{\mathrm{supp}}$. Then $A_{\mathrm{supp}} > 0$ and
\begin{align}
  -A_{\mathrm{supp}}
  \;\le\;
  Y_{\mathrm{supp}}
  \;\le\;
  A_{\mathrm{supp}}.
\end{align}
Since $A_{\mathrm{supp}}$ is strictly positive, it has an inverse square root
$A_{\mathrm{supp}}^{-1/2}$, and conjugating the above inequality yields
\begin{align}
  -\mathbb I_{\mathrm{supp}(A)}
  \;\le\;
  A_{\mathrm{supp}}^{-1/2} Y_{\mathrm{supp}} A_{\mathrm{supp}}^{-1/2}
  \;\le\;
  \mathbb I_{\mathrm{supp}(A)}.
\end{align}
Define
\begin{align}
  K_{\mathrm{supp}}
  \coloneqq
  A_{\mathrm{supp}}^{-1/2} Y_{\mathrm{supp}} A_{\mathrm{supp}}^{-1/2}.
\end{align}
Then $K_{\mathrm{supp}}$ is Hermitian and satisfies
$-\,\mathbb I \le K_{\mathrm{supp}} \le \mathbb I$, so
$\|K_{\mathrm{supp}}\|_\infty \le 1$.
Extend $K_{\mathrm{supp}}$ to all of $\mathcal H$ by
\begin{align}
  K \coloneqq K_{\mathrm{supp}} \oplus 0_{\ker(A)}.
\end{align}
This $K$ is Hermitian, $\|K\|_\infty = \|K_{\mathrm{supp}}\|_\infty \le 1$, and
\begin{align}
  A^{1/2} K A^{1/2}
  = A_{\mathrm{supp}}^{1/2} K_{\mathrm{supp}} A_{\mathrm{supp}}^{1/2}
     \oplus 0_{\ker(A)} = Y_{\mathrm{supp}} \oplus 0_{\ker(A)}
   \;=\; Y.
\end{align}
This proves the lemma.
\end{proof}

\subsection{Concentration bounds for Haar-random states}
\label{subsec:haar-states}
In this subsection we recall two basic properties of Haar-random states that will be used in our proofs:
(i) the distribution of the overlap with a fixed reference state, and
(ii) concentration of the operator norm of reduced density matrices.
For background on Haar measure and random quantum states, see
e.g.~\cite{AubrunSzarek2017AliceBob}.

We denote by $\mu_{\mathrm{Haar}}$ the Haar probability measure on the unit
sphere $\mathbb S^{2d-1} \subset \mathbb C^d$. A random unit vector
$\ket{\Psi} \in \mathbb C^d$ with distribution $\mu_{\mathrm{Haar}}$ will be
called a \emph{Haar-random state}. Throughout, $\mathcal N_d(m,V)$ denotes the
$d$-dimensional real Gaussian distribution with mean
$m \in \mathbb R^d$ and covariance $V \in \mathbb R^{d\times d}$
(see the notation subsection~\ref{sub:notation}).

\begin{definition}[Standard complex Gaussian vector]
\label{de:standard-complex-gaussian}
Let
\[
  Z = (Z_1,\dots,Z_{2d})^T
  \sim \mathcal N_{2d}\bigl(0,\tfrac12 I_{2d}\bigr),
\]
and identify $\mathbb R^{2d} \simeq \mathbb C^d$ via
\[
  g_\alpha \coloneqq Z_{2\alpha-1} + i Z_{2\alpha},
  \qquad \alpha = 1,\dots,d.
\]
We call $g = (g_1,\dots,g_d)^T \in \mathbb C^d$ a \emph{standard complex Gaussian} vector.
Equivalently, each component can be written as $g_\alpha = X_\alpha + i Y_\alpha$ with
$X_\alpha, Y_\alpha \sim \mathcal N_1\bigl(0,\tfrac12\bigr)$ independent for all
$\alpha \in [d]$.
\end{definition}
In particular,
$\abs{g_\alpha}^2 = X_\alpha^2 + Y_\alpha^2$ has the exponential distribution
$\mathrm{Exp}(1)$, i.e., it takes values in $[0,\infty)$ with probability density
$p(t) = e^{-t}$ for $t \ge 0$, and the random variables
$\{\abs{g_\alpha}^2\}_{\alpha=1}^d$ are i.i.d.

\begin{lemma}[Gaussian representation of Haar-random states]
\label{le:haar-gaussian}
Let $g \in \mathbb C^d$ be a standard complex Gaussian vector as in
Definition~\ref{de:standard-complex-gaussian}, and let
$\{\ket{\alpha}\}_{\alpha=1}^d$ be any orthonormal basis of $\mathbb C^d$.
Define
\begin{align}
  \ket{\Psi}
  \coloneqq \frac{1}{\|g\|_2}
             \sum_{\alpha=1}^d g_\alpha \ket{\alpha}.
\end{align}
Then $\ket{\Psi}$ is distributed according to the Haar measure
$\mu_{\mathrm{Haar}}$ on the unit sphere of $\mathbb C^d$.
\end{lemma}

\begin{proof}
Let $U \in \mathrm U(d)$ be arbitrary and consider the transformed random
vector $Ug$. By construction of $g$ and unitary invariance of the complex
Gaussian distribution, $Ug$ has the same distribution as $g$. Therefore
\begin{align}
  U \ket{\Psi}
  = \frac{1}{\|g\|_2}
    \sum_{\alpha=1}^d g_\alpha \, U\ket{\alpha}
  = \frac{1}{\|Ug\|_2}
    \sum_{\alpha=1}^d (Ug)_\alpha \ket{\alpha}
\end{align}
has the same distribution as $\ket{\Psi}$. In other words, the probability
measure of $\ket{\Psi}$ on the unit sphere is invariant under the natural
action of $\mathrm U(d)$. By uniqueness of the $\mathrm U(d)$-invariant probability measure
on the unit sphere $\mathbb S^{2d-1}$, this measure must coincide with
$\mu_{\mathrm{Haar}}$. Hence $\ket{\Psi}$ is Haar distributed.
\end{proof}
The previous lemma shows that a Haar-random state can be generated by
sampling a standard complex Gaussian vector and normalizing it. This
representation allows us to compute \emph{exactly} the distribution of the
overlap between a Haar-random state and any fixed reference vector.
In particular, we obtain the following classical fact, for which we include
a proof for completeness. We remark that large-deviation tools such as
Lévy's lemma~\cite{Popescu_2006,Hayden_2004} would only yield exponential tail bounds, whereas the following argument provides the precise dependence that will be crucial for our purposes (notably, it reproduces the fact that the probability vanishes exactly at $\varepsilon = 1$).

\begin{lemma}[Overlap with a fixed vector is Beta distributed]
\label{le:beta-overlap}
Let $\ket{\Psi}$ be Haar distributed on $\mathbb C^d$ and let
$\ket{v} \in \mathbb C^d$ be a fixed unit vector. Define
$X \coloneqq \bigl|\braket{v}{\Psi}\bigr|^2$. Then $X$ has probability density
\begin{align}
  p_X(x)
  = (d-1)\,(1-x)^{d-2},
  \qquad x \in [0,1],
\end{align}
that is, $X$ follows a Beta distribution with parameters $(1,d-1)$,
which we denote by $X \sim \mathrm{Beta}(1,d-1)$
(see notation subsection~\ref{sub:notation}).
In particular, for every $\varepsilon \in [0,1]$,
\begin{align}
  \Pr_{\Psi \sim \mu_{\mathrm{Haar}}}
  \bigl[\bigl|\braket{v}{\Psi}\bigr|^2 \ge \varepsilon\bigr]
  = (1-\varepsilon)^{d-1}
  \le \exp\bigl(-(d-1)\,\varepsilon\bigr).
\end{align}
\end{lemma}

\begin{proof}
By unitary invariance of the Haar measure we may assume
$\ket{v} = \ket{1}$, the first vector of a fixed orthonormal basis
$\{\ket{\alpha}\}_{\alpha=1}^d$ of $\mathbb C^d$.
Using Lemma~\ref{le:haar-gaussian}, we can write $\ket{\Psi}
  = \frac{1}{\|g\|_2} \sum_{\alpha=1}^d g_\alpha \ket{\alpha},$ where $g = (g_1,\dots,g_d) \in \mathbb C^d$ is a standard complex
Gaussian vector (Definition~\ref{de:standard-complex-gaussian}).
Then
\begin{align}
  X
  = \bigl|\braket{1}{\Psi}\bigr|^2
  = \frac{|g_1|^2}{\sum_{\alpha=1}^d |g_\alpha|^2}.
\end{align}
Set $Y_\alpha \coloneqq |g_\alpha|^2$.
By Definition~\ref{de:standard-complex-gaussian}, the random variables
$Y_1,\dots,Y_d$ are i.i.d.\ $\mathrm{Exp}(1)$.
Consider the random vector
\begin{align}
  (P_1,\dots,P_d)
  \coloneqq
  \frac{1}{\sum_{\alpha=1}^d Y_\alpha}
  \bigl(Y_1,\dots,Y_d\bigr).
\end{align}
It is known (see, e.g., Refs.~\cite[Ch.~49]{JohnsonKotzBalakrishnan2000,DirichletWiki}) that then $P_1 \sim \mathrm{Beta}(1,d-1),$ where $\mathrm{Beta}(1,d-1)$ has probability density  
\begin{align}
  f_X(x)
  = \frac{x^{1-1}(1-x)^{(d-1)-1}}{B(1,d-1)}
  = (d-1)\,(1-x)^{d-2},
  \qquad x \in [0,1],
\end{align}
where $B(\cdot,\cdot)$ denotes the Beta function and we used
$B(1,d-1) = 1/(d-1)$.
Since, $X= P_1,$ this proves the claimed density of $X$.
For the tail bound, we integrate explicitly:
\begin{align}
  \Pr[X \ge \varepsilon]
  &= \int_\varepsilon^1 (d-1)\,(1-x)^{d-2}\,\mathrm dx = \Bigl[ -(1-x)^{d-1} \Bigr]_{x=\varepsilon}^{x=1}
   = (1-\varepsilon)^{d-1}.
\end{align}
Finally, using $\log(1-\varepsilon)\le -\varepsilon$ for $\varepsilon \in [0,1]$,
we obtain
\begin{align}
  (1-\varepsilon)^{d-1}
  = \exp\bigl((d-1)\log(1-\varepsilon)\bigr)
  \le \exp\bigl(-(d-1)\varepsilon\bigr),
\end{align}
which gives the exponential tail bound.
\end{proof}

We now turn to reduced density matrices of Haar-random bipartite states.
Let $\mathcal H_A \simeq \mathbb C^{d_A}$ and
$\mathcal H_B \simeq \mathbb C^{d_B}$, and consider a Haar-random unit
vector $\ket{\Psi} \in \mathcal H_A \otimes \mathcal H_B$.
We write
\begin{align}
  \rho_A \coloneqq \tr_B\bigl(\ketbra{\Psi}{\Psi}\bigr)
  \in \mathcal D(\mathcal H_A)
\end{align}
for the reduced density matrix on system~$A$.

By Lemma~\ref{le:haar-gaussian}, there exists a random matrix
$G \in \mathbb C^{d_A \times d_B}$ with i.i.d.\ standard complex Gaussian
entries (as in Definition~\ref{de:standard-complex-gaussian}) such that
\begin{align}
  \ket{\Psi}
  = \frac{1}{\|G\|_{2}}
    \sum_{i=1}^{d_A}\sum_{j=1}^{d_B} G_{ij}\,\ket{i}_A \otimes \ket{j}_B,
\end{align}
where $\{\ket{i}_A\}_{i=1}^{d_A}$ and $\{\ket{j}_B\}_{j=1}^{d_B}$ are fixed
orthonormal bases and $\|\cdot\|_{\mathrm 2}$ denotes the Frobenius norm.
The reduced state on $A$ is then
\begin{align}
  \rho_A
  \coloneqq \tr_B\bigl(\ketbra{\Psi}{\Psi}\bigr)
  = \frac{G G^\dagger}{\tr(G G^\dagger)}.
\end{align}
In other words, if we view the columns of $G$ as i.i.d.\ samples from a
standard complex Gaussian on $\mathbb C^{d_A}$, then
$W \coloneqq G G^\dagger$ is (up to a scalar factor) the empirical covariance
matrix of these samples (a complex Wishart matrix), and $\rho_A$ is just this
covariance matrix normalized to have unit trace.

The following results can be derived from first principles by analysing the
induced (normalized) Wishart distribution of the reduced states. For
convenience, and to keep the exposition concise, we instead quote a sharp
concentration bound for the maximum Schmidt coefficient of a Haar-random
bipartite pure state, stated as Proposition~6.36 in
Ref.~\cite{AubrunSzarek2017AliceBob}, and translate it into a bound on the
operator norm of the reduced state.

\begin{lemma}[Operator norm of a reduced Haar-random state]
\label{le:random-induced-opnorm}
Let $n \le s$ and let $\ket{\Psi}$ be a Haar-distributed pure state on
$\mathbb C^n \otimes \mathbb C^s$. Let
\begin{align}
  \rho_A
  \coloneqq \tr_B\bigl(\ketbra{\Psi}{\Psi}\bigr)
  \in \mathcal L(\mathbb C^n)
\end{align}
be the reduced state on the first subsystem. Then, for every $\delta \in (0,1)$,
with probability at least $1-\delta$,
\begin{align}
  \|\rho_A\|_\infty
  \;\le\;
  \Biggl(
    \frac{1}{\sqrt n}
    + \frac{
          1 + \sqrt{\tfrac{1}{n}\,\log\tfrac{1}{\delta}}
        }{\sqrt s}
  \Biggr)^2.
  \label{eq:induced-opnorm-delta}
\end{align}
\end{lemma}

\begin{proof}
Let $\lambda_1(\psi)\ge\cdots\ge\lambda_n(\psi)$ denote the Schmidt coefficients
of $\psi$ with respect to the bipartition $\mathbb C^n \otimes \mathbb C^s$. Then the eigenvalues of $\rho_A$ are $\{\lambda_1(\psi)^2,\dots,\lambda_n(\psi)^2\}$, so $ \|\rho_A\|_\infty
  = \lambda_1(\psi)^2.$ By Proposition~6.36 in Ref.~\cite{AubrunSzarek2017AliceBob}, for every
$\varepsilon > 0$,
\begin{align}
  \Pr\left[
    \lambda_1(\psi)
    \;\ge\;
    \frac{1}{\sqrt n}
    + \frac{1+\varepsilon}{\sqrt s}
  \right]
  \;\le\;
  \exp(-n\varepsilon^2).
\end{align}
Set $\varepsilon \coloneqq \sqrt{\tfrac{1}{n}\,\log\tfrac{1}{\delta}}$.
Then $\exp(-n\varepsilon^2) = \delta$, and hence, with probability at least
$1-\delta$,
\begin{align}
  \lambda_1(\psi)
  \;\le\;
  \frac{1}{\sqrt n}
  + \frac{
        1 + \sqrt{\tfrac{1}{n}\,\log\tfrac{1}{\delta}}
      }{\sqrt s}.
\end{align}
Squaring both sides yields \eqref{eq:induced-opnorm-delta}.
\end{proof}

We now specialise this bound to the bipartition relevant for our Choi-state
analysis.

\begin{corollary}[Operator norm of the input marginal in the Choi setting]
\label{co:choi-input-opnorm}
Let
$\mathcal H_{\mathrm{out}} \simeq \mathbb C^{d_{\mathrm{out}}}$,
$\mathcal H_{\mathrm{in}} \simeq \mathbb C^{d_{\mathrm{in}}}$,
$\mathcal H_E \simeq \mathbb C^{k}$,
and let
$\ket{\Psi} \in
  \mathcal H_{\mathrm{out}}
  \otimes \mathcal H_{\mathrm{in}}
  \otimes \mathcal H_E$
be Haar distributed.
Define
\begin{align}
  J
  &\coloneqq \tr_E\bigl(\ketbra{\Psi}{\Psi}\bigr)
   \in \mathcal L(\mathcal H_{\mathrm{out}}\otimes\mathcal H_{\mathrm{in}}),
  \\
  \rho_{\mathrm{in}}
  &\coloneqq \tr_{\mathrm{out},E}\bigl(\ketbra{\Psi}{\Psi}\bigr)
   \in \mathcal L(\mathcal H_{\mathrm{in}}).
\end{align}
Assume that $d_{\mathrm{in}} \le k\,d_{\mathrm{out}}$, which is automatically
satisfied if $k$ is the Kraus rank of a quantum channel
$\Lambda : \mathcal L(\mathcal H_{\mathrm{in}}) \to \mathcal L(\mathcal H_{\mathrm{out}})$
(see Subsection~\ref{subsec:qchannels}).
Then, for every $\delta_{\mathrm{Haar}} \in (0,1)$, with probability at least
$1-\delta_{\mathrm{Haar}}$ over the choice of $\ket{\Psi}$,
\begin{align}
  \|\rho_{\mathrm{in}}\|_\infty
  \le
  \frac{1}{d_{\mathrm{in}}}
  \Biggl(
    1
    + \Bigl(
        1 + \sqrt{\frac{1}{d_{\mathrm{in}}}\,
                     \log\frac{1}{\delta_{\mathrm{Haar}}}}
      \Bigr)
      \sqrt{\frac{d_{\mathrm{in}}}{k\,d_{\mathrm{out}}}}
  \Biggr)^2\le  \frac{1}{d_{\mathrm{in}}}
  \Biggl(
    2+
      \sqrt{\frac{1}{k\,d_{\mathrm{out}}}\,
                     \log\frac{1}{\delta_{\mathrm{Haar}}}}
  \Biggr)^2
  \label{eq:choi-input-opnorm}
\end{align}
\end{corollary}

\begin{proof}
The claim follows from Lemma~\ref{le:random-induced-opnorm} with
$\delta = \delta_{\mathrm{Haar}}$, $n = d_{\mathrm{in}}$ and
$s = k\,d_{\mathrm{out}}$.
Indeed, view $\ket{\Psi}$ as a vector in
$\mathcal H_A \otimes \mathcal H_B$ with
$\mathcal H_A \coloneqq \mathcal H_{\mathrm{in}}$ and
$\mathcal H_B \coloneqq \mathcal H_{\mathrm{out}}\otimes\mathcal H_E$.
Then
$\rho_{\mathrm{in}}
 = \tr_{\mathrm{out},E}(\ketbra{\Psi}{\Psi})
 = \tr_B(\ketbra{\Psi}{\Psi})$,
so Lemma~\ref{le:random-induced-opnorm} applied to this bipartition yields
\eqref{eq:choi-input-opnorm}. In the last step, we use $d_{\mathrm{in}} \le k\,d_{\mathrm{out}}$.
\end{proof}


\subsection{Quantum state tomography}
In this subsection we recall the basic formulation of quantum state tomography,
summarise its optimal sample-complexity scaling, and collect several ingredients
that will be central for us: (i) an optimal pure-state tomography protocol based
on Hayashi's covariant measurement~\cite{Hay98,HayashiHashimotoHoribe2005,Holevo1982},
reviewed in Subsubsection~\ref{subsub:optpure};
(ii) a convenient structural description of its error, namely that the
deviation of the estimator from the true state can be modelled as pointing in a
Haar-random direction orthogonal to the true state
(Subsubsection~\ref{subsub:errorHaar});
(iii) a simple symmetrisation argument showing that any tomography scheme can be
modified, without degrading its sample complexity, so as to satisfy this error
structure (Subsubsection~\ref{subsub:covariantError}); and
(iv) a random purification channel introduced in
Ref.~\cite{tang2025conjugatequerieshelp} which, given copies of a state as
input, outputs copies of a random purification, and which was recently used to
perform optimal mixed-state tomography via a reduction to pure-state tomography
in Ref.~\cite{pelecanos2025mixedstatetomographyreduces}
(Subsubsection~\ref{subsub:reducing}).

We work on a $d$-dimensional Hilbert space $\mathbb C^d$, and write
$\mathcal D(\mathbb C^d)$ for the set of density operators on $\mathbb C^d$.
A \emph{tomography scheme} for a family of states $\mathcal F \subseteq \mathcal D(\mathbb C^d)$ with number of states copies $N$ consists of
a POVM on $(\mathbb C^d)^{\otimes N}$ together with a classical
post-processing map that, on input $N$ copies of an unknown state
$\rho \in \mathcal F$, outputs an estimator
$\hat\rho \in \mathcal D(\mathbb C^d)$.
The accuracy of state tomography is typically measured in trace distance,
$d_{\mathrm{tr}}(\rho,\sigma) \coloneqq \tfrac12 \|\rho - \sigma\|_1$, which is
widely regarded as the most operationally meaningful distance between quantum
states~\cite{WatrousBook,Helstrom1976}.

Given accuracy and failure parameters $\varepsilon,\delta \in (0,1)$, we call a
tomography scheme \emph{$(\varepsilon,\delta)$-accurate on} $\mathcal F$ if,
for every $\rho \in \mathcal F$, the corresponding estimator $\hat\rho$
satisfies
\begin{align}
  \Pr\bigl[d_{\mathrm{tr}}(\hat\rho,\rho) \le \varepsilon\bigr]
  \;\ge\; 1 - \delta .
\end{align}

The \emph{sample complexity} of tomography on $\mathcal F$, denoted
$N_{\mathrm{state}}(\mathcal F,\varepsilon,\delta)$, is the smallest
$N \in \mathbb N$ for which there exists an $(\varepsilon,\delta)$-accurate
scheme using $N$ copies.

Let $\mathcal D_{d,r} \subseteq \mathcal D(\mathbb C^d)$ denote the set of
states of rank at most $r$.
The optimal sample complexity of finite-dimensional state tomography in trace
distance is by now understood up to constant factors, thanks to a series of
works (see, e.g.,~\cite{Hay98,KRT17,Har13,Haah_2017,OW16,SSW25,pelecanos2025mixedstatetomographyreduces}), and it scales as
\begin{align}
  N_{\mathrm{state}}(\mathcal D_{d,r},\varepsilon,\delta)
  = \Theta\Bigl(\frac{rd + \log(1/\delta)}{\varepsilon^2}\Bigr).
\end{align}
In particular, for pure states ($r=1$) one has
$\Theta(d/\varepsilon^2)$, while for arbitrary mixed states ($r=d$) this
becomes $\Theta(d^2/\varepsilon^2)$.
The best currently known upper bounds with this scaling are achieved in
Ref.~\cite{pelecanos2025mixedstatetomographyreduces}, while
Ref.~\cite{SSW25} proves matching information-theoretic lower bounds
$\Omega(rd/\varepsilon^2)$.
In what follows we first focus on the pure-state case and recall a concrete
measurement scheme that achieves this optimal scaling.
\subsubsection{Optimal pure-state tomography}
\label{subsub:optpure}
A canonical tomography scheme achieving the optimal sample-complexity scaling
for pure states is due to Hayashi~\cite{Hay98}. Restricted to the symmetric
subspace
\(
  \vee^N \mathbb C^d \subset (\mathbb C^d)^{\otimes N}
\)
(see, e.g., Ref.~\cite{Har13} for preliminaries on the symmetric subspace),
the scheme is specified by a POVM with continuous outcome space given by pure
states $\ket{u} \in \mathbb C^d$ and POVM density
\begin{align}
  F(u)
  \coloneqq \dimSym(N,d)\,\ketbra{u}{u}^{\otimes N}\,\mathrm d u ,
\end{align}
where
\begin{align}
  \dimSym(N,d)
  \coloneqq \dim\bigl(\vee^N \mathbb C^d\bigr)
  = \binom{d+N-1}{N},
\end{align}
and $\mathrm d u$ denotes the normalized Haar measure on the unit sphere of
$\mathbb C^d$. One checks that
\begin{align}
  \int \ketbra{u}{u}^{\otimes N}\,\mathrm d u
  = \frac{1}{\dimSym(N,d)}\,\Pi_{\mathrm{sym}}^{(N)},
\end{align}
so that $\int F(u)
  = \Pi_{\mathrm{sym}}^{(N)},$ the identity on $\vee^N \mathbb C^d$. Thus $\{F(u)\}_u$ defines a POVM on the
symmetric subspace. If desired, one can extend this to a POVM on the full space
$(\mathbb C^d)^{\otimes N}$ by adding a dummy outcome with POVM element
$\Pi_{\mathrm{sym}}^{(N)\perp}$; since all input states are of the form
$\ket{\psi}^{\otimes N}$ and hence lie in $\vee^N \mathbb C^d$, this extension
does not affect the measurement statistics.

Given $N$ copies of an unknown pure state $\ket{\psi} \in \mathbb C^d$, we
apply the POVM $\{F(u)\}_u$ described above. The outcome space is the unit
sphere of $\mathbb C^d$, parametrised by unit vectors $u$, and probabilities
are described by a probability density $p(\cdot \mid \psi)$ with respect to the
Haar measure $\mathrm d u$. For fixed $\ket{\psi}$, the Born rule gives
\begin{align}
  p(u \mid \psi)\,\mathrm d u
  \coloneqq \tr\bigl[F(u)\,\ketbra{\psi}{\psi}^{\otimes N}\bigr]
  = \dimSym(N,d)\,\bigl|\braket{u}{\psi}\bigr|^{2N}\,\mathrm d u.
\end{align}
The associated estimator simply identifies the outcome with the estimate, i.e.\
if the measurement returns the outcome $u$ we set $\ket{v} \coloneqq \ket{u}$.

Let $X \coloneqq \bigl|\braket{v}{\psi}\bigr|^2$. By
Lemma~\ref{le:beta-overlap}, if $u$ is Haar-distributed on the unit sphere of
$\mathbb C^d$ then $Y \coloneqq \bigl|\braket{u}{\psi}\bigr|^2$ has density
$(d-1)(1-x)^{d-2}$ on $[0,1]$ with respect to $\mathrm d x$, i.e.\
$Y \sim \mathrm{Beta}(1,d-1)$. In Hayashi's tomography scheme, the probability
of obtaining an outcome $u$ is proportional to $|\braket{u}{\psi}|^{2N} = Y^N$
times its Haar probability. Hence, the probability that $Y$ lies in a small
interval $[x,x+\mathrm d x]$ is proportional to
$x^N (d-1)(1-x)^{d-2}\,\mathrm d x$. After normalising, this is exactly the
density of a $\mathrm{Beta}(N+1,d-1)$ random variable, so
$Y \sim \mathrm{Beta}(N+1,d-1)$. Since the estimator is defined by
$\ket{v} \coloneqq \ket{u}$, the same holds for
$X = \bigl|\braket{v}{\psi}\bigr|^2$, and we conclude that
\begin{align}
  X \sim \mathrm{Beta}(N+1,d-1).
\end{align}
In particular, this gives an exact expression for the probability density of
the overlap between the true state and Hayashi's estimator $\ket{v}$.

Thus, for any $\eta \in (0,1)$, the probability that Hayashi's scheme achieves
squared fidelity (squared overlap) at least $1-\eta$ is
\begin{align}
  \Pr\bigl[\,|\braket{v}{\psi}|^2 \ge 1-\eta\,\bigr]
  = \Pr\bigl[X \ge 1-\eta\bigr]
  = \int_{1-\eta}^1
      \frac{x^{N}(1-x)^{d-2}}{B(N+1,d-1)}\,\mathrm d x,
\end{align}
where $B(\cdot,\cdot)$ denotes the Beta function (see the notation subsection~\ref{sub:notation}).

Using this explicit overlap distribution, Ref.~\cite{pelecanos2025mixedstatetomographyreduces}
derives tail bounds for Hayashi's estimator and, in particular, its resulting
sample complexity, which we summarise next.

\begin{lemma}[Hayashi's pure-state tomography with high probability]
\label{le:hayashi-tomography}
Let $\ket{\psi} \in \mathbb C^d$ be an unknown pure state, and let
$\ket{v} \in \mathbb C^d$ denote the estimate returned by Hayashi's measurement
when applied to $\ket{\psi}^{\otimes N}$. Then, for all $\eta,\delta \in (0,1)$,
if
\begin{align}
  N \;\ge\; 4\,\frac{d + \log(1/\delta)}{\eta} ,
\end{align}
one has
\begin{align}
  \Pr\bigl[\,|\braket{v}{\psi}|^2 \ge 1 - \eta\,\bigr]
  \;\ge\; 1 - \delta .
\end{align}
In particular, a careful inspection of the proof of
Ref.~\cite[Proposition~5.1]{pelecanos2025mixedstatetomographyreduces}
shows that the universal constant in the pure-state bound can be taken to be $4$.
\end{lemma}

For pure states, trace distance and fidelity are related by
$d_{\mathrm{tr}}(\ketbra{\psi}{\psi},\ketbra{v}{v})
 = \sqrt{1 - |\braket{v}{\psi}|^2}$.
Hence Lemma~\ref{le:hayashi-tomography} implies that, with probability at least
$1-\delta$, one has
$d_{\mathrm{tr}}(\ketbra{\psi}{\psi},\ketbra{v}{v}) \le \varepsilon$
whenever
\begin{align}
  N \;\ge\; 4\,\frac{d + \log(1/\delta)}{\varepsilon^2} ,
\end{align}
obtained by applying the lemma with $\eta = \varepsilon^2$.

\subsubsection{Distribution of the error state}
\label{subsub:errorHaar}
An important feature of Hayashi's pure-state tomography scheme, reviewed in the
previous subsection, is the distribution of its error state. In particular, we
will show that one can decompose the outcome as
\begin{align}
  \ket{v}
  = \sqrt{1-\varepsilon^2}\,\ket{\psi} + \varepsilon\,\ket{\psi_{\mathrm{err}}},
\end{align}
for some random error parameter $\varepsilon \in [0,1]$ (for Hayashi's scheme $1-\varepsilon^2$ follows a $\mathrm{Beta}(N+1,d-1)$ distribution), and
such that the error direction $\ket{\psi_{\mathrm{err}}}$ is Haar-distributed
in the subspace orthogonal to $\ket{\psi}$. This ultimately follows from the
covariance property satisfied by Hayashi's
scheme~\cite{Hay98,HayashiHashimotoHoribe2005,Holevo1982}, which we now define.

We first formalise covariance for a general pure-state tomography scheme.

\begin{definition}[Covariant pure-state tomography]
\label{de:covariant}
Let $\mathcal A$ be a pure-state tomography scheme on $\mathbb C^d$ that, on
input $\ket{\psi}^{\otimes N}$, outputs an estimate $\ket{v} \in \mathbb C^d$
with outcome density $p_{\mathcal A}(v \mid \psi)$. We say that $\mathcal A$ is
\emph{covariant} if for every unitary $U \in \mathrm U(d)$ and every unit
vector $\ket{\psi}$ one has
\begin{align}
  p_{\mathcal A}(Uv \mid U\psi) = p_{\mathcal A}(v \mid \psi) .
\end{align}
\end{definition}
A covariant scheme has the convenient feature that its performance is the same
for all input states $\ket{\psi}$: in particular, the distribution of any
unitarily invariant error metric (such as the trace distance to the true state)
does not depend on $\ket{\psi}$, so its worst-case performance coincides with
its average performance over Haar-random inputs.

Hayashi's POVM is covariant in this sense. Indeed, for his scheme
$p(v \mid \psi) = \tr[F(v)\ketbra{\psi}{\psi}^{\otimes N}]$, and using
$F(Uv) = U^{\otimes N}F(v)U^{\otimes N\dagger}$ and
$\ketbra{U\psi}{U\psi}^{\otimes N}
 = U^{\otimes N}\ketbra{\psi}{\psi}^{\otimes N}U^{\otimes N\dagger}$, we obtain
\begin{align}
  p(Uv \mid U\psi)
  = \tr[F(Uv)\ketbra{U\psi}{U\psi}^{\otimes N}]
  = \tr[F(v)\ketbra{\psi}{\psi}^{\otimes N}]
  = p(v \mid \psi).
\end{align}

We can now describe the error direction for any covariant scheme, and in
particular for Hayashi's scheme.

\begin{lemma}[Isotropic error direction for covariant tomography schemes]
\label{le:hayashi-haar-error}
Let $\ket{\psi} \in \mathbb C^d$ be a fixed pure state, and let
$\ket{v} \in \mathbb C^d$ denote the outcome of a covariant pure-state
tomography scheme applied to $\ket{\psi}^{\otimes N}$. Define the random error
amplitude $\varepsilon \in [0,1]$ by
\begin{align}
  \varepsilon^2 \coloneqq 1 - \bigl|\braket{v}{\psi}\bigr|^2 .
\end{align}
Then there exists a random unit vector $\ket{\psi_{\mathrm{err}}}$ orthogonal
to $\ket{\psi}$, Haar-distributed on the unit sphere in the subspace orthogonal
to $\ket{\psi}$ and independent of $\varepsilon$, such that
\begin{align}
\label{eq:covariant-error-decomp}
  \ket{v}
  = \sqrt{1-\varepsilon^2}\,\ket{\psi} + \varepsilon\,\ket{\psi_{\mathrm{err}}}.
\end{align}
In particular, if $\ket{\psi_{\mathrm{Haar}}}$ is Haar-distributed on the unit
sphere of $\mathbb C^d$, then $\ket{\psi_{\mathrm{err}}}$ has the same
distribution as
\begin{align}
  \ket{\psi_{\mathrm{err}}}
  = \frac{\bigl(\mathbb I - \ketbra{\psi}{\psi}\bigr)\ket{\psi_{\mathrm{Haar}}}}
         {\bigl\|\bigl(\mathbb I - \ketbra{\psi}{\psi}\bigr)\ket{\psi_{\mathrm{Haar}}}\bigr\|_2} \, .
\end{align}
\end{lemma}

\begin{proof}
Set $X \coloneqq \bigl|\braket{v}{\psi}\bigr|^2 = 1-\varepsilon^2$. We assume
$\varepsilon>0$, otherwise the statement is trivial. Write
$\ket{v}
 = \braket{\psi}{v}\,\ket{\psi}
   + \bigl(\ket{v} - \braket{\psi}{v}\,\ket{\psi}\bigr)$
and define
\[
  \ket{\Phi}
  \coloneqq
  \frac{\ket{v} - \braket{\psi}{v}\,\ket{\psi}}
       {\bigl\|\ket{v} - \braket{\psi}{v}\,\ket{\psi}\bigr\|_2} ,
\]
so that $\ket{\Phi}$ is a unit vector orthogonal to $\ket{\psi}$ and
$\bigl\|\ket{v} - \braket{\psi}{v}\,\ket{\psi}\bigr\|_2^2 = 1-X = \varepsilon^2$.
Writing $\braket{\psi}{v} = e^{i\theta}\sqrt{X}$ for some
$\theta \in [0,2\pi)$, we obtain $\ket{v}
  = \sqrt{X}\,\ket{\psi} + \sqrt{1-X}\,e^{i\theta}\ket{\Phi}.$
Absorbing the phase into the orthogonal component by setting
$\ket{\psi_{\mathrm{err}}} \coloneqq e^{i\theta}\ket{\Phi}$ and using
$\varepsilon^2 = 1-X$ yields
\[
  \ket{v}
  = \sqrt{1-\varepsilon^2}\,\ket{\psi} + \varepsilon\,\ket{\psi_{\mathrm{err}}}.
\]

We now determine the distribution of $\ket{\psi_{\mathrm{err}}}$. Let
$\mathrm{Stab}_\psi \coloneqq \{U \in \mathrm U(d) : U\ket{\psi} = \ket{\psi}\}$.
By covariance (Definition~\ref{de:covariant}), for every
$U \in \mathrm{Stab}_\psi$ the random vectors $\ket{v}$ and $U\ket{v}$ have the
same distribution, i.e.\ $p(Uv \mid \psi) = p(v \mid \psi)$. Applying the
decomposition~\eqref{eq:covariant-error-decomp} to $U\ket{v}$ shows that the
associated pair $(\varepsilon,\ket{\psi_{\mathrm{err}}})$ has the same
distribution as $(\varepsilon,U\ket{\psi_{\mathrm{err}}})$ for all
$U \in \mathrm{Stab}_\psi$.

Fix $e \in [0,1]$ and let $\mu_e$ be the conditional distribution of
$\ket{\psi_{\mathrm{err}}}$ given $\varepsilon=e$. The invariance just
established implies that $\mu_e$ is invariant under the action of
$\mathrm{Stab}_\psi$ on the subspace orthogonal to $\ket{\psi}$; this action is
equivalent to the standard action of $\mathrm U(d-1)$ on the unit sphere in
$\mathbb C^{d-1}$. By uniqueness of the unitarily invariant probability measure
on that sphere, $\mu_e$ must coincide with the Haar measure there. In particular,
$\mu_e$ does not depend on $e$, which implies that $\ket{\psi_{\mathrm{err}}}$
is Haar-distributed on the unit sphere orthogonal to $\ket{\psi}$ and
independent of $\varepsilon$.

Finally, we justify the ``in particular'' statement. Let $\ket{\psi_{\mathrm{Haar}}}$
be Haar-distributed on the unit sphere of $\mathbb C^d$, and define
\[
  \ket{\phi}
  \coloneqq
  \frac{\bigl(\mathbb I - \ketbra{\psi}{\psi}\bigr)\ket{\psi_{\mathrm{Haar}}}}
       {\bigl\|\bigl(\mathbb I - \ketbra{\psi}{\psi}\bigr)\ket{\psi_{\mathrm{Haar}}}\bigr\|_2} .
\]
Then $\ket{\phi}$ is a random unit vector orthogonal to $\ket{\psi}$. For every
$U \in \mathrm{Stab}_\psi$ we have $U\ket{\psi_{\mathrm{Haar}}}$ Haar-distributed
on $\mathbb C^d$, and
\[
  \bigl(\mathbb I - \ketbra{\psi}{\psi}\bigr)U\ket{\psi_{\mathrm{Haar}}}
  = U\bigl(\mathbb I - \ketbra{\psi}{\psi}\bigr)\ket{\psi_{\mathrm{Haar}}}.
\]
Hence the distribution of $\ket{\phi}$ is invariant under the action of
$\mathrm{Stab}_\psi$ on the unit sphere in the subspace orthogonal to
$\ket{\psi}$. As above, this action is equivalent to the standard
$\mathrm U(d-1)$ action on the unit sphere in $\mathbb C^{d-1}$, and by uniqueness
of the unitarily invariant probability measure on that sphere, the distribution
of $\ket{\phi}$ coincides with the Haar measure there. Since we already showed
that $\ket{\psi_{\mathrm{err}}}$ is Haar-distributed on the same sphere, it
follows that $\ket{\psi_{\mathrm{err}}}$ has the same distribution as
$\ket{\phi}$.
\end{proof}

Lemma~\ref{le:hayashi-haar-error} describes the error state as a Haar-random
direction on the sphere orthogonal to $\ket{\psi}$. For our later analysis it
will be more convenient to work instead with a globally Haar-random vector in
$\mathbb C^d$. The next corollary shows that we can equivalently express the
outcome $\ket{v}$ as a linear combination of the true state $\ket{\psi}$ and a
global Haar state, with explicit (random) coefficients.

\begin{corollary}[Error state as a global Haar state]
\label{co:haar-two-dim}
Let $\ket{\psi} \in \mathbb C^d$ be a fixed pure state, and let
$\ket{v} \in \mathbb C^d$ denote the outcome of a covariant pure-state
tomography scheme applied to $\ket{\psi}^{\otimes N}$. Define the random error
amplitude $\varepsilon \in [0,1]$ by
\begin{align}
  \varepsilon^2 \coloneqq 1 - \bigl|\braket{v}{\psi}\bigr|^2 .
\end{align}
Let $\ket{\psi_{\mathrm{Haar}}}$ be Haar-distributed on the unit sphere of
$\mathbb C^d$, independent of $\varepsilon$, and set
\begin{align}
  f_{\mathrm{Haar}} \coloneqq \braket{\psi}{\psi_{\mathrm{Haar}}} \in \mathbb C.
\end{align}
Then $\ket{v}$ admits the decomposition
\begin{align}
  \ket{v} = a\,\ket{\psi} + b\,\ket{\psi_{\mathrm{Haar}}},
\end{align}
with coefficients
\begin{align}
  a &= \sqrt{1-\varepsilon^2}
       \;-\; \frac{\varepsilon\,f_{\mathrm{Haar}}}{\sqrt{1 - |f_{\mathrm{Haar}}|^2}}, \\
  b &= \frac{\varepsilon}{\sqrt{1 - |f_{\mathrm{Haar}}|^2}} .
\end{align}
\end{corollary}

\begin{proof}
By Lemma~\ref{le:hayashi-haar-error}, there exists a unit vector $\ket{\psi_{\mathrm{err}}} \perp \ket{\psi}$,
independent of $\varepsilon$, such that
\begin{align}
  \ket{v}
  = \sqrt{1-\varepsilon^2}\,\ket{\psi}
    + \varepsilon\,\ket{\psi_{\mathrm{err}}}.
\end{align}
Moreover, the lemma implies that for a Haar-distributed
$\ket{\psi_{\mathrm{Haar}}}$ on the unit sphere of $\mathbb C^d$,
\begin{align}
  \ket{\psi_{\mathrm{err}}}
  = \frac{(\mathbb I - \ketbra{\psi}{\psi})\ket{\psi_{\mathrm{Haar}}}}
         {\bigl\|(\mathbb I - \ketbra{\psi}{\psi})\ket{\psi_{\mathrm{Haar}}}\bigr\|_2}.
\end{align}
Writing
$f_{\mathrm{Haar}} \coloneqq \braket{\psi}{\psi_{\mathrm{Haar}}}$, we have
\begin{align}
  (\mathbb I - \ketbra{\psi}{\psi})\ket{\psi_{\mathrm{Haar}}}
  &= \ket{\psi_{\mathrm{Haar}}} - f_{\mathrm{Haar}}\ket{\psi}, \\
  \bigl\|(\mathbb I - \ketbra{\psi}{\psi})\ket{\psi_{\mathrm{Haar}}}\bigr\|_2^2
  &= 1 - |f_{\mathrm{Haar}}|^2,
\end{align}
and hence
\begin{align}
  \ket{\psi_{\mathrm{err}}}
  = \frac{\ket{\psi_{\mathrm{Haar}}} - f_{\mathrm{Haar}}\ket{\psi}}
         {\sqrt{1 - |f_{\mathrm{Haar}}|^2}}.
\end{align}
Substituting into the expression for $\ket{v}$ gives
\begin{align}
  \ket{v}
  &= \sqrt{1-\varepsilon^2}\,\ket{\psi}
     + \varepsilon\,\frac{\ket{\psi_{\mathrm{Haar}}} - f_{\mathrm{Haar}}\ket{\psi}}
                         {\sqrt{1 - |f_{\mathrm{Haar}}|^2}} \\
  &= \left(\sqrt{1-\varepsilon^2}
           - \frac{\varepsilon\,f_{\mathrm{Haar}}}{\sqrt{1 - |f_{\mathrm{Haar}}|^2}}\right)
      \ket{\psi}
     + \frac{\varepsilon}{\sqrt{1 - |f_{\mathrm{Haar}}|^2}}\ket{\psi_{\mathrm{Haar}}}.
\end{align}
This is the claimed decomposition with the stated coefficients $a$ and $b$.
\end{proof}

We also note that $f_{\mathrm{Haar}}$ introduced above is typically very small.
Indeed, by Lemma~\ref{le:beta-overlap}, if $\ket{\psi_{\mathrm{Haar}}} \in \mathbb C^d$ is
Haar-distributed then $|f_{\mathrm{Haar}}|^2 = |\braket{\psi}{\psi_{\mathrm{Haar}}}|^2$
has a $\mathrm{Beta}(1,d-1)$ distribution with mean $1/d$,
and strong concentration around this value. In particular, with high probability
one has $|f_{\mathrm{Haar}}|^2 = O(1/d)$, so the correction terms involving
$f_{\mathrm{Haar}}$ in the coefficients $a$ and $b$ are typically very small in
high dimensions. Nonetheless, we will not rely on this heuristic and will
instead analyze the dependence on $f_{\mathrm{Haar}}$ explicitly in the proof
of our main theorem.

\subsubsection{Enforcing Haar-distributed error for pure-state tomography}
\label{subsub:covariantError}
We have seen that Hayashi's algorithm satisfies the covariance property, which
in turn implies that the corresponding error direction is Haar-distributed in
the orthogonal complement of the true state
(Lemma~\ref{le:hayashi-haar-error}). We now show that covariance is not a
restriction: starting from an arbitrary pure-state tomography scheme
$\mathcal A$, one can construct a new scheme $\mathcal A'$ that is covariant
and has the same sample-complexity guarantees with respect to any unitarily
invariant error metric. In particular, by
Lemma~\ref{le:hayashi-haar-error}, $\mathcal A'$ has a Haar-distributed error
direction.

The construction uses a simple Haar-twirling procedure.

\begin{lemma}[Haar twirling yields a covariant scheme without degrading performance]
\label{le:haar-twirl-covariant}
Let $\mathcal A$ be a pure-state tomography scheme on $\mathbb C^d$ that, on
input $\ket{\psi}^{\otimes N}$, outputs an estimate $\ket{v} \in \mathbb C^d$
with outcome density $p_{\mathcal A}(v \mid \psi)$. Define a new scheme
$\mathcal A'$ as follows:
\begin{enumerate}
  \item Sample a Haar-random unitary $U \in \mathrm U(d)$.
  \item Apply $U$ to each input copy, obtaining $(U\ket{\psi})^{\otimes N}$.
  \item Run $\mathcal A$ on $(U\ket{\psi})^{\otimes N}$ and obtain an estimate
        $\ket{w} \in \mathbb C^d$.
  \item Output $\ket{v'} \coloneqq U^\dagger \ket{w}$.
\end{enumerate}
Then the following hold:
\begin{enumerate}[(i)]
  \item The outcome density of $\mathcal A'$ is
  \begin{align}
    p_{\mathcal A'}(v \mid \psi)
    = \int_{\mathrm U(d)} p_{\mathcal A}(U v \mid U\psi)\,\mathrm dU ,
    \label{eq:A-prime-density}
  \end{align}
  where $\mathrm dU$ denotes Haar measure on $\mathrm U(d)$.
  \item $\mathcal A'$ is covariant in the sense of Definition~\ref{de:covariant},
  i.e.\ $p_{\mathcal A'}(Uv \mid U\psi) = p_{\mathcal A'}(v \mid \psi)$ for all
  $U \in \mathrm U(d)$.
  \item Let $d(\cdot,\cdot)$ be any unitarily invariant distance on pure states.
  If for some $N,\varepsilon,\delta \in (0,1)$ the scheme $\mathcal A$ satisfies
  \begin{align}
    \Pr_{\mathcal A}\bigl[d(\ket{\psi},\ket{v_{\mathcal A}}) > \varepsilon
                          \,\bigm|\, \ket{\psi}^{\otimes N}\bigr]
    \le \delta
    \quad \text{for all pure states } \ket{\psi} \in \mathbb C^d ,
    \label{eq:A-uniform-accuracy}
  \end{align}
  where $\ket{v_{\mathcal A}}$ denotes the (random) output of $\mathcal A$, then
  the Haar-twirled scheme $\mathcal A'$ also satisfies
  \begin{align}
    \Pr_{\mathcal A'}\bigl[d(\ket{\psi},\ket{v_{\mathcal A'}}) > \varepsilon
                           \,\bigm|\, \ket{\psi}^{\otimes N}\bigr]
    \le \delta
    \quad \text{for all pure states } \ket{\psi} \in \mathbb C^d ,
    \label{eq:A-prime-uniform-accuracy}
  \end{align}
  where $\ket{v_{\mathcal A'}}$ is the output of $\mathcal A'$. In particular,
  Haar twirling does not increase the sample complexity required to achieve
  $(\varepsilon,\delta)$-accurate tomography with respect to $d$.
  \item Consequently, since $\mathcal A'$ is covariant, the conclusion of
  Lemma~\ref{le:hayashi-haar-error} and Corollary~\ref{co:haar-two-dim} apply to $\mathcal A'$: there exists a
  random variable $E \in [0,1]$ and a random unit vector
  $\ket{\psi_{\mathrm{err}}} \perp \ket{\psi}$, Haar-distributed on the unit
  sphere in the subspace orthogonal to $\ket{\psi}$ and independent of $E$,
  such that the outcome of $\mathcal A'$ can be written as
  \begin{align}
    \ket{v_{\mathcal A'}}
    = \sqrt{1-E}\,\ket{\psi} + \sqrt{E}\,\ket{\psi_{\mathrm{err}}}.
  \end{align}
\end{enumerate}
\end{lemma}

\begin{proof}
For fixed $\ket{\psi}$ and $\ket{v}$, the scheme $\mathcal A'$ outputs $\ket{v}$
if and only if, for the sampled $U$, the original scheme $\mathcal A$ outputs
$\ket{w} = U\ket{v}$ when run on $(U\ket{\psi})^{\otimes N}$. Conditional on
$U$, this happens with density $p_{\mathcal A}(U v \mid U\psi)$, so by the law
of total probability we obtain~\eqref{eq:A-prime-density} by averaging over $U$,
which proves (i).

For (ii), fix $U_0 \in \mathrm U(d)$ and compute
\begin{align}
  p_{\mathcal A'}(U_0 v \mid U_0\psi)
  &= \int_{\mathrm U(d)} p_{\mathcal A}(U (U_0 v) \mid U (U_0\psi))\,\mathrm dU \\
  &= \int_{\mathrm U(d)} p_{\mathcal A}(W v \mid W \psi)\,\mathrm dW
   = p_{\mathcal A'}(v \mid \psi),
\end{align}
where we used the change of variables $W \coloneqq U U_0$ and left invariance
of the Haar measure. Thus $\mathcal A'$ is covariant.

For (iii), let $\ket{v_{\mathcal A'}}$ denote the random output of $\mathcal A'$
on input $\ket{\psi}^{\otimes N}$, and let $\ket{v_{\mathcal A}}$ be the output
of $\mathcal A$ on input $(U\ket{\psi})^{\otimes N}$ for the same Haar-random
$U$. By construction, $\ket{v_{\mathcal A'}} = U^\dagger \ket{v_{\mathcal A}}$,
so unitary invariance of $d$ gives
\begin{align}
  d(\ket{\psi},\ket{v_{\mathcal A'}})
  = d\bigl(\ket{\psi},U^\dagger\ket{v_{\mathcal A}}\bigr)
  = d\bigl(U\ket{\psi},\ket{v_{\mathcal A}}\bigr).
\end{align}
Thus, conditional on $U$, the error of $\mathcal A'$ at $\ket{\psi}$ has the
same distribution as the error of $\mathcal A$ at the rotated state
$U\ket{\psi}$, and hence
\begin{align}
  \Pr_{\mathcal A'}\bigl[d(\ket{\psi},\ket{v_{\mathcal A'}}) > \varepsilon
                         \,\bigm|\, U\bigr]
  = \Pr_{\mathcal A}\bigl[d(U\ket{\psi},\ket{v_{\mathcal A}}) > \varepsilon
                          \,\bigm|\, (U\ket{\psi})^{\otimes N}\bigr].
\end{align}
By the uniform accuracy assumption~\eqref{eq:A-uniform-accuracy}, the right-hand
side is bounded by $\delta$ for all $U$ and $\ket{\psi}$, because $U\ket{\psi}$
ranges over all pure states. Taking the expectation over the Haar-random $U$
yields~\eqref{eq:A-prime-uniform-accuracy} for all $\ket{\psi}$.

Finally, (iv) follows immediately from (ii) and
Lemma~\ref{le:hayashi-haar-error}, applied to the covariant scheme $\mathcal A'$.
\end{proof}

In summary, Lemma~\ref{le:haar-twirl-covariant} shows that, for the purposes of
sample-complexity analysis with respect to any unitarily invariant metric, we
may without loss of generality restrict attention to covariant pure-state
tomography schemes whose error direction is Haar-distributed in the orthogonal
complement of the true state. In what follows we will work with such
a covariant, Haar-symmetric scheme, and exploit the decomposition
\eqref{co:haar-two-dim} as a key structural ingredient in our
channel-learning algorithm.

\subsubsection{Optimal mixed-state tomography via purification}
\label{subsub:reducing}

A recent work has shown that mixed-state tomography can be reduced to
pure-state tomography via a suitable purification channel
\cite{tang2025conjugatequerieshelp,pelecanos2025mixedstatetomographyreduces}. Given copies of a mixed state
$\rho \in \mathcal D_{d,r}$, one first applies a channel that maps
$\rho^{\otimes N}$ to copies of a random purification $\ket{\rho}$ of $\rho$,
and then runs an optimal pure-state tomography scheme (such as Hayashi's POVM)
on the purified system. We record the purification step as a lemma.

\begin{lemma}[Purification channel~\cite{pelecanos2025mixedstatetomographyreduces,tang2025conjugatequerieshelp}]
\label{le:purification-channel}
For every $d,r \in \mathbb N$ and every $N \ge 1$ there exists a quantum
channel
\begin{align}
  \mathcal P_{d,r}^{(N)} :
  \mathcal D(\mathbb C^d)^{\otimes N}
  \to \mathcal D\bigl((\mathbb C^d \otimes \mathbb C^r)^{\otimes N}\bigr)
\end{align}
with the following properties.
Let $\rho \in \mathcal D_{d,r}$ and let
$\ket{\rho} \in \mathbb C^d \otimes \mathbb C^r$ be any purification of $\rho$.
Then
\begin{align}
  \mathcal P_{d,r}^{(N)}\bigl(\rho^{\otimes N}\bigr)
  = \mathbb E_{\ket{\rho'}}\bigl[\ketbra{\rho'}{\rho'}^{\otimes N}\bigr],
\end{align}
where the expectation is over purifications $\ket{\rho'}$ distributed according
to the unitarily invariant (Haar) measure on the purification register.
Moreover, for every accuracy parameter $\delta > 0$ the channel
$\mathcal P_{d,r}^{(N)}$ admits a circuit implementation that
$\delta$-approximates the above map in diamond norm and has size
$\mathrm{poly}\bigl(N,\log d,\log(1/\delta)\bigr)$.
\end{lemma}

Operationally, Lemma~\ref{le:hayashi-tomography} and Lemma~\ref{le:purification-channel}
allow us to view sample-optimal mixed-state tomography~\cite{pelecanos2025mixedstatetomographyreduces} as a two-step procedure:
\emph{purify, then learn}. First, one applies $\mathcal P_{d,r}^{(N)}$ to
produce copies of a random purification of $\rho$; second, one performs
pure-state tomography on this purification using an optimal covariant
pure-state scheme with Haar-distributed error direction.

\section{An algorithm for learning quantum channels in diamond norm}
\label{sec:main-diamond}

We now turn to our central task: learning an unknown quantum channel in
diamond distance. In Subsection~\ref{subsec:problem-diamond} we formalise
the learning task. In
Subsection~\ref{subsec:regularise-cptp} we explain how to regularise an
intermediate estimator to a CPTP map without worsening the guarantees by
more than a factor of two. In
Subsection~\ref{subsec:coherence-copies} we explain our algorithm and the role of coherence
between different uses of the channel in our construction.
 Finally, Subsection~\ref{subsec:main-theorem}
then states and proves our main upper bound,
Theorem~\ref{th:main-diamond-tomography} (and its corollaries), which shows that
\(
  O\bigl(d_{\mathrm{in}} d_{\mathrm{out}} k/\varepsilon^2\bigr)
\)
uses of the channel suffice to learn $\Lambda$ to diamond-distance accuracy
$\varepsilon$ with any fixed success probability.

\subsection{The problem of quantum channel tomography}
\label{subsec:problem-diamond}

We consider an unknown quantum channel
\(
  \Lambda : \mathcal L(\mathcal H_{\mathrm{in}})
  \to \mathcal L(\mathcal H_{\mathrm{out}})
\)
with
\(d_{\mathrm{in}} \coloneqq \dim(\mathcal H_{\mathrm{in}})\)
and
\(d_{\mathrm{out}} \coloneqq \dim(\mathcal H_{\mathrm{out}})\).
Throughout we assume that $\Lambda$ has Kraus rank at most $k$, and we set
\(
  d_{\mathrm{tot}} \coloneqq d_{\mathrm{in}} d_{\mathrm{out}} k,
\)
the Hilbert-space dimension of a purification of its normalised Choi state.

Our goal is to estimate $\Lambda$ to a prescribed accuracy in
\emph{diamond distance}. For an estimator $\hat\Lambda$, we write
\(
  d_\diamond(\hat\Lambda,\Lambda) \coloneqq \tfrac12 \|\hat\Lambda - \Lambda\|_\diamond,
\)
which equals the maximum total-variation distance between the outcome
distributions of any single-use experiment involving $\Lambda$, possibly
with an entangled reference system and followed by an optimal measurement.

A \emph{channel learning scheme} with sample size $N$ is any quantum
strategy that
\begin{itemize}
  \item prepares input states on $\mathcal H_{\mathrm{in}}$ together with
    auxiliary systems,
  \item makes up to $N$ calls to the black-box channel $\Lambda$, possibly
    adaptively and interleaved with free (channel-independent) operations,
  \item performs a final measurement and classical post-processing to output
    a description of an estimator $\hat\Lambda$.
\end{itemize}
We say that such a scheme achieves accuracy $\varepsilon \in (0,1)$ and
failure probability $\delta \in (0,1)$ in the \emph{worst-case} sense if,
for every channel
\(
  \Lambda : \mathcal L(\mathcal H_{\mathrm{in}})
  \to \mathcal L(\mathcal H_{\mathrm{out}})
\)
with Kraus rank at most~$k$,
\begin{align}
  \Pr\bigl[d_\diamond(\hat\Lambda,\Lambda) \le \varepsilon\bigr]
  \;\ge\; 1 - \delta,
\end{align}
where the probability is over all internal randomness of the protocol and
the measurement outcomes.

The \emph{sample complexity} of diamond-distance learning is the minimal
number of channel uses required to achieve accuracy~$\varepsilon$ and
confidence~$1-\delta$ in this worst-case sense. Formally, we define
\begin{align}
  N_{\mathrm{chan}}^\star(
    \varepsilon,\delta; d_{\mathrm{in}},d_{\mathrm{out}},k
  )
  \;\coloneqq\;
  \inf\Bigl\{
    N \in \mathbb N :
    \exists\ \text{$N$-use scheme achieving accuracy } (\varepsilon,\delta)
    \text{ in the worst-case sense}
  \Bigr\}.
\end{align}

Our main upper bound,
Theorem~\ref{th:main-diamond-tomography}, is realised by a concrete
non-adaptive tomography procedure
(Algorithm~\ref{alg:diamond-tomography}) and shows that, in the relevant regime
$\delta \ge 4\exp(-d_{\mathrm{tot}})$ where the $\delta$-dependence takes
a particularly simple form,
\begin{align}
  N_{\mathrm{chan}}^\star(
    \varepsilon,\delta; d_{\mathrm{in}},d_{\mathrm{out}},k
  )
  \;=\;
  O\Bigl(
    \frac{
      d_{\mathrm{in}} d_{\mathrm{out}} k
      + d_{\mathrm{in}}\log(1/\delta)
    }{\varepsilon^2}
  \Bigr).
\end{align}

\subsubsection{Regularising the estimator to a CPTP map}
\label{subsec:regularise-cptp}

A tomographic procedure with sample size $N$ typically outputs a linear map
$\Lambda^{\mathrm{est}}_N$ that is close to the true channel in diamond norm,
but need not be CPTP for finite $N$. In our setting it is natural to enforce
complete positivity and trace preservation for the final estimate.

We do so by projecting $\Lambda^{\mathrm{est}}_N$ onto the convex set of CPTP
maps in diamond norm. Let $\mathsf{CPTP}$ be the set of all channels
$\Phi : \mathcal L(\mathcal H_{\mathrm{in}}) \to \mathcal L(\mathcal H_{\mathrm{out}})$,
and define the regularised estimator by
\[
  \hat\Lambda_N \in
  \arg\min_{\Phi \in \mathsf{CPTP}}
  \bigl\|\Phi - \Lambda^{\mathrm{est}}_N\bigr\|_\diamond.
\]
Using the standard SDP representation of the diamond norm in terms of Choi operators (see, e.g.,~\cite{WatrousBook,SkrzypczykCavalcantiSDP}), this optimisation can be implemented as a semidefinite-program post-processing step whose size is polynomial in the Hilbert-space dimension. Of course, for many-body systems this may still be prohibitive in the number of qudits.

\begin{lemma}[Diamond-norm projection onto CPTP maps]
\label{le:cptp-regularisation}
Let $\Lambda$ be a CPTP map and let $\Lambda^{\mathrm{est}}$ be a
Hermiticity-preserving map such that
$\|\Lambda^{\mathrm{est}} - \Lambda\|_\diamond \le \varepsilon$.
Let $\hat\Lambda$ be any solution of
$\hat\Lambda \in \arg\min_{\Phi \in \mathsf{CPTP}}
 \|\Phi - \Lambda^{\mathrm{est}}\|_\diamond$.
Then $\|\hat\Lambda - \Lambda\|_\diamond \le 2\varepsilon$.
\end{lemma}

\begin{proof}
By optimality of $\hat\Lambda$ and feasibility of $\Lambda$ we have
$\|\hat\Lambda - \Lambda^{\mathrm{est}}\|_\diamond
 = \inf_{\Phi \in \mathsf{CPTP}}\|\Phi - \Lambda^{\mathrm{est}}\|_\diamond
 \le \|\Lambda - \Lambda^{\mathrm{est}}\|_\diamond$.
Combining this with $\|\Lambda^{\mathrm{est}} - \Lambda\|_\diamond \le \varepsilon$
and using the triangle inequality,
\[
  \|\hat\Lambda - \Lambda\|_\diamond
  \le
  \|\hat\Lambda - \Lambda^{\mathrm{est}}\|_\diamond
  +
  \|\Lambda^{\mathrm{est}} - \Lambda\|_\diamond
  \le
  \|\Lambda - \Lambda^{\mathrm{est}}\|_\diamond
  +
  \|\Lambda^{\mathrm{est}} - \Lambda\|_\diamond
  \le 2\varepsilon,
\]
as claimed.
\end{proof}


\subsection{Our channel tomography protocol}
\label{subsec:coherence-copies}

Our channel-learning scheme in Algorithm~\ref{alg:diamond-tomography} is
closely aligned with the mixed-state tomography framework of
Ref.~\cite{pelecanos2025mixedstatetomographyreduces}, where fixed-rank
mixed-state tomography is reduced to pure-state tomography via the random
\emph{purification channel}~\cite{tang2025conjugatequerieshelp}. Given $N$ copies of a rank-$r$ state
$\rho \in \mathcal D_{d,r}$, their protocol first applies a purification
channel $\mathcal P_{d,r}^{(N)}$ (of the form in
Lemma~\ref{le:purification-channel}) to $\rho^{\otimes N}$, thereby
producing $N$ copies of a purification $\ket{\rho}$, and then performs
pure-state tomography on $\ket{\rho}^{\otimes N}$ before tracing out the
purifying register. A key conceptual point is that \emph{coherence between
copies is required only for the purification step}: once a purification is
available, one may use a pure-state tomography scheme that acts
independently on each copy. An example is the projected least-squares (PLS)
scheme of Ref.~\cite{GutaKahnKuengTropp2020}, which uses only
single-copy measurements yet achieves optimal sample complexity
\(N_{\mathrm{pure}}(d,\varepsilon,\delta)
  = \Theta\bigl((d + \log(1/\delta))/\varepsilon^2\bigr)\)
in trace distance.

Our channel protocol has precisely the same architecture. For a rank-$k$
channel
\(
  \Lambda : \mathcal L(\mathcal H_{\mathrm{in}})
  \to \mathcal L(\mathcal H_{\mathrm{out}})
\)
with Choi state $\Phi_c = J(\Lambda)$ of rank at most $k$, we proceed as
follows:
\begin{enumerate}
  \item \emph{Parallel Choi-state preparation.}
        Use $N$ calls to $\Lambda$ in parallel to prepare $N$ copies of the
        Choi state $\Phi_c$, by applying $\Lambda^{\otimes N}$ to $N$
        maximally entangled inputs. In particular, our use of the black-box
        channel is completely non-adaptive.
  \item \emph{Purification.}
        Apply a purification channel
        \(
          \mathcal P^{(N)} :
          \mathcal D\bigl(
            (\mathcal H_{\mathrm{out}}\otimes\mathcal H_{\mathrm{in}})^{\otimes N}
          \bigr)
          \to
          \mathcal D\bigl(
            (\mathcal H_{\mathrm{out}}\otimes
             \mathcal H_{\mathrm{in}}\otimes \mathcal H_E)^{\otimes N}
          \bigr),
        \)
        which maps $\Phi_c^{\otimes N}$ to $N$ copies of a fixed purification
        $\ket{\Phi_c}$ with $\dim(\mathcal H_E)=k$
        (Lemma~\ref{le:purification-channel} with
        $d = d_{\mathrm{out}} d_{\mathrm{in}}$ and $r = k$). This is the
        \emph{only} step that requires coherent operations across different
        copies.
  \item \emph{Pure-state tomography.}
        Perform pure-state tomography on $\ket{\Phi_c}^{\otimes N}$ using an
        optimal pure-state tomography algorithm $\mathcal A$ on
        $\mathbb C^{d_{\mathrm{tot}}}$, where
        \(d_{\mathrm{tot}} = d_{\mathrm{in}} d_{\mathrm{out}} k\),
        and post-process its output into an estimate of the Choi state
        $\Phi_c$ and hence of the channel $\Lambda$. If desired, one can then
        regularise the resulting estimate to a CPTP map as in
        Subsection~\ref{subsec:regularise-cptp}.
\end{enumerate}

Theorem~\ref{th:main-diamond-tomography} shows that the diamond-norm sample
complexity of this channel-learning protocol (for any constant success probability $\delta$) is governed entirely by the
pure-state sample complexity
\(N_{\mathrm{pure}}(\mathcal A,d_{\mathrm{tot}},
                   \varepsilon_{\mathrm{pure}},\delta)\)
in trace distance, evaluated at
\(\varepsilon_{\mathrm{pure}} = \Theta(\varepsilon)\),
where $\varepsilon$ is the target diamond-norm error. The analysis does not
use any feature specific to $\mathcal A$ beyond covariance: by
Lemma~\ref{le:haar-twirl-covariant}, any pure-state scheme can be
covariantised without worsening its worst-case performance.

In particular, one may instantiate $\mathcal A$ with a PLS-type scheme
as in Ref.~\cite{GutaKahnKuengTropp2020}, which uses only
single-copy measurements. Theorem~\ref{th:main-diamond-tomography} then remains unchanged:
at the level of channel uses, the protocol is fully non-adaptive and
parallel, all entanglement between copies is confined to the purification
stage, and the tomography stage itself can be implemented with single-copy
measurements.

For the sake of obtaining explicit constants of our bound, we will
instantiate Theorem~\ref{th:main-diamond-tomography} (later in Theorem~\ref{th:th1Mainapp}) with Hayashi’s optimal pure-state tomography scheme (see Lemma~\ref{le:hayashi-tomography}), which uses collective multi-copy
measurements.

\subsubsection{Main theorem: diamond-norm guarantees}
\label{subsec:main-theorem}

We now combine the ingredients developed so far into a complete tomography
scheme that learns an unknown quantum channel in diamond norm by reducing
the task to pure-state tomography on a purification of its normalised Choi
state. Let $\varepsilon_{\mathrm{pure}}
= \varepsilon_{\mathrm{pure}}(N,\delta,d_{\mathrm{tot}})$ denote the
trace-distance accuracy parameter of the underlying covariant pure-state
tomography scheme $\mathcal A$: by assumption, when $\mathcal A$ is run on
$N$ copies of an unknown pure state in $\mathbb C^{d_{\mathrm{tot}}}$, it
returns $|\hat\psi\rangle$ with $d_{\mathrm{tr}}(\psi,\hat\psi)
\le \varepsilon_{\mathrm{pure}}$ with probability at least $1-\delta/2$.
For instance, if $\mathcal A$ is instantiated by Hayashi's covariant scheme~\ref{le:hayashi-tomography},
then
\begin{align}
  \varepsilon_{\mathrm{pure}}
  \;\le\;
  \sqrt{4\,\frac{d_{\mathrm{tot}} + \log(2/\delta)}{N}}.
\end{align}

\begin{theorem}[Diamond-distance tomography for rank-$k$ channels]
\label{th:main-diamond-tomography}
Let $\varepsilon \in (0,1)$ and $\delta \in (0,1)$.
Let
\(
  \Lambda : \mathcal L(\mathcal H_{\mathrm{in}})
  \to \mathcal L(\mathcal H_{\mathrm{out}})
\)
be a quantum channel with
\(
  d_{\mathrm{in}} \coloneqq \dim(\mathcal H_{\mathrm{in}})
\)
and
\(
  d_{\mathrm{out}} \coloneqq \dim(\mathcal H_{\mathrm{out}})
\),
and assume:
\begin{itemize}
  \item $\Lambda$ has Kraus rank at most $k$, and we set $d_{\mathrm{tot}} \coloneqq d_{\mathrm{in}} d_{\mathrm{out}} k;$

  \item $\mathcal A$ is a covariant pure-state tomography scheme on
        $\mathbb C^{d_{\mathrm{tot}}}$ such that, for every pure state
        $\ket{\psi} \in \mathbb C^{d_{\mathrm{tot}}}$, when $\mathcal A$ is
        run on $N$ copies of $\ket{\psi}$ it outputs a pure state
        $|\hat\psi\rangle$ with $d_{\mathrm{tr}}(\psi,\hat\psi)
        \le \varepsilon_{\mathrm{pure}}$ with probability at least
        $1 - \delta/2$.

\end{itemize}

Consider the channel-tomography procedure of
Algorithm~\ref{alg:diamond-tomography}, which, given $N$ uses of $\Lambda$
and access to $\mathcal A$, outputs a channel estimate $\hat\Lambda$ (which
is CPTP). Then, with probability at least $1 - \delta$,
\begin{align}
  \frac12\|\hat\Lambda - \Lambda\|_\diamond
  \;\le\;
  \Bigl(
    2 + \sqrt{\tfrac{1}{k d_{\mathrm{out}}}\log\tfrac{4}{\delta}}
  \Bigr)\,
  \bigl[4 + S_\delta(d_{\mathrm{tot}})\bigr]\,
  \varepsilon_{\mathrm{pure}},
\end{align}
where $S_\delta(d_{\mathrm{tot}}) \ge 0$ is a subleading function in $d_{\mathrm{tot}}$ defined
explicitly in Eq.~\eqref{eq:S-def} in the proof below.
\end{theorem}

\begin{proof}
Let $\Phi_c \coloneqq J(\Lambda)$ be the normalised Choi state of $\Lambda$.
By the Choi-state preparation step and the random purification map of
Lemma~\ref{le:purification-channel}, after $N$ uses of the channel we obtain
$N$ purified Choi-state copies $\ket{\Phi_c}^{\otimes N}$, where each copy
lives in a Hilbert space of dimension $d_{\mathrm{tot}}$. Applying
$\mathcal A$ to $\ket{\Phi_c}^{\otimes N}$ yields an estimate
$|\hat\Phi_c\rangle$. By the assumption on $\mathcal A$, with probability at
least $1-\delta/2$ we have
$d_{\mathrm{tr}}(\Phi_c,\hat\Phi_c) \le \varepsilon_{\mathrm{pure}}$.
For pure states,
\(
  d_{\mathrm{tr}}(\Phi_c,\hat\Phi_c)
  = \sqrt{1 - |\langle \hat\Phi_c | \Phi_c \rangle|^2},
\)
so on this event
\begin{align}
  \sqrt{1 - \bigl|\langle \hat\Phi_c | \Phi_c \rangle\bigr|^2}
  \;\le\; \varepsilon_{\mathrm{pure}}.
\end{align}

Let $\Lambda^{\mathrm{est}}$ be the CP map whose normalised Choi state is
$\tr_E(|\hat\Phi_c\rangle\langle\hat\Phi_c|)$; this is the intermediate
estimate produced in Algorithm~\ref{alg:diamond-tomography} before CPTP
regularisation. By Lemma~\ref{le:cptp-regularisation} there exists a CPTP
map $\hat\Lambda$ such that
\begin{align}
  \tfrac12\|\hat\Lambda - \Lambda\|_\diamond
  \;\le\; \|\Lambda^{\mathrm{est}} - \Lambda\|_\diamond.
  \label{eq:hatLambda-reduction}
\end{align}
It is therefore enough to bound $\|\Lambda^{\mathrm{est}} - \Lambda\|_\diamond$.

By Lemma~\ref{le:hayashi-haar-error} and
Corollary~\ref{co:haar-two-dim} (applied to
$\ket{\psi}=\ket{\Phi_c}$ and $|\hat\Phi_c\rangle$),
there exists a Haar-distributed vector
$\ket{\Psi_{\mathrm{Haar}}} \in \mathbb C^{d_{\mathrm{tot}}}$,
independent of $\varepsilon_{\mathrm{pure}}$, such that
\begin{align}
  |\hat\Phi_c\rangle
  = a\,\ket{\Phi_c} + b\,\ket{\Psi_{\mathrm{Haar}}},
\end{align}
where $f_{\mathrm{Haar}} \coloneqq \braket{\Phi_c}{\Psi_{\mathrm{Haar}}}$
and $a,b$ are the coefficients from
Corollary~\ref{co:haar-two-dim}.
Set $\ket{\Phi_1} \coloneqq \ket{\Phi_c}$ and
$\ket{\Phi_2} \coloneqq \ket{\Psi_{\mathrm{Haar}}}$, and for
$i,j\in\{1,2\}$ define
\begin{align}
  K_{ij}
  \coloneqq
  \tr_E\bigl(\ketbra{\Phi_i}{\Phi_j}\bigr).
\end{align}
Then
\begin{align}
  J(\Lambda^{\mathrm{est}} - \Lambda)
  &= \tr_E\Bigl(
       |\hat\Phi_c\rangle\langle\hat\Phi_c|
       - \ketbra{\Phi_c}{\Phi_c}
     \Bigr) \\
  &= (|a|^2 - 1)\,K_{11}
     + |b|^2 K_{22}
     + a b^* K_{12}
     + a^* b K_{21}.
\end{align}
Writing $a b^* = |a b| e^{i\theta}$ for some $\theta\in\mathbb R$, we obtain
\begin{align}
  a b^* K_{12} + a^* b K_{21}
  = |a b|\bigl(e^{i\theta} K_{12} + e^{-i\theta} K_{21}\bigr),
\end{align}
and hence
\begin{align}
  \|\Lambda^{\mathrm{est}} - \Lambda\|_\diamond
  &= \|J(\Lambda^{\mathrm{est}} - \Lambda)\|_\diamond\\
  &\le |\,|a|^2 - 1\,|\,\|K_{11}\|_\diamond
       + |b|^2 \|K_{22}\|_\diamond 
       + |a b|\,
         \bigl\|e^{i\theta} K_{12} + e^{-i\theta} K_{21}\bigr\|_\diamond.
  \label{eq:diamond-est-bound}
\end{align}

We now bound the blocks $K_{ij}$. Since $K_{11}$ is the Choi operator of
$\Lambda$, $\|K_{11}\|_\diamond = \|\Lambda\|_\diamond = 1$. Set
$\delta_{\mathrm{Haar}} \coloneqq \delta/2$.
For $K_{22}$, note that
\(
  K_{22}
  = \tr_E(\ketbra{\Psi_{\mathrm{Haar}}}{\Psi_{\mathrm{Haar}}})
\)
is the Choi operator of a CP map with (normalised) Choi state
$\ket{\Psi_{\mathrm{Haar}}}$. By Lemma~\ref{le:diamond-positive-choi},
\begin{align}
  \|K_{22}\|_\diamond
  = d_{\mathrm{in}}\,
    \|\rho_{\mathrm{in}}^{(\mathrm{Haar})}\|_\infty,
  \qquad
  \rho_{\mathrm{in}}^{(\mathrm{Haar})}
  =
  \tr_{\mathrm{out},E}(\ketbra{\Psi_{\mathrm{Haar}}}{\Psi_{\mathrm{Haar}}}).
\end{align}
Corollary~\ref{co:choi-input-opnorm}, applied with failure probability
$\delta_{\mathrm{Haar}}/2 = \delta/4$ and
$d_{\mathrm{tot}} = d_{\mathrm{in}} d_{\mathrm{out}} k$, implies that, with
probability at least $1-\delta_{\mathrm{Haar}}/2$,
\begin{align}
  \|K_{22}\|_\diamond
  \le C(\delta)^2,
  \label{eq:K22-bound}
\end{align}
where we define
\begin{align}
  C(\delta)
  \coloneqq
  2 + \sqrt{\frac{1}{k\,d_{\mathrm{out}}}
               \log\frac{2}{\delta_{\mathrm{Haar}}}}
  \;=\;
  2 + \sqrt{\frac{1}{k\,d_{\mathrm{out}}}
               \log\frac{4}{\delta}}.
\end{align}

For the off-diagonal term, apply the diamond Cauchy--Schwarz inequality
(Lemma~\ref{le:diamond-CS}) to the block matrix with off-diagonals
$e^{\pm i\theta} K_{12},K_{21}$:
\begin{align}
  \bigl\|e^{i\theta} K_{12} + e^{-i\theta} K_{21}\bigr\|_\diamond
  &\le 2\sqrt{\|K_{11}\|_\diamond\,\|K_{22}\|_\diamond}
   \le 2C(\delta),
  \label{eq:offdiag-bound}
\end{align}
on the same event.

Next we control the overlap $f_{\mathrm{Haar}}$.
By Lemma~\ref{le:beta-overlap}, $|f_{\mathrm{Haar}}|^2$ has the
$\mathrm{Beta}(1,d_{\mathrm{tot}}-1)$ distribution. Set
\begin{align}
  \varepsilon_{\mathrm{ov}}(\delta)
  \coloneqq 1-\left(\frac{\delta_{\mathrm{Haar}}}{2}\right)^{\frac{1}{d_{\mathrm{tot}}-1}},
\end{align}
and apply Lemma~\ref{le:beta-overlap} with failure probability
$\delta_{\mathrm{Haar}}/2 = \delta/4$ to obtain
\begin{align}
  \Pr\bigl[\,|f_{\mathrm{Haar}}|^2 \ge \varepsilon_{\mathrm{ov}}(\delta)\bigr]
  = (1-\varepsilon_{\mathrm{ov}}(\delta))^{d_{\mathrm{tot}}-1}
  = \frac{\delta_{\mathrm{Haar}}}{2}.
\end{align}
Thus, with probability $1-\delta_{\mathrm{Haar}}/2$ we have
$|f_{\mathrm{Haar}}|^2 \le \varepsilon_{\mathrm{ov}}(\delta)$.
Define
\begin{align}
  c_{\mathrm{ov}}(\delta)
  \coloneqq
  \sqrt{\frac{\varepsilon_{\mathrm{ov}}(\delta)}
             {1-\varepsilon_{\mathrm{ov}}(\delta)}}.
\end{align}
Using the explicit expressions for $a$ and $b$ from
Corollary~\ref{co:haar-two-dim}, a direct estimate gives
\begin{align}
  |b|^2
  &\le \frac{\varepsilon_{\mathrm{pure}}^2}
             {1-\varepsilon_{\mathrm{ov}}(\delta)},\\
  \bigl||a|^2 - 1\bigr|
  &\le \bigl(1 + 2c_{\mathrm{ov}}(\delta) + c_{\mathrm{ov}}(\delta)^2\bigr)\,
       \varepsilon_{\mathrm{pure}},\\
  |a b|
  &\le \frac{1 + c_{\mathrm{ov}}(\delta)}
             {\sqrt{1-\varepsilon_{\mathrm{ov}}(\delta)}}\,
       \varepsilon_{\mathrm{pure}}.
\end{align}
Combining these bounds with
\eqref{eq:diamond-est-bound}–\eqref{eq:offdiag-bound}, we obtain
\begin{align}
  \|\Lambda^{\mathrm{est}} - \Lambda\|_\diamond
  &\le \bigl(1 + 2c_{\mathrm{ov}}(\delta) + c_{\mathrm{ov}}(\delta)^2\bigr)\,
        \varepsilon_{\mathrm{pure}}
       + \frac{C(\delta)^2}{1-\varepsilon_{\mathrm{ov}}(\delta)}\,
         \varepsilon_{\mathrm{pure}}^2
       + \frac{2\bigl(1+c_{\mathrm{ov}}(\delta)\bigr)}
              {\sqrt{1-\varepsilon_{\mathrm{ov}}(\delta)}}\,
        C(\delta)\,\varepsilon_{\mathrm{pure}}.
  \label{eq:raw-full-bound}
\end{align}

Without loss of generality, we may assume $C(\delta)^2 \varepsilon_{\mathrm{pure}}^2 \le 1$.
Indeed, if $C(\delta)^2 \varepsilon_{\mathrm{pure}}^2 > 1$, then the right-hand side of the
target inequality in the theorem statement exceeds $1$, whereas for any two valid quantum channels we
always have $\tfrac12\|\hat\Lambda - \Lambda\|_\diamond \le 1$, so the bound is trivially
satisfied. Therefore, in the non-trivial regime, we have
\begin{align}
  C(\delta)^2 \varepsilon_{\mathrm{pure}}^2
  \le C(\delta)\,\varepsilon_{\mathrm{pure}}.
\end{align}
Using this in \eqref{eq:raw-full-bound} yields
\begin{align}
  \|\Lambda^{\mathrm{est}} - \Lambda\|_\diamond
  &\le \bigl(1 + 2c_{\mathrm{ov}}(\delta) + c_{\mathrm{ov}}(\delta)^2\bigr)\,
        \varepsilon_{\mathrm{pure}}
       + \frac{C(\delta)}{1-\varepsilon_{\mathrm{ov}}(\delta)}\,
         \varepsilon_{\mathrm{pure}}
       + \frac{2\bigl(1+c_{\mathrm{ov}}(\delta)\bigr)}
              {\sqrt{1-\varepsilon_{\mathrm{ov}}(\delta)}}\,
        C(\delta)\,\varepsilon_{\mathrm{pure}}.
  \label{eq:est-small-pre}
\end{align}

Since $C(\delta)\ge 1$, we can factor it out and write
\begin{align}
  \|\Lambda^{\mathrm{est}} - \Lambda\|_\diamond
  &\le C(\delta)
       \Bigl[
          1 + 2c_{\mathrm{ov}}(\delta) + c_{\mathrm{ov}}(\delta)^2
          + \frac{1}{1-\varepsilon_{\mathrm{ov}}(\delta)}
          + \frac{2\bigl(1+c_{\mathrm{ov}}(\delta)\bigr)}
                 {\sqrt{1-\varepsilon_{\mathrm{ov}}(\delta)}}
       \Bigr] \varepsilon_{\mathrm{pure}}.
  \label{eq:est-small-bracket}
\end{align}
We now define
\begin{align}
  S_\delta(d_{\mathrm{tot}})
  &\coloneqq
  \Bigl(\tfrac{1}{1-\varepsilon_{\mathrm{ov}}(\delta)} - 1\Bigr)
  + 4c_{\mathrm{ov}}(\delta)
  + c_{\mathrm{ov}}(\delta)^2
  + 2\bigl(1+c_{\mathrm{ov}}(\delta)\bigr)
     \Bigl(\tfrac{1}{\sqrt{1-\varepsilon_{\mathrm{ov}}(\delta)}} - 1\Bigr),
  \label{eq:S-def}
\end{align}
so that the bracket in \eqref{eq:est-small-bracket} is exactly
$4 + S_\delta(d_{\mathrm{tot}})$. We therefore obtain
\begin{align}
  \|\Lambda^{\mathrm{est}} - \Lambda\|_\diamond
  \le
  C(\delta)\,
  \bigl[4 + S_\delta(d_{\mathrm{tot}})\bigr]\,
  \varepsilon_{\mathrm{pure}}.
\end{align}
Combining this with \eqref{eq:hatLambda-reduction} gives the target inequality of the theorem.

Finally, we account for failure probabilities. The tomography guarantee
fails with probability at most $\delta/2$, the bound on
$\|K_{22}\|_\diamond$ with probability at most
$\delta_{\mathrm{Haar}}/2 = \delta/4$, and the overlap bound with
probability at most $\delta_{\mathrm{Haar}}/2 = \delta/4$. By a union
bound, all three events hold simultaneously with probability at least
$1-\delta$. This completes the proof.
\end{proof}

To make the dependence on $d_{\mathrm{in}}$, $d_{\mathrm{out}}$, $k$, and
$\delta$ fully explicit, and to compute an explicit leading constant, we now
track the constants in Theorem~\ref{th:main-diamond-tomography} more
carefully. Suppose that
\begin{align}
  \varepsilon_{\mathrm{pure}}
  \;\le\;
  \sqrt{c\,\frac{d_{\mathrm{tot}} + \log(2/\delta)}{N}}
\end{align}
for some constant $c > 0$ (for Hayashi's scheme,
Lemma~\ref{le:hayashi-tomography}, one may take $c = 4$) like any sample-optimal pure-state tomography algorithm. Then
Theorem~\ref{th:main-diamond-tomography} implies
\begin{align}
  \frac12\|\hat\Lambda - \Lambda\|_\diamond
  &\le
  \Bigl(
    2 + \sqrt{\tfrac{1}{k d_{\mathrm{out}}}\log\tfrac{4}{\delta}}
  \Bigr)
  \bigl[4 + S_\delta(d_{\mathrm{tot}})\bigr]\,
  \sqrt{c\,\frac{d_{\mathrm{tot}} + \log(2/\delta)}{N}}
  \nonumber\\[0.4em]
  &=
  \bigl[4 + S_\delta(d_{\mathrm{tot}})\bigr]\,
  \sqrt{
    c\,\frac{d_{\mathrm{tot}} + \log(2/\delta)}{N}\,
    \Bigl(
      4
      + \frac{1}{k d_{\mathrm{out}}}\log\frac{4}{\delta}
      + 4\sqrt{\frac{1}{k d_{\mathrm{out}}}\log\frac{4}{\delta}}
    \Bigr)
  }.
  \label{eq:semplHA}
\end{align}
Expanding the term under the square root and using
$d_{\mathrm{tot}} = k d_{\mathrm{in}} d_{\mathrm{out}}$ yields
\begin{align}
  \frac{c}{N}\Bigl(
    4 k d_{\mathrm{in}} d_{\mathrm{out}}
    + d_{\mathrm{in}}\log\frac{4}{\delta}
    + 4 d_{\mathrm{in}}\sqrt{k d_{\mathrm{out}}\log\frac{4}{\delta}}
    + 4\log\frac{2}{\delta}
    + \frac{\log(2/\delta)}{k d_{\mathrm{out}}}\log\frac{4}{\delta}
    + 4\log\frac{2}{\delta}\sqrt{\frac{1}{k d_{\mathrm{out}}}\log\frac{4}{\delta}}
  \Bigr).
\end{align}

In the regime where $\delta$ is not extremely small relative to the total
dimension, i.e.\ when $\delta > 4 \exp(-d_{\mathrm{tot}})$, one can verify that
\begin{align}
  S_\delta(d_{\mathrm{tot}})
  = O\left(\sqrt{\frac{\log(1/\delta)}{d_{\mathrm{tot}}}}\right)
  \le O(1).
\end{align}
Combining this with the constraint $k \ge d_{\mathrm{in}}/d_{\mathrm{out}}$
(Lemma~\ref{le:dimen-constraint-kraus}) shows that, beyond the leading-order
contribution, the bound simplifies to
\begin{align}
  \frac12\|\hat\Lambda - \Lambda\|_\diamond
  \;\le\;
  \sqrt{
    \frac{
      64c\,d_{\mathrm{in}} d_{\mathrm{out}}k
      + O\bigl(d_{\mathrm{in}}\log\tfrac{1}{\delta}\bigr)
    }{N}
  }.
\end{align}
In particular, to guarantee accuracy $\varepsilon$ in diamond distance it is
sufficient to take
\begin{align}
  N
  \;=\;
  64c\,\frac{d_{\mathrm{in}} d_{\mathrm{out}} k}{\varepsilon^{2}}
  + O\left(\frac{d_{\mathrm{in}}\log(1/\delta)}{\varepsilon^{2}}\right)
\end{align}
in the regime $\delta > 4 \exp(-d_{\mathrm{tot}})$, which is the parameter
regime of primary interest. For Hayashi's scheme,
Lemma~\ref{le:hayashi-tomography}, we have $c = 4$, so this specialises to
\begin{align}
  N
  \;=\;
  256\,\frac{d_{\mathrm{in}} d_{\mathrm{out}} k}{\varepsilon^{2}}
  + O\left(\frac{d_{\mathrm{in}}\log(1/\delta)}{\varepsilon^{2}}\right).
\end{align}

Combining the above estimates, we obtain the following theorem.

\begin{theorem}[Diamond-distance tomography for rank-$k$ channels]
\label{th:th1Mainapp}
Let 
\(
  \Lambda : \mathcal{L}(\mathbb{C}^{d_{\mathrm{in}}}) 
  \to \mathcal{L}(\mathbb{C}^{d_{\mathrm{out}}})
\)
be a quantum channel with Kraus rank $k$, and set
\( d_{\mathrm{tot}} \coloneqq d_{\mathrm{in}} d_{\mathrm{out}} k \).
Then, for any $0 < \varepsilon \le 1$ and any failure probability $\delta$
satisfying
\(
  \delta > 4 \exp(-d_{\mathrm{tot}})
\)
(i.e.\ $\delta$ is not extremely small relative to the total dimension),
there exists a quantum process-tomography algorithm that uses
\begin{align}
  N
  = 256\,\frac{d_{\mathrm{in}}\, d_{\mathrm{out}}\, k}{\varepsilon^{2}}
    + O\left(\frac{d_{\mathrm{in}}\log(1/\delta)}{\varepsilon^{2}}\right)
\end{align}
invocations of $\Lambda$ and outputs a classical description of a quantum
channel estimate $\hat{\Lambda}$ satisfying
\begin{align}
  \frac{1}{2}\|\hat{\Lambda}-\Lambda \|_\diamond \le \varepsilon
\end{align}
with probability at least $1 - \delta$.
\end{theorem}

\begin{remark}[Very small failure probabilities]
\label{re:boosting-delta}
In the statement of Theorem~\ref{th:th1Mainapp} and in the simplified bounds
that follow (for channels and, subsequently, for states, isometries, and
POVMs), we work in the regime
\begin{align}
  \delta
  \;>\;
  4 \exp(-d_{\mathrm{tot}}),
\end{align}
where $d_{\mathrm{tot}} = d_{\mathrm{in}} d_{\mathrm{out}} k$ is the total
dimension of the Choi purification in
Theorem~\ref{th:main-diamond-tomography}. In other words, we assume that the
target failure probability $\delta$ is not extremely small compared to the
total dimension. In this parameter range, Theorem~\ref{th:th1Mainapp} shows
that, for any $0 < \varepsilon \le 1$, there exists a process-tomography
algorithm which, given
\begin{align}
  N
  \;\ge\;
  256\,\frac{d_{\mathrm{in}} d_{\mathrm{out}} k}{\varepsilon^{2}}
  + O\left(\frac{d_{\mathrm{in}}\log(1/\delta)}{\varepsilon^{2}}\right)
\end{align}
invocations of $\Lambda$, outputs a channel estimate $\hat\Lambda$ satisfying $\frac{1}{2}\|\Lambda - \hat\Lambda\|_\diamond \le \varepsilon$ with probability at least $1-\delta$.  

If one wishes to work with \emph{extremely} small failure probabilities
$\delta$, there are two natural options:
\begin{itemize}
  \item[(i)]
    One can return to the full statement of
    Theorem~\ref{th:main-diamond-tomography} (or to Eq.~\eqref{eq:semplHA}) and use the unsimplified estimate there,
    keeping the explicit dependence on $\delta$ through the function
    $S_\delta(d_{\mathrm{tot}})$ and the logarithmic terms, without imposing
    any restriction of the form $\delta > 4 \exp(-d_{\mathrm{tot}})$.
  \item[(ii)]
    Alternatively, one can first work in the regime of constant failure
    probability and then boost the success probability by repetition.
    Fix, for example, a constant $\delta_0 \in (0,1/2)$, say $\delta_0 = 0.49$.
    Running the tomography scheme with target failure probability $\delta_0$
    uses
    \begin{align}
      N_0
      \;=\;
      256\,\frac{d_{\mathrm{in}} d_{\mathrm{out}} k}{\varepsilon^{2}}
      + O\left(\frac{d_{\mathrm{in}}}{\varepsilon^{2}}\right)
    \end{align}
    channel uses (since $\log(1/\delta_0)$ is a constant) and
    produces an estimator $\hat\Lambda$ such that
    \begin{align}
      \Pr\bigl[
        \tfrac{1}{2}\|\hat\Lambda - \Lambda\|_\diamond \le \varepsilon
      \bigr]
      \;\ge\; 1 - \delta_0.
    \end{align}
    Repeating this procedure $T$ times independently and combining the
    resulting estimators $\hat\Lambda^{(1)},\dots,\hat\Lambda^{(T)}$ via a
    standard boosting scheme (e.g., see Ref.~\cite[Proposition~2.4]{HaahKothariODonnellTang2023}),
    one can achieve any target failure probability $\delta \in (0,1)$ with
    \begin{align}
      T
      \;=\;
      O\bigl(\log(1/\delta)\bigr).
    \end{align}
    The corresponding total number of channel uses is
    \begin{align}
      N_{\mathrm{tot}}
      \;=\;
      T\,N_0
      \;=\;
      O\Bigl(
        \frac{d_{\mathrm{in}} d_{\mathrm{out}} k}{\varepsilon^{2}}
        \,\log\frac{1}{\delta}
      \Bigr),
    \end{align}
    where the additional logarithmic factor in the failure probability is the
    usual overhead incurred when boosting from constant to arbitrarily small
    error.
\end{itemize}
\end{remark}

\subsection{Applications: states, isometries, and POVMs}
\label{sec:applicationsAPP}
From a structural viewpoint, quantum channels form a unifying framework that
contains as special cases quantum states (via $d_{\mathrm{in}} = 1$),
isometries (via $k = 1$) and, in particular, unitaries (via
$d_{\mathrm{in}} = d_{\mathrm{out}}$ and $k = 1$), as well as POVMs (for
instance, binary POVMs via $d_{\mathrm{out}} = 2$). Accordingly, Theorem~\ref{th:main-diamond-tomography} and its
consequences simultaneously capture and extend several learning
results for these primitives.

In this section we specialise our general bounds to three basic tasks:
trace-distance tomography of finite-dimensional states, diamond-distance
learning of isometries (including unitary channels), and operator-norm
learning of binary POVMs. Throughout, we work in the relevant parameter regime
where the target failure probability is not required to be exponentially small
in the relevant dimension, so that the simplified bound of
Theorem~\ref{th:th1Main} applies (see Remark~\ref{re:boosting-delta} for a
discussion of this assumption). In particular, for any fixed constant success probability,
our results entail that:
\begin{itemize}
  \item in the case $d_{\mathrm{in}} = 1$, we recover the optimal
    $\Theta(d k / \varepsilon^{2})$ scaling for trace-distance tomography of
    rank-$k$ states;
  \item for Kraus-rank-one channels, we obtain (to the best of our knowledge)
    the first essentially optimal $\Theta(d_{\mathrm{in}} d_{\mathrm{out}} /
    \varepsilon^{2})$ guarantees for learning arbitrary isometries in diamond
    distance; for unitary channels, combining our Choi-state-based learner
    with the boosting scheme of Ref.~\cite{HaahKothariODonnellTang2023}
    further yields $\Theta(d_{\mathrm{in}} d_{\mathrm{out}} / \varepsilon)$
    scaling in the accuracy parameter; and
  \item for binary POVMs, we obtain $O(d^{2} / \varepsilon^{2})$ bounds for
    operator-norm learning, which (to the best of our knowledge) are
    state-of-the-art in their dimensional dependence.
\end{itemize}

Thus, our results place previously separate guarantees for states, unitaries,
isometries, and measurements within a single, channel-unified framework.

\subsubsection{Trace-distance tomography of states}
\label{subsec:state-tomography-application}

Our general result, Theorem~\ref{th:main-diamond-tomography}, provides
diamond-distance guarantees for learning rank-$k$ quantum channels. As a
sanity check, we now specialise to the case $d_{\mathrm{in}} = 1$,
where a channel is simply a state-preparation map and the diamond norm reduces
to the usual trace distance between states, thereby recovering the optimal
sample-complexity scaling for rank-$k$ state tomography~\cite{pelecanos2025mixedstatetomographyreduces,OW16,Haah_2017}.

When $d_{\mathrm{in}} = 1$ we have $\mathcal H_{\mathrm{in}} \simeq \mathbb C$
and $\mathcal L(\mathcal H_{\mathrm{in}}) \simeq \mathbb C$. A channel
$\Lambda : \mathcal L(\mathcal H_{\mathrm{in}}) \to \mathcal L(\mathcal H_{\mathrm{out}})$
is completely specified by $\Lambda(1) = \rho$ with
$\rho \in \mathcal D(\mathcal H_{\mathrm{out}})$, so channels with
$d_{\mathrm{in}} = 1$ are precisely state-preparation maps
$z \mapsto z\,\rho$ for some fixed state $\rho$. The (normalised) Choi state
also reduces to $\rho$, and for two such channels with outputs $\rho_1,\rho_2$
one has $\|\Lambda_1 - \Lambda_2\|_\diamond = \|\rho_1 - \rho_2\|_1$.

We can therefore specialise our channel-learning bounds to state tomography.

\begin{corollary}[Trace-distance tomography of a rank-$r$ state]
\label{co:state-prep-tomography}
Let $\mathcal H \simeq \mathbb C^{d}$ and let
$\rho \in \mathcal D(\mathcal H)$ be an unknown state of rank at most $r$.
Fix $\varepsilon \in (0,1)$ and $\delta \in (0,1)$, and set
$d_{\mathrm{tot}} \coloneqq d r$. Assume that
$\delta > 4 \exp(-d_{\mathrm{tot}})$, i.e.\ that the target failure
probability is not exponentially small in $d r$. Then there exists a
tomography scheme which, given
\begin{align}
  N
  = 256\,\frac{rd}{\varepsilon^{2}}
    + O\left(\frac{\log(1/\delta)}{\varepsilon^{2}}\right),
  \label{eq:state-prep-N-explicit}
\end{align}
copies of $\rho$, outputs an estimator $\hat\rho$ such that
$\Pr\bigl[\tfrac{1}{2}\|\hat\rho - \rho\|_1 \le \varepsilon\bigr] \ge 1 - \delta$.
\end{corollary}

\begin{proof}
View $\rho$ as the output of the state-preparation channel
$\Lambda : \mathcal L(\mathbb C) \to \mathcal L(\mathcal H)$ defined by
$\Lambda(1) \coloneqq \rho$. The Kraus rank of $\Lambda$ equals the rank of
$\rho$, so the rank assumption is exactly the Kraus-rank assumption in
Theorem~\ref{th:th1Mainapp} with $d_{\mathrm{in}} = 1$,
$d_{\mathrm{out}} = d$ and $d_{\mathrm{tot}} = d r$. Applying
Theorem~\ref{th:th1Mainapp} under the condition
$\delta > 4 \exp(-d_{\mathrm{tot}})$ yields a tomography algorithm which,
given $N$ as in \eqref{eq:state-prep-N-explicit}, produces an estimator
$\hat\Lambda$ with
$\Pr\bigl[\tfrac{1}{2}\|\hat\Lambda - \Lambda\|_\diamond \le \varepsilon\bigr]
\ge 1 - \delta$. For $d_{\mathrm{in}} = 1$ each channel use prepares one copy
of $\rho$, and
$\tfrac{1}{2}\|\hat\Lambda - \Lambda\|_\diamond
 = \tfrac{1}{2}\|\hat\rho - \rho\|_1$, which gives the claim.
\end{proof}

\noindent
The bound \eqref{eq:state-prep-N-explicit} matches the optimal scaling
$N = \Theta(r d  / \varepsilon^{2})$ for $r$-rank quantum state tomography in
trace distance, with an explicit leading constant $256$ in the regime where
$\delta$ is not exponentially small in $r d $ (for instance, for constant
failure probability).
\subsubsection{Diamond-distance learning of isometries}
\label{subsec:isometry-learning}

We next specialise our main result to channels of Kraus rank~$1$. Recall that a
channel $\Lambda : \mathcal L(\mathcal H_{\mathrm{in}})
\to \mathcal L(\mathcal H_{\mathrm{out}})$ has Kraus rank~$1$ if and only if
it is of the form $\Lambda(\rho) = V \rho V^\dagger$ for an isometry
$V : \mathcal H_{\mathrm{in}} \to \mathcal H_{\mathrm{out}}$, so necessarily
$d_{\mathrm{in}} \le d_{\mathrm{out}}$. In particular, when
$d_{\mathrm{in}} = d_{\mathrm{out}} = d$, such channels are exactly unitary
channels. In this regime our general theorem yields diamond-norm tomography
with sample complexity scaling linearly in $d_{\mathrm{in}} d_{\mathrm{out}}$.

\begin{corollary}[Diamond-distance tomography for isometries]
\label{co:isometry-diamond}
Let $\Lambda(\rho) = V \rho V^\dagger$ be an isometric channel with
$V : \mathcal H_{\mathrm{in}} \to \mathcal H_{\mathrm{out}}$,
$d_{\mathrm{in}} \coloneqq \dim(\mathcal H_{\mathrm{in}})$ and
$d_{\mathrm{out}} \coloneqq \dim(\mathcal H_{\mathrm{out}})$, so that
$d_{\mathrm{in}} \le d_{\mathrm{out}}$. Fix $\varepsilon \in (0,1)$ and
$\delta \in (0,1)$, and set
$d_{\mathrm{tot}} \coloneqq d_{\mathrm{in}} d_{\mathrm{out}}$. Assume that
$\delta > 4 \exp(-d_{\mathrm{tot}})$, i.e.\ that the target failure probability
is not exponentially small in $d_{\mathrm{in}} d_{\mathrm{out}}$. Then there
exists a tomography scheme which, given
\begin{align}
  N
  = 256\,\frac{d_{\mathrm{in}} d_{\mathrm{out}}}{\varepsilon^{2}}
    + O\left(\frac{d_{\mathrm{in}}\log(1/\delta)}{\varepsilon^{2}}\right),
  \label{eq:isometry-sample-complexity}
\end{align}
uses of $\Lambda$, outputs an estimator $\hat\Lambda$ such that
\[
  \Pr\bigl[\tfrac{1}{2}\|\hat\Lambda - \Lambda\|_\diamond \le \varepsilon\bigr]
  \;\ge\; 1 - \delta.
\]
In particular, for unitary channels ($d_{\mathrm{in}} = d_{\mathrm{out}} = d$),
we obtain
\[
  N
  = 256\,\frac{d^{2}}{\varepsilon^{2}}
    + O\left(\frac{d\,\log(1/\delta)}{\varepsilon^{2}}\right),
\]
which simplifies to $N = \Theta(d^{2}/\varepsilon^{2})$ for any fixed constant
success probability, thus matching the optimal $d^{2}$-dependence. The
$1/\varepsilon^{2}$ dependence can be improved to Heisenberg scaling
$1/\varepsilon$ for unitary channels by combining our learner with the boosting
strategy of Ref.~\cite{HaahKothariODonnellTang2023}.
\end{corollary}

\begin{proof}
For a Kraus-rank-one channel we have $k = 1$, so
$d_{\mathrm{tot}} = d_{\mathrm{in}} d_{\mathrm{out}} k
= d_{\mathrm{out}} d_{\mathrm{in}}$. Applying
Theorem~\ref{th:th1Mainapp} with $k = 1$ and
$\delta > 4 \exp(-d_{\mathrm{tot}})$ yields a tomography algorithm which,
given
\[
  N
  = 256\,\frac{d_{\mathrm{in}} d_{\mathrm{out}}}{\varepsilon^{2}}
    + O\left(\frac{d_{\mathrm{in}}\log(1/\delta)}{\varepsilon^{2}}\right),
\]
produces an estimator $\hat\Lambda$ such that
$\Pr\bigl[\tfrac{1}{2}\|\hat\Lambda - \Lambda\|_\diamond \le \varepsilon\bigr]
\ge 1 - \delta$. This is exactly the claimed statement.
\end{proof}

\noindent
To the best of our knowledge, this is the first explicit tomography scheme that
learns arbitrary isometric channels in diamond distance with query complexity
$O(d_{\mathrm{in}} d_{\mathrm{out}} / \varepsilon^{2})$. This dependence on both
the dimensions and the accuracy parameter is optimal: recent lower bounds rule out Heisenberg scaling
$1/\varepsilon$ for generic isometries, in contrast to the unitary
case~\cite{ISOMEyoshida2025quantumadvantagestorageretrieval}. Moreover, in the
isometric (in particular unitary) case the Choi state is already pure, so the
purification step from our general channel-learning construction is not needed:
one can obtain the guarantees of Corollary~\ref{co:isometry-diamond} simply by
preparing many copies of the Choi state and applying a sample-optimal
single-copy pure-state tomography scheme such as the algorithm of Ref.~\cite{GutaKahnKuengTropp2020}. In this regime, our protocol
is therefore fully non-adaptive and does not require any coherent operations
across different Choi-state copies, while still achieving the essentially
optimal scaling in $d_{\mathrm{in}} d_{\mathrm{out}}$ and $1/\varepsilon^{2}$.
In the special case of unitary channels ($d_{\mathrm{in}} = d_{\mathrm{out}} = d$),
this gives a concrete $d^{2}/\varepsilon^{2}$-scaling Choi-state-based learner
which can be plugged directly into the boosting scheme of
Ref.~\cite{HaahKothariODonnellTang2023} to obtain Heisenberg scaling
$N = \Theta(d^{2}/\varepsilon)$.

\subsubsection{Operator-norm learning of POVMs}
\label{subsec:povm-learning}

We now specialise our general channel-tomography guarantees to the tomography
of measurement devices.

A binary POVM on \(\mathcal H\) is specified by a single POVM element
\(E\in\mathcal L(\mathcal H)\) with \(0\le E\le \mathbb I\); the two outcomes
correspond to \(\{E,\mathbb I-E\}\). Such a measurement induces the
classical-output channel
\begin{align}
  \Lambda_E : \mathcal L(\mathcal H) \to \mathcal L(\mathbb C^2),
  \qquad
  \Lambda_E(\rho)
  \coloneqq
  \operatorname{tr}(E\rho)\ketbra{0}{0}
  +
  \operatorname{tr}((\mathbb I-E)\rho)\ketbra{1}{1}.
\end{align}
This fits our general channel framework with input dimension \(d\) and output
dimension \(d_{\mathrm{out}}=2\).

\begin{lemma}[Diamond norm vs.\ operator norm for binary POVMs]
\label{le:binary-povm-diamond}
Let \(E,F\in\mathcal L(\mathcal H)\) with \(0\le E,F\le\mathbb I\), and let
\(\Lambda_E,\Lambda_F\) be the associated binary POVM channels. Then
\begin{align}
  \|\Lambda_E-\Lambda_F\|_\diamond
  =
  2\|E-F\|_\infty .
\end{align}
\end{lemma}

\begin{proof}
Set \(\Delta\coloneqq\Lambda_E-\Lambda_F\) and \(G\coloneqq E-F\). For any
input state \(\rho_{AR}\in\mathcal D(\mathcal H\otimes\mathcal H_R)\), where
the channel acts on the \(A\)-system,
\begin{align}
  (\Delta\otimes\mathrm{id}_R)(\rho_{AR})
  =
  \ketbra{0}{0}\otimes X
  -
  \ketbra{1}{1}\otimes X,
  \qquad
  X\coloneqq
  \operatorname{tr}_A\!\left[(G\otimes\mathbb I_R)\rho_{AR}\right].
\end{align}
The operator \(X\) is Hermitian. Indeed, using cyclicity of the partial trace
over operators acting on the \(A\)-system,
\begin{align}
  X^\dagger
  =
  \operatorname{tr}_A\!\left[\rho_{AR}(G\otimes\mathbb I_R)\right]
  =
  \operatorname{tr}_A\!\left[(G\otimes\mathbb I_R)\rho_{AR}\right]
  =
  X .
\end{align}
Since the two output blocks are orthogonal,
\(\|(\Delta\otimes\mathrm{id}_R)(\rho_{AR})\|_1=2\|X\|_1\). Moreover, by
duality of the trace norm for Hermitian operators,
\begin{align}
  \|X\|_1
  &=
  \max_{\substack{H=H^\dagger\\ \|H\|_\infty\le 1}}
  \left|\operatorname{tr}(HX)\right|  \\
  &=
  \max_{\substack{H=H^\dagger\\ \|H\|_\infty\le 1}}
  \left|\operatorname{tr}\!\left[(G\otimes H)\rho_{AR}\right]\right|
  \le
  \max_{\substack{H=H^\dagger\\ \|H\|_\infty\le 1}}
  \|G\otimes H\|_\infty
  \le
  \|G\|_\infty .
\end{align}
Thus \(\|\Lambda_E-\Lambda_F\|_\diamond\le 2\|E-F\|_\infty\). For the reverse
inequality, take a product input \(\rho_{AR}=\rho_A\otimes\rho_R\), with
\(\rho_A\) an eigenstate of \(G\) corresponding to an eigenvalue of maximal
absolute value. Then
\begin{align}
  \bigl\|(\Delta\otimes\mathrm{id}_R)(\rho_{AR})\bigr\|_1
  =
  2|\operatorname{tr}(G\rho_A)|
  =
  2\|G\|_\infty .
\end{align}
Taking the supremum over inputs proves the claim.
\end{proof}

The next lemma is the step needed to pass from a generic channel estimate to
an estimate of a POVM element. This is necessary because the channel estimator
produced by Theorem~\ref{th:th1Mainapp} need not itself be a binary POVM
channel.

\begin{lemma}[Extracting a POVM element from a channel estimate]
\label{le:extract-povm-effect}
Let \(E\) be a POVM element on \(\mathcal H\), and let \(\Lambda_E\) be the
associated binary POVM channel. Let
\(\widehat\Lambda:\mathcal L(\mathcal H)\to\mathcal L(\mathbb C^2)\) be any
CPTP map, and define
\begin{align}
  \widehat E\coloneqq \widehat\Lambda^\ast(\ketbra{0}{0}) .
\end{align}
Then \(0\le \widehat E\le\mathbb I\), and
\begin{align}
  \|\widehat E-E\|_\infty
  \le
  \frac12\|\widehat\Lambda-\Lambda_E\|_\diamond .
\end{align}
\end{lemma}

\begin{proof}
Since \(\widehat\Lambda\) is CPTP, its adjoint \(\widehat\Lambda^\ast\) is
completely positive and unital; hence \(0\le \widehat E\le\mathbb I\). Set
\(\Delta\coloneqq\widehat\Lambda-\Lambda_E\). For every
\(\rho\in\mathcal D(\mathcal H)\),
\begin{align}
  \operatorname{tr}((\widehat E-E)\rho)
  =
  \operatorname{tr}\!\left[\ketbra{0}{0}\Delta(\rho)\right].
\end{align}
The operator \(\Delta(\rho)\) is Hermitian and traceless, because both
\(\widehat\Lambda\) and \(\Lambda_E\) are trace preserving. If \(X\) is
traceless Hermitian and \(0\le M\le\mathbb I\), then
\(|\operatorname{tr}(MX)|\le \|X\|_1/2\). Indeed, writing
\(X=X_+-X_-\), tracelessness gives
\(\operatorname{tr}X_+=\operatorname{tr}X_-=\|X\|_1/2\), while
\(0\le \operatorname{tr}(MX_\pm)\le\operatorname{tr}X_\pm\). Applying this
with \(X=\Delta(\rho)\) and \(M=\ketbra{0}{0}\), we get
\begin{align}
  |\operatorname{tr}((\widehat E-E)\rho)|
  \le
  \frac12\|\Delta(\rho)\|_1
  \le
  \frac12\|\Delta\|_\diamond .
\end{align}
Taking the supremum over states \(\rho\), and using that \(\widehat E-E\) is
Hermitian, gives the claim.
\end{proof}

We now specialise our diamond-distance tomography result to
\(d_{\mathrm{out}}=2\). Any channel
\(\Lambda:\mathcal L(\mathbb C^d)\to\mathcal L(\mathbb C^{d_{\mathrm{out}}})\)
admits a Kraus representation with at most \(d\,d_{\mathrm{out}}\) operators.
Thus, for binary POVM channels we may use \(k=2d\), and the Choi purification
dimension in Theorem~\ref{th:th1Mainapp} is
\(d_{\mathrm{tot}}=d_{\mathrm{in}}d_{\mathrm{out}}k=d\cdot2\cdot2d=4d^2\).

\begin{corollary}[Learning a binary POVM in operator norm]
\label{co:binary-povm-learning}
Let \(E\in\mathcal L(\mathcal H)\) be a POVM element with
\(0\le E\le\mathbb I\) on \(\mathcal H\simeq\mathbb C^d\), and let
\(\Lambda_E\) be the associated binary POVM channel. Fix
\(\varepsilon_{\mathrm{POVM}}\in(0,1)\) and \(\delta\in(0,1)\), and assume
\(\delta>4\exp(-4d^2)\). Then there exists a tomography scheme which, given
\begin{align}
  N
  =
  1024\,\frac{d^2}{\varepsilon_{\mathrm{POVM}}^2}
  +
  O\!\left(
    \frac{d\log(1/\delta)}{\varepsilon_{\mathrm{POVM}}^2}
  \right)
  \label{eq:binary-povm-sample-complexity}
\end{align}
uses of \(\Lambda_E\), outputs a POVM element \(\widehat E\) such that
\begin{align}
  \Pr\!\left[
    \|\widehat E-E\|_\infty\le \varepsilon_{\mathrm{POVM}}
  \right]
  \ge
  1-\delta .
\end{align}
\end{corollary}

\begin{proof}
The channel \(\Lambda_E\) has \(d_{\mathrm{in}}=d\),
\(d_{\mathrm{out}}=2\), and Kraus rank at most \(2d\). Applying
Theorem~\ref{th:th1Mainapp} with \(k=2d\) is valid, and the assumption
\(\delta>4\exp(-4d^2)\) puts us in the simplified regime. Hence, using
\begin{align}
  N
  =
  256\,\frac{d_{\mathrm{in}}d_{\mathrm{out}}k}
            {\varepsilon_{\mathrm{POVM}}^2}
  +
  O\!\left(
    \frac{d_{\mathrm{in}}\log(1/\delta)}
         {\varepsilon_{\mathrm{POVM}}^2}
  \right)
  =
  1024\,\frac{d^2}{\varepsilon_{\mathrm{POVM}}^2}
  +
  O\!\left(
    \frac{d\log(1/\delta)}{\varepsilon_{\mathrm{POVM}}^2}
  \right),
\end{align}
we obtain a CPTP channel estimator \(\widehat\Lambda\) such that
\begin{align}
  \Pr\!\left[
    \frac12\|\widehat\Lambda-\Lambda_E\|_\diamond
    \le
    \varepsilon_{\mathrm{POVM}}
  \right]
  \ge
  1-\delta .
\end{align}
Define \(\widehat E\coloneqq \widehat\Lambda^\ast(\ketbra{0}{0})\). By
Lemma~\ref{le:extract-povm-effect}, \(\widehat E\) is a valid POVM element and, on
the above success event,
\begin{align}
  \|\widehat E-E\|_\infty
  \le
  \frac12\|\widehat\Lambda-\Lambda_E\|_\diamond
  \le
  \varepsilon_{\mathrm{POVM}} .
\end{align}
This proves the claim.
\end{proof}

\noindent
The bound \eqref{eq:binary-povm-sample-complexity} gives an
\(O(d^2/\varepsilon_{\mathrm{POVM}}^2)\) query complexity for learning a
generic binary POVM in operator norm. Compared with the incoherent
non-adaptive measurement-tomography guarantees of
Refs.~\cite{POVMnetzambrano2025fastquantummeasurementtomography,barberarodriguez2025boostingprojectivemethodsquantum},
this improves the dimensional scaling by using the coherent channel-tomography
primitive developed in this work.

Moreover, the quadratic dependence on \(d\) is optimal at constant accuracy.
Indeed, combining Corollary~\ref{co:binary-povm-learning} with
Proposition~\ref{pr:binary-povm-lower-bound} in the lower-bound section shows
that the \(\Theta(d^2)\) scaling for binary POVM learning is necessary and
sufficient at constant \(\varepsilon_{\mathrm{POVM}}\).

\section{Lower bounds for quantum channel tomography}
\label{sec:lower-bound}

In this section we establish lower bounds on the number of channel uses required
to learn an unknown quantum channel in diamond distance. We first prove a
lower bound which is dimension-optimal for constant accuracy and constant success
probability. We then prove a complementary lower bound with explicit dependence
on the failure probability~$\delta$, showing in particular that a separate
\(\Omega(1/\varepsilon^2)\) contribution is unavoidable for channels with
non-minimal Kraus rank.

The proofs rely on two reductions, which we briefly outline before giving the
formal statements. The dimension-dependent lower bound is obtained by reducing
tomography to a channel-identification problem. Namely, if a family of channels
is pairwise separated in diamond distance, then any tomography protocol with
small enough error can be used to identify which channel from the family was
given. We therefore upper bound the success probability of arbitrary
\(N\)-query identification protocols, and combine this with a large
diamond-distance packing of Kraus-rank-\(k\) channels. The packing is built by
packing Stinespring isometries and quotienting out the environment-unitary
freedom, since Stinespring isometries related by an environment unitary define
the same channel. The second lower bound is based on a different construction:
we embed the biases of independent classical coins into a family of channels,
so that channel tomography would imply coin-bias estimation.

\begin{theorem}
\label{thm:lower-bound}
Any tomography protocol that, for every channel
\(\Lambda:\mathcal L(\mathbb C^{d_{\mathrm{in}}})\to
\mathcal L(\mathbb C^{d_{\mathrm{out}}})\) of Kraus rank at most \(k\),
uses \(N\) queries to \(\Lambda\) and outputs an estimate \(\hat\Lambda\)
satisfying $\|\hat\Lambda-\Lambda\|_\diamond\le \varepsilon$ with constant success probability must use
\begin{align}
  N=\Omega\bigl(d_{\mathrm{in}}\,d_{\mathrm{out}}\,k\bigr)
\end{align}
channel queries.
The same lower bound holds even if the protocol is fully adaptive, is granted
access at each query to a fixed Stinespring dilation of \(\Lambda\), with the
environment register retained, and no definite causal order is imposed among
the \(N\) uses.
\end{theorem}

The proof of Theorem~\ref{thm:lower-bound} follows from the
packing-and-identification reduction described above, and is given below in
Theorem~\ref{thm:lower-boundPROOF}, where we also obtain an explicit, although
generally non-optimal, dependence on the accuracy parameter~\(\varepsilon\).

As mentioned, we also prove a complementary lower bound based on a
subfamily of channels whose learning reduces to estimating the biases of
independent classical coins. Combining the packing-identification lower bound with the coin-bias lower bound gives the following theorem.

\begin{theorem}[Query lower bounds for channel tomography]
\label{prop:combined-lb-delta-general-k}
Assume \(d_{\mathrm{out}}\ge 2\). Let
\(k_{\min}\coloneqq\left\lceil d_{\mathrm{in}}/d_{\mathrm{out}}\right\rceil\)
and assume \(k\ge k_{\min}\), since otherwise the class of Kraus-rank-\(\le k\)
channels is empty. There exist constants \(\varepsilon_0,\delta_0>0\) such that,
for every \(0<\varepsilon\le\varepsilon_0\) and
\(0<\delta\le\delta_0\), the following holds.

Define
\begin{align}
  m_{\mathrm{coin}}
  &\coloneqq
  \max\left\{
    r\in\{0,\ldots,d_{\mathrm{in}}\}:
    2r+
    \left\lceil\frac{d_{\mathrm{in}}-r}{d_{\mathrm{out}}}\right\rceil
    \le k
  \right\},
  \qquad
  m
  \coloneqq
  \max\left\{
    m_{\mathrm{coin}},
    \mathbf 1_{\{k\ge k_{\min}+1\}}
  \right\}.
  \label{eq:def-m-theorem}
\end{align}
In the worst case over all channels
\(\Lambda:\mathcal L(\mathbb C^{d_{\mathrm{in}}})\to
\mathcal L(\mathbb C^{d_{\mathrm{out}}})\) of Kraus rank at most \(k\), any
tomography protocol that outputs an estimator \(\widehat\Lambda\) satisfying
\[
  \frac12\|\widehat\Lambda-\Lambda\|_\diamond\le \varepsilon
\]
with probability at least \(1-\delta\) must use a number of channel queries
\(N\) obeying
\begin{align}
  N
  =
  \Omega\!\left(
    \frac{d_{\mathrm{out}}d_{\mathrm{in}}k}{\varepsilon^{\beta}}
    +
    \frac{m}{\varepsilon^{2}}\log\!\frac{m}{\delta}
  \right),
  \label{eq:combined-delta-lb-main}
\end{align}
where the second term is understood to be absent when \(m=0\), and
\[
  \beta
  \coloneqq
  \frac{
    2d_{\mathrm{in}}d_{\mathrm{out}}k-d_{\mathrm{in}}^2-k^2
  }{
    2\bigl(d_{\mathrm{in}}d_{\mathrm{out}}k-1\bigr)
  } .
\]
In particular, for every non-minimal Kraus-rank
\(k\ge k_{\min}+1\), we have \(m\ge1\), and hence the second term yields
\(N=\Omega(\varepsilon^{-2}\log(1/\delta))\). Moreover, when
\(k\ge 2d_{\mathrm{in}}\), we have \(m_{\mathrm{coin}}=m=d_{\mathrm{in}}\), so
the second term becomes
\[
  \Omega\!\left(
    \frac{d_{\mathrm{in}}}{\varepsilon^{2}}
    \log\frac{d_{\mathrm{in}}}{\delta}
  \right).
\]
\end{theorem}

The rest of this section makes these arguments precise. In
Subsection~\ref{subsec:identification}, we reduce channel tomography to
\emph{channel identification} and upper bound the success probability of the
identification task. In channel identification, the unknown channel is promised
to be drawn uniformly from a finite family
\(\{\Lambda_x\}_{x=1}^M\), and the goal is to output the correct label \(x\).
In Subsection~\ref{subsec:packing}, we construct a large family of
Kraus-rank-\(k\) channels that are pairwise separated in diamond norm. The
construction proceeds by packing Stinespring isometries and quotienting out the
environment-unitary gauge freedom. In Subsection~\ref{subsec:proof-lower-bound},
we combine these ingredients to prove Theorem~\ref{thm:lower-boundPROOF}, and
hence Theorem~\ref{thm:lower-bound}. We also comment on the scope of the
method: for general Kraus-rank-\(k\) channels, it captures the optimal
\emph{dimension} dependence but is not, in general, tight in the accuracy
parameter~\(\varepsilon\); for important special cases, notably pure-state
tomography in trace distance and unitary-channel tomography in diamond norm,
the same framework recovers the optimal scaling in \emph{both} the dimension
parameters and the target accuracy. In the same section, we also show a
dimension-matching lower bound for learning binary POVMs in operator norm.
Finally, in Subsection~\ref{subsec:delta-lb-coins}, we prove the
$N=\Omega(\varepsilon^{-2}\log(1/\delta))$ lower bound in
Theorem~\ref{prop:combined-lb-delta-general-k} via the reduction to estimating
biases of classical coin flips, extending it to binary POVMs as well.

\subsection{A channel-identification reduction for the dimension lower bound 
$\Omega(d_{\mathrm{in}}d_{\mathrm{out}}k)$}
\label{subsec:identification}

Let $\{\Lambda_x\}_{x=1}^M$ be a finite family of channels.
In \emph{channel identification}, the unknown channel $\Lambda$ is promised to be one of the $\Lambda_x$'s and is chosen
uniformly at random from this family.
A $Q$-query identification protocol may access $\Lambda$ exactly $Q$ times (possibly adaptively) and then outputs a guess
$\hat x\in[M]$.
Its performance is measured by the average success probability
\begin{align}
p_{\mathrm{succ}}
\coloneqq
\frac{1}{M}\sum_{x=1}^M \Pr[\hat x=x \mid \Lambda=\Lambda_x].
\end{align}

Tomography is a stronger task than identification: if one can estimate $\Lambda$ in diamond norm, then one can certainly
identify it among well separated candidates.
The next lemma formalizes this reduction.

\begin{lemma}[Tomography implies identification]
\label{lem:tomo-implies-ident}
Fix $\varepsilon>0$ and $\delta\in(0,1)$.
Suppose there exists a (possibly adaptive) $Q$-query tomography procedure $\mathsf T$ such that for every channel
$\Lambda:\mathcal L(\mathbb C^{d_{\mathrm{in}}})\to \mathcal L(\mathbb C^{d_{\mathrm{out}}})$ its output $\hat\Lambda$
satisfies
\begin{align}
\Pr\bigl[\|\hat\Lambda-\Lambda\|_\diamond \le \varepsilon\bigr]\;\ge\; 1-\delta .
\end{align}
Let $\{\Lambda_x\}_{x=1}^M$ be any family with
\begin{align}
\|\Lambda_x-\Lambda_y\|_\diamond \;>\; 2\varepsilon
\qquad\forall\,x\neq y.
\end{align}
Then there exists a (possibly adaptive) $Q$-query identification protocol such that
\begin{align}
p_{\mathrm{succ}}
\;\coloneqq\;
\frac{1}{M}\sum_{x=1}^M \Pr[\hat x=x\mid \Lambda=\Lambda_x]
\;\ge\; 1-\delta .
\end{align}
\end{lemma}

\begin{proof}
Given oracle access to $\Lambda\in\{\Lambda_x\}_{x=1}^M$, run $\mathsf T$ and obtain an estimate $\hat\Lambda$.
Output any index $\hat x \in \arg\min_{x\in[M]} \|\hat\Lambda-\Lambda_x\|_\diamond$.
Fix $x\in[M]$ and suppose $\Lambda=\Lambda_x$.
On the event $\|\hat\Lambda-\Lambda_x\|_\diamond\le \varepsilon$, for every $y\neq x$ the triangle inequality yields
\begin{align}
\|\hat\Lambda-\Lambda_y\|_\diamond
\;\ge\;
\|\Lambda_x-\Lambda_y\|_\diamond-\|\hat\Lambda-\Lambda_x\|_\diamond
\;>\;
\varepsilon,
\end{align}
whereas $\|\hat\Lambda-\Lambda_x\|_\diamond\le \varepsilon$.
Thus $x$ is the unique minimizer, hence $\hat x=x$ on this event.
Therefore,
\begin{align}
\Pr[\hat x=x\mid \Lambda=\Lambda_x]
\;\ge\;
\Pr\bigl[\|\hat\Lambda-\Lambda_x\|_\diamond\le \varepsilon\bigr]
\;\ge\;
1-\delta .
\end{align}
Averaging over $x$ proves the claim.
\end{proof}

We now upper bound the success probability of channel identification.
\begin{proposition}[Upper bound on identification success probability]
\label{prop:succ-kraus}
Let $\Lambda_1,\dots,\Lambda_M$ be CPTP maps
$\Lambda_x:\mathcal L(\mathbb C^{d_{\mathrm{in}}})\to \mathcal L(\mathbb C^{d_{\mathrm{out}}})$,
each of Kraus rank at most $k$.
Consider any $Q$-use identification protocol, allowing full adaptivity, and even granting the protocol purified
access to the channel (i.e.\ each use is provided via a Stinespring isometry and the environment register may be retained).
Then the average success probability satisfies
\begin{align}
p_{\mathrm{succ}}
\coloneqq
\frac{1}{M}\sum_{x=1}^M \Pr[\hat x=x\mid \Lambda=\Lambda_x]
\;\le\;
\frac{1}{M}\binom{Q + D - 1}{D - 1},
\qquad
D := k\, d_{\mathrm{in}} d_{\mathrm{out}}.
\end{align}
\end{proposition}
\begin{proof}
Fix an arbitrary $Q$-query identification protocol.
For each $x\in[M]$, choose a Stinespring isometry
$V_x:\mathbb C^{d_{\mathrm{in}}}\to \mathbb C^{d_{\mathrm{out}}}\otimes E$ with $E\simeq \mathbb C^{k}$ such that
$\Lambda_x(\cdot)=\Tr_E\bigl[V_x(\cdot)V_x^\dagger\bigr]$.
Consider the modified protocol in which each query applies $V_x$ and the environment register $E$ is retained.
This can only increase the distinguisher's information (tracing out $E$ recovers the original protocol), so it suffices
to bound the success probability in this purified model.

By purification and deferred measurement, we may assume the protocol is coherent until a final measurement:
there exist isometries $W_0,\dots,W_Q$ (independent of $x$), an initial unit vector $\ket{\psi_0}$ in some workspace, and a final
POVM $\{M_x\}_{x=1}^M$ on the final register $\mathcal H_{\mathrm{fin}}$ such that the final pure state equals
\begin{align}
\ket{\Phi_x}
\;=\;
W_Q\,(V_x\otimes I)\,W_{Q-1}\cdots W_1\,(V_x\otimes I)\,W_0\,\ket{\psi_0}
\;\in\;\mathcal H_{\mathrm{fin}} .
\label{eq:Phi-product-form-rewrite}
\end{align}
Let $\ket{v_x}\coloneqq \mathrm{vec}(V_x)\in\mathbb C^{D}$.
Because the right-hand side of \eqref{eq:Phi-product-form-rewrite} is linear in each of the $Q$ occurrences of $V_x$,
there exists a linear map $\widetilde L:(\mathbb C^{D})^{\otimes Q}\to\mathcal H_{\mathrm{fin}}$, depending only on the fixed
interleaving isometries $W_0,\dots,W_Q$, such that
\begin{align}
\ket{\Phi_x}
\;=\;
\widetilde L\bigl(\ket{v_x}^{\otimes Q}\bigr).
\label{eq:Phi-as-Ltilde}
\end{align}
Moreover, for every $\ket{v}\in\mathbb C^{D}$ one has $\ket{v}^{\otimes Q}\in \mathrm{Sym}^Q(\mathbb C^{D})$ (i.e., the symmetric subspace~\cite{Har13}), hence only the
restriction $L\coloneqq \widetilde L\big|_{\mathrm{Sym}^Q(\mathbb C^{D})}$ matters and we can write
\begin{align}
\ket{\Phi_x}
\;=\;
L\bigl(\ket{v_x}^{\otimes Q}\bigr),
\qquad
L:\mathrm{Sym}^Q(\mathbb C^{D})\to \mathcal H_{\mathrm{fin}} .
\label{eq:Phi-as-L}
\end{align}
In particular, all final states lie in the subspace
\begin{align}
\mathcal S
\;\coloneqq\;
L\bigl(\mathrm{Sym}^Q(\mathbb C^{D})\bigr)
\;\subseteq\;
\mathcal H_{\mathrm{fin}},
\qquad
\dim(\mathcal S)
\;\le\;
\dim\mathrm{Sym}^Q(\mathbb C^{D})
\;=\;
\binom{Q+D-1}{D-1},
\label{eq:dimS-rewrite}
\end{align}
where the inequality follows because $\mathcal S$ is the image of a linear map (and a linear map cannot increase dimension).

Let $\rho_x\coloneqq \ket{\Phi_x}\bra{\Phi_x}$ and let $\Pi_{\mathcal S}$ be the orthogonal projector onto $\mathcal S$.
Since $\rho_x$ is supported on $\mathcal S$, we have $\rho_x=\Pi_{\mathcal S}\rho_x\Pi_{\mathcal S}$ and $0\preceq\rho_x\preceq I_{\mathcal S}$,
where $I_{\mathcal S}$ denotes the identity on $\mathcal S$.
Define the compressed POVM elements $M'_x\coloneqq \Pi_{\mathcal S}M_x\Pi_{\mathcal S}$, so that $\sum_x M'_x=I_{\mathcal S}$ and
$\Tr(M_x\rho_x)=\Tr(M'_x\rho_x)$ for all $x$.
Using $0\preceq\rho_x\preceq I_{\mathcal S}$ we obtain
\begin{align}
\sum_{x=1}^M \Pr[\hat x=x\mid \Lambda=\Lambda_x]
&=
\sum_{x=1}^M \Tr(M_x\rho_x)
=
\sum_{x=1}^M \Tr(M'_x\rho_x)
\;\le\;
\sum_{x=1}^M \Tr(M'_x I_{\mathcal S})
=
\sum_{x=1}^M \Tr(M'_x)
=
\Tr(I_{\mathcal S})
=
\dim(\mathcal S).
\end{align}
Dividing by $M$ and invoking \eqref{eq:dimS-rewrite}, we conclude the proof.
\end{proof}

\paragraph{Alternative proof and robustness beyond causal order.}
\label{subsubsec:alt-proof-unordered}

Proposition~\ref{prop:succ-kraus} was proved above in an arguably elementary way within a very general $Q$-use black-box model,
allowing full adaptivity between the $Q$ uses of the channel (and, in fact, even granting purified access via a fixed Stinespring dilation).

We now give an alternative proof, following the representation-theoretic strategy of
Ref.~\cite{HaahKothariODonnellTang2023}, which enjoys an additional advantage: it does not rely on the assumption that the $Q$ calls to
the unknown channel are arranged in a definite time order.
Informally, ``dropping the causal-order constraint'' means that we do not insist on a sequential implementation where one call is
completed and processed before deciding how to feed the next call.
Instead, we allow the $Q$ uses to be wired together in a single global procedure, potentially incompatible with any fixed
ordering of the calls \cite{Oreshkov_2012,Chiribella_2013,Ara_jo_2015,Hoffreumon_2021,Chiribella_2008,Chiribella_2009}.

Therefore, the resulting success-probability bound applies not only to ordinary physical (causally ordered) adaptive strategies,
but also to these more permissive causally unordered models.
In particular, this strengthens our optimality claim for diamond-norm learning: the dimension-dependent scaling
$\Theta(d_{\mathrm{in}}d_{\mathrm{out}}k)$ of our protocol cannot be improved by any physically realizable strategy, nor even by allowing such
unphysical---and in principle more powerful---forms of access.

\begin{proposition}[Upper bound on identification success probability beyond causal order]
\label{prop:succ-krausUNPHYS}
Let $\Lambda_1,\dots,\Lambda_M$ be CPTP maps
$\Lambda_x:\mathcal L(\mathbb C^{d_{\mathrm{in}}})\to \mathcal L(\mathbb C^{d_{\mathrm{out}}})$,
each of Kraus rank at most $k$.
Consider any $Q$-query identification procedure, without assuming any definite causal order among the $Q$ uses.
Then the average success probability satisfies
\begin{align}
p_{\mathrm{succ}}
\coloneqq
\frac{1}{M}\sum_{x=1}^M \Pr[\hat x=x\mid \Lambda=\Lambda_x]
\;\le\;
\frac{1}{M}\binom{Q + D - 1}{D - 1},
\qquad
D := k\, d_{\mathrm{in}} d_{\mathrm{out}}.
\end{align}
\end{proposition}

\begin{proof}
Let $|\Phi\rangle:=\sum_{i=1}^{d_{\mathrm{in}}}|i\rangle|i\rangle$ and define the Choi operator
\begin{align}
J(\Lambda)\;\coloneqq\;(\mathrm{id}\otimes \Lambda)\bigl(|\Phi\rangle\langle\Phi|\bigr).
\end{align}
A general representation theorem~\cite[Sec.~C]{Bavaresco_2022} for $Q$-use channel experiments implies that there exist operators $T_x\succeq 0$ such that,
for every CPTP map $\Lambda$,
\begin{align}
\Pr[\hat x=x\mid \Lambda]=\Tr\bigl(T_x\,J(\Lambda)^{\otimes Q}\bigr),
\qquad
\Tr\bigl(T_{\mathrm{sum}}\,J(\Lambda)^{\otimes Q}\bigr)=1,
\qquad
T_{\mathrm{sum}}\coloneqq\sum_{x=1}^M T_x .
\label{eq:tester}
\end{align}
This includes the usual time-ordered tester/comb formalism \cite{Chiribella_2008,Chiribella_2009} and also causally unordered
model, see e.g.\ \cite[Sec.~C]{Bavaresco_2022}
and references therein.
In what follows we will only use positivity and the normalization in \eqref{eq:tester}; in particular we do not invoke the additional linear constraints on $\{T_x\}$ that encode a definite causal order \cite{Chiribella_2008,Chiribella_2009,Bavaresco_2022}.

Fix an environment $E\simeq\mathbb C^k$ and set $D_{\mathrm{out}}\coloneqq d_{\mathrm{out}}k$, so
$\mathbb C^{d_{\mathrm{out}}}\otimes E\simeq\mathbb C^{D_{\mathrm{out}}}$ and $D=d_{\mathrm{in}}D_{\mathrm{out}}$.
Choose Stinespring isometries $V_x:\mathbb C^{d_{\mathrm{in}}}\to\mathbb C^{d_{\mathrm{out}}}\otimes E\simeq\mathbb C^{D_{\mathrm{out}}}$
with $\Lambda_x(\rho)=\Tr_E(V_x\rho V_x^\dagger)$, and fix $V_0:=V_1$.
Then $V_x=U_xV_0$ for some unitary $U_x\in U(D_{\mathrm{out}})$.

Introduce the lifted isometry channel $\widetilde\Lambda_V(\rho)\coloneqq V\rho V^\dagger$ (whose output is on $\mathbb C^{D_{\mathrm{out}}}$).
By linearity of the Choi map and of the partial trace,
\begin{align}
J(\Lambda_x)^{\otimes Q}
=
\Tr_{E^{\otimes Q}}\bigl(J(\widetilde\Lambda_{V_x})^{\otimes Q}\bigr).
\label{eq:lift}
\end{align}
Let $\mathcal H\coloneqq \mathbb C^{d_{\mathrm{in}}}\otimes\mathbb C^{D_{\mathrm{out}}}$ (so $\dim\mathcal H=D$) and define
$|\psi_V\rangle\coloneqq (I\otimes V)|\Phi\rangle\in\mathcal H$, so that
$J(\widetilde\Lambda_V)=|\psi_V\rangle\langle\psi_V|$.
Set $|\psi\rangle\coloneqq |\psi_{V_0}\rangle^{\otimes Q}\in \mathrm{Sym}^Q(\mathcal H)$.

Let us consider the group $G:=U(D_{\mathrm{out}})$ and let it act on $\mathrm{Sym}^Q(\mathcal H)$ by the unitary representation
$R(U)\coloneqq (I_{\mathbb C^{d_{\mathrm{in}}}}\otimes U)^{\otimes Q}\big|_{\mathrm{Sym}^Q(\mathcal H)}$.
Define the Haar twirl
\begin{align}
\bar C_{\mathrm{iso}}
\;\coloneqq\;
\int_{U(D_{\mathrm{out}})} R(U)\,|\psi\rangle\langle\psi|\,R(U)^\dagger\,dU,
\qquad
\bar C\;\coloneqq\;\Tr_{E^{\otimes Q}}(\bar C_{\mathrm{iso}}),
\label{eq:twirl}
\end{align}
where $dU$ is normalized Haar measure.
A general ``twirl'' upper bound \cite[App.~F, Lem.~2]{Bavaresco_2022} (stated below for completeness in Lemma~\ref{lem:twirl-domination}) gives
\begin{align}
|\psi\rangle\langle\psi|
\;\preceq\;
\dim\bigl(\mathrm{Sym}^Q(\mathcal H)\bigr)\,\bar C_{\mathrm{iso}}
=
\binom{Q+D-1}{D-1}\,\bar C_{\mathrm{iso}} .
\label{eq:psi-dom}
\end{align}

Using $V_x=U_xV_0$, we have
$J(\widetilde\Lambda_{V_x})^{\otimes Q}=R(U_x)\,|\psi\rangle\langle\psi|\,R(U_x)^\dagger$.
Since $\bar C_{\mathrm{iso}}$ is $R(U)$-invariant by construction, \eqref{eq:psi-dom} implies
$J(\widetilde\Lambda_{V_x})^{\otimes Q}\preceq \binom{Q+D-1}{D-1}\,\bar C_{\mathrm{iso}}$.
Applying $\Tr_{E^{\otimes Q}}$ and \eqref{eq:lift} yields
\begin{align}
J(\Lambda_x)^{\otimes Q}\;\preceq\;\binom{Q+D-1}{D-1}\,\bar C .
\label{eq:choi-dom}
\end{align}

Combining \eqref{eq:tester} and \eqref{eq:choi-dom} we obtain
\begin{align}
p_{\mathrm{succ}}
&=
\frac{1}{M}\sum_{x=1}^M \Tr\bigl(T_x\,J(\Lambda_x)^{\otimes Q}\bigr)
\;\le\;
\frac{1}{M}\binom{Q+D-1}{D-1}\sum_{x=1}^M \Tr(T_x\bar C) \\
&=
\frac{1}{M}\binom{Q+D-1}{D-1}\Tr(T_{\mathrm{sum}}\bar C)
=
\frac{1}{M}\binom{Q+D-1}{D-1},
\end{align}
where $\Tr(T_{\mathrm{sum}}\bar C)=1$ follows by averaging the normalization in \eqref{eq:tester} over the family of channels
$\Lambda_{UV_0}(\rho)\coloneqq \Tr_E(UV_0\rho V_0^\dagger U^\dagger)$:
since each $\Lambda_{UV_0}$ is CPTP, \eqref{eq:tester} gives
$\Tr\bigl(T_{\mathrm{sum}}J(\Lambda_{UV_0})^{\otimes Q}\bigr)=1$ for all $U$,
and integrating over Haar measure yields $\Tr(T_{\mathrm{sum}}\bar C)=1$.
\end{proof}

\medskip

\begin{lemma}[Bounding by Haar twirling~\cite{Bavaresco_2022}]
\label{lem:twirl-domination}
Let $G$ be a compact group with a unitary representation $R$ on a finite-dimensional Hilbert space $\mathsf K$, and let
$|\psi\rangle\in\mathsf K$ be a unit vector. Define
\begin{align}
\mathcal{T}(|\psi\rangle\langle\psi|) \;\coloneqq\; \int_G R(g)\,|\psi\rangle\langle\psi|\,R(g)^\dagger\,dg ,
\end{align}
with $dg$ normalized Haar measure. Then
\begin{align}
|\psi\rangle\langle\psi| \;\preceq\; \gamma_G\,\mathcal{T}(|\psi\rangle\langle\psi|),
\qquad
\gamma_G \;=\; \sum_{\mu} d_\mu\,\min\{m_\mu,d_\mu\},
\end{align}
where the sum runs over the irreducible components of $R$ on $\mathsf K$, with irrep dimension $d_\mu$ and multiplicity $m_\mu$.
In particular, $\gamma_G\le \sum_\mu d_\mu m_\mu=\dim(\mathsf K)$.
This is a special case of \cite[App.~F, Lem.~2]{Bavaresco_2022} (see also the references therein).
\end{lemma}

\subsubsection{Packing Kraus-rank-$k$ channels separated in diamond norm}
\label{subsec:packing}
In this subsection, we construct a large set of Kraus-rank-$k$ quantum channels that are well separated in
diamond norm, i.e.\ a diamond-norm packing of the set of Kraus-rank-$k$ channels, that we will later use to obtain our query complexity lower bound for learning channels.

Here, we start by briefly reviewing the basic notions of covering nets and packings, and collect
the standard bounds we utilize for the unitary group and Stiefel manifold (isometries)~\cite{AubrunSzarek2017AliceBob,Vershynin_2018,NETHOMszarek1997metricentropyhomogeneousspaces}.
Our packing construction for channels then proceeds via Stinespring dilations. Any Kraus-rank-$k$ channel admits a Stinespring
isometry from the input space into an output space tensored with an environment of dimension $k$. Two
Stinespring isometries describe the same channel exactly when they differ by a unitary acting only on the
environment. We therefore construct a packing of Stinespring isometries modulo this natural $U(k)$ gauge
action. Finally, we use standard relations between the resulting distance on Stinespring
dilations and the diamond-norm distance of the induced channels~\cite{STINdistkretschmann2007continuitytheoremstinespringsdilation},
which allow us to transfer the packing to diamond norm.

\subsubsubsection{Preliminaries on packing and covering nets}
We briefly recall the basic definitions; see~\cite{AubrunSzarek2017AliceBob,Vershynin_2018} for further
background.
Let $(X,d)$ be a metric space and $\varepsilon>0$.
An \emph{$\varepsilon$-net} (or \emph{$\varepsilon$-cover}) is a set $\mathcal N\subseteq X$ such that for every
$x\in X$ there exists $y\in\mathcal N$ with $d(x,y)\le \varepsilon$.
The corresponding \emph{covering number} is the smallest possible cardinality of such a set:
\begin{align}
N(X,d,\varepsilon)
\;\coloneqq\;
\min\Bigl\{|\mathcal N|:\ \mathcal N\subseteq X\text{ is an }\varepsilon\text{-net for }X\Bigr\}.
\end{align}
An \emph{$\varepsilon$-packing} is a set $\mathcal P\subseteq X$ with $d(x,y)>\varepsilon$ for all distinct
$x,y\in\mathcal P$; the \emph{packing number} is the largest possible cardinality of such a set:
\begin{align}
M(X,d,\varepsilon)
\;\coloneqq\;
\max\Bigl\{|\mathcal P|:\ \mathcal P\subseteq X\text{ is an }\varepsilon\text{-packing}\Bigr\}.
\end{align}
Covering and packing numbers are equivalent up to constant factors in $\varepsilon$: for all $\varepsilon>0$,
\begin{align}
M(X,d,2\varepsilon)\ \le\ N(X,d,\varepsilon)\ \le\ M(X,d,\varepsilon).
\label{eq:pack-cover}
\end{align}

We will use covering/packing bounds for the unitary group and the so-called complex Stiefel manifold with the operator-norm metric $d_{\infty}(X,Y)\coloneqq\|X-Y\|_{\infty}$.
Recall that the complex Stiefel manifold is the set of isometries $\mathbb C^p\to\mathbb C^n$,
\begin{align}
\mathrm{St}_{\mathbb C}(n,p)
\;\coloneqq\;
\bigl\{V\in\mathbb C^{n\times p}:\ V^\dagger V=I_p\bigr\}.
\end{align}

\begin{lemma}[Operator-norm nets for $U(n)$ and $\mathrm{St}_{\mathbb C}(n,p)$~\cite{NETHOMszarek1997metricentropyhomogeneousspaces}]
\label{lem:unitary-stiefel-nets}
There exist constants $c_U,C_U,c_S,C_S>0$ such that for all $n\ge 1$, $1\le p\le n$, and
$\varepsilon\in(0,1]$,
\begin{align}
\Bigl(\frac{c_S}{\varepsilon}\Bigr)^{2np-p^2}
\ \le\
N\bigl(\mathrm{St}_{\mathbb C}(n,p),d_{\infty},\varepsilon\bigr)
\ \le\
\Bigl(\frac{C_S}{\varepsilon}\Bigr)^{2np-p^2}.
\label{eq:cover-St}
\end{align}
and (in particular for $p=n$)
\begin{align}
\Bigl(\frac{c_U}{\varepsilon}\Bigr)^{n^2}
\ \le\
N\bigl(U(n),d_{\infty},\varepsilon\bigr)
\ \le\
\Bigl(\frac{C_U}{\varepsilon}\Bigr)^{n^2},
\label{eq:cover-U}
\end{align}
By Eq.~\eqref{eq:pack-cover}, the corresponding packing numbers satisfy the same bounds up to rescaling
$\varepsilon$.
\end{lemma}

\begin{proof}[Proof sketch with pointers to the literature]
The operator-norm covering-number bound for the unitary group is given in
\cite[Theorem~7]{NETHOMszarek1997metricentropyhomogeneousspaces}.
For the Stiefel manifold, note that any $V\in\mathrm{St}_{\mathbb C}(n,p)$ can be realized as the first $p$
columns of a unitary $U\in U(n)$, and that two unitaries correspond to the same $V$ if and only if they differ by
right multiplication by a block-diagonal unitary $\mathrm{diag}(I_p,W)$ with $W\in U(n-p)$. Hence
$\mathrm{St}_{\mathbb C}(n,p)$ can be identified with the quotient $U(n)/U(n-p)$ (see, e.g.,
\cite[Sec.~B.4]{AubrunSzarek2017AliceBob}). Theorem~~11 in~\cite{NETHOMszarek1997metricentropyhomogeneousspaces} provides upper and lower bounds on the
covering numbers of such quotients equipped with the so-called
\emph{quotient metric} induced by $\|\cdot\|_\infty$ on $U(n)$.
As noted in \cite[Sec.~2]{NETHOMszarek1997metricentropyhomogeneousspaces}, this quotient metric is equivalent (up to
absolute constants) to the operator-norm metric.
Applying these bounds to the quotient $U(n)/U(n-p)$ yields the claimed covering-number estimates,
with exponent
\begin{align}
\dim_{\mathbb R}\mathrm{St}_{\mathbb C}(n,p)
=
\dim_{\mathbb R}\bigl(U(n)/U(n-p)\bigr)
=
\dim_{\mathbb R}U(n)-\dim_{\mathbb R}U(n-p)
=
n^2-(n-p)^2
=
2np-p^2.
\end{align}
Alternatively, one may obtain~\eqref{eq:cover-St} by combining the operator-norm covering estimates for $U(n)$
and for the complex Grassmannian reported in Table~5.1 of~\cite{AubrunSzarek2017AliceBob}, together with the
standard relation between the Stiefel manifold, the Grassmannian, and $U(p)$ (see~\cite[Sec.~B.4]{AubrunSzarek2017AliceBob}).

For an explicit net construction in the real case (adaptable to the complex case), see also
\cite[Lemma~4.1]{REALhinrichs2016entropynumbersembeddingsschatten}.
\end{proof}

Let $\mathcal V=\mathrm{St}_{\mathbb C}(d_{\mathrm{out}}k,d_{\mathrm{in}})$.
Since Kraus-rank-$k$ channels are invariant under unitary changes of the environment, we work modulo
environment unitaries: we declare $V,W\in\mathcal V$ equivalent if $W=(I_B\otimes U)V$ for some $U\in U(k)$, and
we write $\widetilde{\mathcal V}\coloneqq \mathcal V/U(k)$ for the corresponding set of equivalence classes.
We equip $\widetilde{\mathcal V}$ with the distance (see Ref.~\cite{STINdistkretschmann2007continuitytheoremstinespringsdilation})
\begin{align}
\label{eq:dStin}
d_{\mathrm{Stin}}(V,W)
\coloneqq
\inf_{U\in U(k)} \bigl\|(I_{B}\otimes U)V - W\bigr\|_{\infty}.
\end{align}
The right-hand side depends only on the equivalence classes of $V$ and $W$, and hence defines a genuine metric
on $\widetilde{\mathcal V}$.
Intuitively, $d_{\mathrm{Stin}}(V,W)$ is the smallest operator-norm distance achievable after choosing the best
environment basis.

\begin{lemma}[A packing lower bound modulo the $U(k)$ gauge for isometries]
\label{lem:gauge-quotient-cover}
There exists a constant $c>0$ such that for all $\eta\in(0,1]$,
\begin{align}
M\bigl(\widetilde{\mathcal V},d_{\mathrm{Stin}},\eta\bigr)
\ \ge\
N\bigl(\widetilde{\mathcal V},d_{\mathrm{Stin}},\eta\bigr)
\ \ge\
\frac{N\bigl(\mathcal V,d_{\infty},2\eta\bigr)}{N\bigl(U(k),d_{\infty},\eta\bigr)}
\ \ge\
\Bigl(\frac{c}{\eta}\Bigr)^{\,2k\,d_{\mathrm{in}}d_{\mathrm{out}}-d_{\mathrm{in}}^2-k^2}.
\label{eq:gauge-quotient-cover}
\end{align}
\end{lemma}

\begin{proof}
The first inequality is due to Eq.~\eqref{eq:pack-cover}.

For the second inequality, let $K\coloneqq N(\widetilde{\mathcal V},d_{\mathrm{Stin}},\eta)$.
Choose representatives $V_1,\dots,V_K\in\mathcal V$ of $K$ equivalence classes such that for every $V\in\mathcal V$
there exists $t\in[K]$ with $d_{\mathrm{Stin}}(V_t,V)\le \eta$.
Let $L\coloneqq N(U(k),d_\infty,\eta)$ and choose $U_1,\dots,U_L\in U(k)$ such that for every $U\in U(k)$ there exists
$s\in[L]$ with $\|U-U_s\|_\infty\le \eta$.

We claim that the $KL$ points $\{(I_B\otimes U_s)V_t\}_{t\in[K],\,s\in[L]}$ form a $2\eta$-net for $(\mathcal V,d_\infty)$.
Indeed, fix $V\in\mathcal V$ and pick $t$ with $d_{\mathrm{Stin}}(V_t,V)\le \eta$.
Then there exists $U\in U(k)$ such that $\|(I_B\otimes U)V_t - V\|_\infty\le \eta$.
Pick $s$ with $\|U-U_s\|_\infty\le \eta$. Using $\|V_t\|_\infty=1$, we obtain
\begin{align}
\|(I_B\otimes U_s)V_t - V\|_\infty
&\le
\|(I_B\otimes (U_s-U))V_t\|_\infty + \|(I_B\otimes U)V_t - V\|_\infty \\
&\le
\|U_s-U\|_\infty\,\|V_t\|_\infty + \eta
\le 2\eta.
\end{align}
Hence $N(\mathcal V,d_\infty,2\eta)\le KL$, which yields the second inequality in~\eqref{eq:gauge-quotient-cover}.

Finally, we apply Lemma~\ref{lem:unitary-stiefel-nets} to $\mathcal V=\mathrm{St}_{\mathbb C}(d_{\mathrm{out}}k,d_{\mathrm{in}})$
and $U(k)$, and absorb all constants into $c$.
\end{proof}

\subsubsubsection{Packing number lower bound for quantum channels}
\label{subsec:packing2}
We now combine the previous ingredients to obtain a diamond-norm packing of Kraus-rank-$k$ channels.
Let $B\simeq \mathbb C^{d_{\mathrm{out}}}$ and $E\simeq \mathbb C^{k}$, and set
$\mathcal V\coloneqq \mathrm{St}_{\mathbb C}(d_{\mathrm{out}}k,d_{\mathrm{in}})$.
For $V\in\mathcal V$ define the channel
\begin{align}
\Lambda_V(\rho)\;\coloneqq\;\Tr_E\bigl[V\rho V^\dagger\bigr].
\end{align}
Then $\Lambda_V$ has Kraus rank at most $k$, and $\Lambda_{(I_B\otimes U)V}=\Lambda_V$ for all $U\in U(k)$.
Moreover, \cite[Eq.~(3)]{STINdistkretschmann2007continuitytheoremstinespringsdilation} implies
\begin{align}
\|\Lambda_V-\Lambda_W\|_\diamond \ \ge\ d_{\mathrm{Stin}}(V,W)^2 .
\label{eq:diamond-ge-stin}
\end{align}

\begin{proposition}[Large diamond-separated family of Kraus-rank-$k$ channels]
\label{prop:hard-family-diamond-kraus}
There exists a constant $c>0$ such that for every $\varepsilon\in(0,1]$ there exist channels
$\Lambda_1,\dots,\Lambda_M:\mathcal L(\mathbb C^{d_{\mathrm{in}}})\to \mathcal L(\mathbb C^{d_{\mathrm{out}}})$
of Kraus rank at most $k$ with $\|\Lambda_i-\Lambda_j\|_\diamond \ge \varepsilon$ for all $i\neq j,$
and
\begin{align}
M
\ \ge\
\left(\frac{c}{\sqrt{\varepsilon}}\right)^{2d_{\mathrm{in}} d_{\mathrm{out}} k-d_{\mathrm{in}}^2-k^2}.
\end{align}
\end{proposition}

\begin{proof}
Let $\eta\in(0,1]$. By Lemma~\ref{lem:gauge-quotient-cover} there exists an $\eta$-packing
$\{V_1,\dots,V_M\}\subset\mathcal V$ with respect to $d_{\mathrm{Stin}}$ such that $M
\ \ge\
\left(c_1/\eta\right)^{2d_{\mathrm{in}} d_{\mathrm{out}} k-d_{\mathrm{in}}^2-k^2},$ for a constant $c_1>0$.
Set $\Lambda_i\coloneqq \Lambda_{V_i}$. Then for $i\neq j$ we have $d_{\mathrm{Stin}}(V_i,V_j)>\eta$, and hence
\eqref{eq:diamond-ge-stin} gives
\begin{align}
\|\Lambda_i-\Lambda_j\|_\diamond
\ \ge\
d_{\mathrm{Stin}}(V_i,V_j)^2
\ >\ \eta^2.
\end{align}
Choosing $\eta\coloneqq \sqrt{\varepsilon}$ and absorbing constants
into $c$ yields the claim.
\end{proof}
In summary, the exponent $2d_{\mathrm{in}} d_{\mathrm{out}} k-d_{\mathrm{in}}^2-k^2$
has a natural geometric interpretation. It is the effective dimension of the
Stinespring-isometry family after quotienting out the environment-unitary gauge
freedom. Informally, the Stinespring-isometry manifold contributes
\(2d_{\mathrm{in}} d_{\mathrm{out}} k-d_{\mathrm{in}}^2\) real dimensions: the
term \(2d_{\mathrm{in}} d_{\mathrm{out}} k\) comes from the ambient complex
matrix space, while the \(d_{\mathrm{in}}^2\) term comes from the isometry
constraints. The quotient by environment unitaries removes a further \(k^2\)
dimensions.

This exponent is nonnegative throughout the admissible parameter regime. Indeed,
one can rewrite
\[
2 d_{\mathrm{in}} d_{\mathrm{out}} k - d_{\mathrm{in}}^2 - k^2
=
k(d_{\mathrm{in}} d_{\mathrm{out}}-k)
+
d_{\mathrm{in}}(d_{\mathrm{out}}k-d_{\mathrm{in}}).
\]
Both terms on the right-hand side are nonnegative for every admissible Kraus
rank \(k\): the first because \(k\le d_{\mathrm{in}}d_{\mathrm{out}}\), and the
second because \(k\ge k_{\min}=\lceil d_{\mathrm{in}}/d_{\mathrm{out}}\rceil\)
implies \(d_{\mathrm{out}}k\ge d_{\mathrm{in}}\)
(see Lemma~\ref{le:dimen-constraint-kraus}). It is strictly positive except in
the trivial case \(d_{\mathrm{out}}=1\) and \(k=d_{\mathrm{in}}\),
where the output space is one-dimensional and there is no nontrivial packing
problem.

\subsubsection{Converting the channel packing into a query lower bound}
\label{subsec:proof-lower-bound}
We now combine the identification upper bound from Subsection~\ref{subsec:identification} with the packing
construction from Subsection~\ref{subsec:packing} to prove the query complexity lower bound for channel tomography.

\begin{theorem}[Query lower bound for channel tomography]
\label{thm:lower-boundPROOF}
Let $\varepsilon\in(0,1)$ and let
$\Lambda:\mathcal L(\mathbb C^{d_{\mathrm{in}}})\to\mathcal L(\mathbb C^{d_{\mathrm{out}}})$
be a quantum channel of Kraus rank at most $k$.
there exists absolute constants $c_1,c_2>0$ such that any $Q$-query protocol that, with probability at least $2/3$, outputs an estimate
$\hat\Lambda$ satisfying $\|\hat\Lambda-\Lambda\|_\diamond\le \varepsilon$
(even allowing fully adaptive strategies and without imposing a definite causal order among the $Q$ uses)
must satisfy
\begin{align}
Q\ \ge\ c_1\,\frac{d_{\mathrm{in}}d_{\mathrm{out}}k}{\varepsilon^{\beta(d_{\mathrm{in}},d_{\mathrm{out}},k)}}
\ \ge\ c_2\,\frac{d_{\mathrm{in}}d_{\mathrm{out}}k}{\varepsilon^{3/8}},
\label{eq:thm-lb-eps}
\end{align}
where
\begin{align}
\beta(d_{\mathrm{in}},d_{\mathrm{out}},k)
\;\coloneqq\;
\frac{2d_{\mathrm{in}}d_{\mathrm{out}}k-d_{\mathrm{in}}^2-k^2}{2\bigl(d_{\mathrm{in}}d_{\mathrm{out}}k-1\bigr)}.
\end{align}
In particular, for any fixed constant accuracy $\varepsilon\in(0,1)$, this implies
$Q=\Omega\bigl(d_{\mathrm{in}}d_{\mathrm{out}}k\bigr)$.
\end{theorem}

\begin{proof}
Fix $\delta:=1/3$ and set $D:=k\,d_{\mathrm{in}}d_{\mathrm{out}}$.
Let $\{\Lambda_x\}_{x=1}^M$ be the family from Proposition~\ref{prop:hard-family-diamond-kraus} with separation $2\varepsilon$.
Thus $\|\Lambda_x-\Lambda_y\|_\diamond\ge 2\varepsilon$ for all $x\neq y$, and for a constant $c_1>0$,
\begin{align}
M\ \ge\ \left(\frac{c_1}{\sqrt{2\varepsilon}}\right)^{2d_{\mathrm{in}} d_{\mathrm{out}} k-d_{\mathrm{in}}^2-k^2}.
\label{eq:M-lb-explicit}
\end{align}

Assume there exists a (possibly adaptive) $Q$-query tomography procedure $\mathsf T$ such that every Kraus-rank-$k$
channel $\Lambda$ is estimated with
$\Pr[\|\hat\Lambda-\Lambda\|_\diamond\le \varepsilon]\ge 1-\delta$.
By Lemma~\ref{lem:tomo-implies-ident}, $\mathsf T$ induces a $Q$-query identification protocol for
$\{\Lambda_x\}_{x=1}^M$ with average success probability $p_{\mathrm{succ}}\ge 1-\delta$.
On the other hand, Proposition~\ref{prop:succ-krausUNPHYS} implies $p_{\mathrm{succ}} \le
\frac{1}{M}\binom{Q+D-1}{D-1}.$
Hence
\begin{align}
(1-\delta)M\ \le\ \binom{Q+D-1}{D-1}.
\label{eq:key-ineq-explicit}
\end{align}
Using $\binom{n}{m}\le (en/m)^m$ with $n=Q+D-1$ and $m=D-1$, we obtain
\begin{align}
Q\ \ge\ (D-1)\left(\frac{1}{e}\bigl((1-\delta)M\bigr)^{\frac{1}{D-1}}-1\right).
\label{eq:Q-lb-in-terms-of-M}
\end{align}
Substituting \eqref{eq:M-lb-explicit} into \eqref{eq:Q-lb-in-terms-of-M} gives
\begin{align}
Q
\ \ge\
(D-1)\left(
\frac{1}{e}(1-\delta)^{\frac{1}{D-1}}
\left(\frac{c_1}{\sqrt{2\varepsilon}}\right)^{\frac{2d_{\mathrm{in}} d_{\mathrm{out}} k-d_{\mathrm{in}}^2-k^2}{D-1}}
-1
\right).
\label{eq:Q-lb-expanded}
\end{align}

We may assume $d_{\mathrm{out}}\ge 2$, since the case $d_{\mathrm{out}}=1$ is trivial. Moreover, $d_{\mathrm{in}}\le d_{\mathrm{out}}k$ is necessary for the existence of a Stinespring isometry $\mathbb C^{d_{\mathrm{in}}}\to\mathbb C^{d_{\mathrm{out}}}\otimes\mathbb C^k$, and always $k\le d_{\mathrm{in}}d_{\mathrm{out}}$. Writing $r:=k/d_{\mathrm{in}}\in[1/d_{\mathrm{out}},\,d_{\mathrm{out}}]$, we have
$d_{\mathrm{in}}^2+k^2=d_{\mathrm{in}}k(r+1/r)\le d_{\mathrm{in}}k(d_{\mathrm{out}}+1/d_{\mathrm{out}})
=(1+1/d_{\mathrm{out}}^2)D$, and therefore
\begin{align}
2d_{\mathrm{in}} d_{\mathrm{out}} k-d_{\mathrm{in}}^2-k^2
\ \ge\
D\left(1-\frac{1}{d_{\mathrm{out}}^2}\right).
\label{eq:alpha-lb}
\end{align}
Using \eqref{eq:alpha-lb} in \eqref{eq:Q-lb-expanded}, together with $(1-\delta)^{1/(D-1)}\ge 1-\delta$ (since $D\ge 2$), yields
\begin{align}
Q
\ \ge\
(D-1)\left(
\frac{1-\delta}{e}
\left(\frac{c_1}{\sqrt{2\varepsilon}}\right)^{\left(1-\frac{1}{d_{\mathrm{out}}^2}\right)\cdot \frac{1}{1-\frac{1}{D}}}
-1
\right).
\label{eq:Q-lb-final-deltaAA}
\end{align}
Since $d_{\mathrm{out}}\ge 2$ implies $1-1/d_{\mathrm{out}}^2\ge 3/4$ and $1/(1-1/D)\ge 1$, we get the scaling
$\varepsilon^{-3/8}$ by lower bounding the exponent accordingly, hence \eqref{eq:thm-lb-eps} follows.
\end{proof}

From the proof of Theorem~\ref{thm:lower-boundPROOF} (in particular, \eqref{eq:Q-lb-in-terms-of-M}) we also obtain:

\begin{lemma}[Query lower bound from an $\varepsilon$-separated family]
\label{lem:Q-lb-M}
Let $\varepsilon\in(0,1)$ and let
$\Lambda_1,\dots,\Lambda_M:\mathcal L(\mathbb C^{d_{\mathrm{in}}})\to\mathcal L(\mathbb C^{d_{\mathrm{out}}})$
be channels of Kraus rank at most $k$ such that
$\|\Lambda_i-\Lambda_j\|_\diamond>2\varepsilon$ for all $i\neq j$.
Set $D\coloneqq k\,d_{\mathrm{in}}d_{\mathrm{out}}\ge2$.
Then any $Q$-query tomography protocol that succeeds with probability at least $2/3$
must satisfy
\begin{align}
Q\ \ge\ (D-1)\left(\frac{1}{e}\left(\frac{2M}{3}\right)^{\frac{1}{D-1}}-1\right).
\label{eq:Q-lb-M}
\end{align}
The same conclusion holds even if the protocol is allowed to be fully adaptive, purified, and without imposing a definite causal order
among the $Q$ uses.

Consequently, if $M$ (the number of $2\varepsilon$-separated channels) satisfies a bound of the form
\begin{align}
M\ \ge\ \left(\frac{c_p}{\varepsilon}\right)^{\alpha},
\end{align}
for some $c_p>0$ and $\alpha>0$, then
\begin{align}
Q\ \ge\ (D-1)\left(
\frac{2}{3e}
\left(\frac{c_p}{\varepsilon}\right)^{\frac{\alpha}{D-1}}
-1\right)
\ =\ 
\Omega\left(
D\,\varepsilon^{-\frac{\alpha}{D-1}}
\right),
\label{eq:Q-lb-from-packing}
\end{align}
where the $\Omega(\cdot)$ notation hides only absolute constants.
\end{lemma}


\noindent

\subsubsubsection{Recovering optimal lower bounds for pure-state and unitary tomography}
\label{subsubsec:pure-state-packing}
For general channels, Lemma~\ref{lem:Q-lb-M} combined with our Stinespring packing yields the optimal dimensional dependence (i.e., in $D$),
but the resulting $\varepsilon$-dependence is not tight in general.
However, as we note in the following, for two important subclasses—pure states (trace distance) and unitary channels (diamond norm)—the same lemma, combined with the
corresponding packing estimates, recovers the optimal dependence in both $d$ and $\varepsilon$~\cite{Hay98,pelecanos2025mixedstatetomographyreduces,HaahKothariODonnellTang2023}.

We begin with pure states. Since Lemma~\ref{lem:Q-lb-M} takes as input the size of an $\varepsilon$-separated family,
we record a standard covering/packing estimate for pure states in trace distance. The exponent $2d-2$ has the expected
parameter-count interpretation: a unit vector in $\mathbb C^d$ has $2d-1$ real degrees of freedom, and quotienting out
the global phase leaves $2d-2$ degrees of freedom for the projector $|\psi\rangle\langle\psi|$.

\begin{lemma}[Pure-state packing in trace distance]
\label{lem:pure-state-packing-trace}
Let
\begin{align}
\mathcal P_d
\;\coloneqq\;
\bigl\{|\psi\rangle\langle\psi|:\ |\psi\rangle\in\mathbb C^d,\ \langle\psi|\psi\rangle=1\bigr\}
\end{align}
be the set of pure states on $\mathbb C^d$, equipped with
$d_{\mathrm{tr}}(\rho,\sigma)\coloneqq \tfrac12\|\rho-\sigma\|_1$.
Then there exist constants $c,C>0$ such that for all $\eta\in(0,1]$,
\begin{align}
\left(\frac{c}{\eta}\right)^{2d-2}
\ \le\
N\bigl(\mathcal P_d,d_{\mathrm{tr}},\eta\bigr)
\ \le\
\left(\frac{C}{\eta}\right)^{2d-2}.
\label{eq:pure-state-pack-tr}
\end{align}
By \eqref{eq:pack-cover}, the corresponding packing numbers satisfy the same bounds up to rescaling $\eta$.
\end{lemma}

\begin{proof}[Proof sketch]
Since pure states are rank-one projectors, it suffices to control covering numbers of the so-called complex Grassmannian
$\mathrm{Gr}_{\mathbb C}(1,d)=U(d)/(U(1)\times U(d-1))$, which can be viewed as taking the first column of a unitary matrix and
modding out the global phase; see \cite{AubrunSzarek2017AliceBob}. The bound \eqref{eq:pure-state-pack-tr} follows from the
general covering estimate \cite[Eq.~(5.20)]{AubrunSzarek2017AliceBob} together with the parameters for
$\mathrm{Gr}_{\mathbb C}(1,d)$ listed in \cite[Table~5.1]{AubrunSzarek2017AliceBob} (in particular $\dim_{\mathbb R}\mathrm{Gr}_{\mathbb C}(1,d)=2d-2$).
The bounds in \cite[Table~5.1]{AubrunSzarek2017AliceBob} are stated for the so-called quotient metric induced by the standard
(called there ``extrinsic'') Schatten-$p$ metric; by \cite[Sec.~5.1.5]{AubrunSzarek2017AliceBob} these quotient metrics are
equivalent up to absolute constants to the corresponding Schatten-$p$ distance, so the covering
bounds transfer after a constant rescaling of~$\eta$. Finally, since $\rho-\sigma$ has rank at most $2$ for
$\rho,\sigma\in\mathcal P_d$, all Schatten norms of $\rho-\sigma$ are equivalent up to absolute constants.
\end{proof}

\begin{corollary}[Pure-state tomography lower bound]
\label{cor:pure-state-lb-d-eps2}
Let $d\ge 2$ and $\varepsilon\in(0,1)$.
Any protocol that, given $Q$ copies of an unknown pure state $\rho\in\mathcal P_d$, outputs $\hat\rho$ with
$d_{\mathrm{tr}}(\hat\rho,\rho)\le \varepsilon$ with probability at least $2/3$ must use
\begin{align}
Q\ =\ \Omega\left(\frac{d}{\varepsilon^{2}}\right).
\end{align}
This matches the upper bound $Q=\Theta(d/\varepsilon^{2})$~\cite{Hay98,pelecanos2025mixedstatetomographyreduces}.
\end{corollary}

\begin{proof}
View the unknown state $\rho$ as a state-preparation channel
$\Lambda_\rho:\mathcal L(\mathbb C)\to\mathcal L(\mathbb C^d)$ defined by $\Lambda_\rho(X)=\Tr(X)\rho$.
Then one query returns one copy of~$\rho$, and for any $\rho,\sigma\in\mathcal P_d$,
$\|\Lambda_\rho-\Lambda_\sigma\|_\diamond=\|\rho-\sigma\|_1=2\,d_{\mathrm{tr}}(\rho,\sigma)$.
By Lemma~\ref{lem:pure-state-packing-trace}, there exists an $\varepsilon$-separated family of pure states
$\{\rho_i\}_{i=1}^M\subset\mathcal P_d$ in $d_{\mathrm{tr}}$ with $M\ge (c_p/\varepsilon)^{2d-2}$ for a constant $c_p>0$,
so the corresponding preparation channels satisfy $\|\Lambda_{\rho_i}-\Lambda_{\rho_j}\|_\diamond>2\varepsilon$.
Applying Lemma~\ref{lem:Q-lb-M} with $d_{\mathrm{in}}=1$, $d_{\mathrm{out}}=d$, $k=1$ (so $D=d$) yields
\begin{align}
Q
\ \ge\
(d-1)\left(\frac{1}{e}\left(\frac{2M}{3}\right)^{\frac{1}{d-1}}-1\right)\ \ge\
(d-1)\left(\frac{2}{3e}\left(\frac{c_p}{\varepsilon}\right)^2-1\right),
\end{align}
which implies $Q=\Omega(d/\varepsilon^2)$.
\end{proof}

We next record the analogous ingredient for \emph{unitary} channels, namely a covering/packing bound in diamond norm.

\begin{lemma}[Covering/packing of unitary channels in diamond norm]
\label{lem:unitary-channel-cover}
Let
\begin{align}
\mathsf{U}_d
\;\coloneqq\;
\bigl\{\mathcal U_U:\mathcal L(\mathbb C^d)\to\mathcal L(\mathbb C^d),\ \mathcal U_U(\rho)=U\rho U^\dagger,\ U\in U(d)\bigr\}
\end{align}
equipped with $d_\diamond(\mathcal U,\mathcal V)\coloneqq\|\mathcal U-\mathcal V\|_\diamond$.
Then there exist constants $c,C>0$ such that for all $d\ge 2$ and $\varepsilon\in(0,1]$,
\begin{align}
\left(\frac{c}{\varepsilon}\right)^{d^2-1}
\ \le\
N\bigl(\mathsf{U}_d,d_\diamond,\varepsilon\bigr)
\ \le\
\left(\frac{C}{\varepsilon}\right)^{d^2-1}.
\label{eq:unitary-channel-cover}
\end{align}
By \eqref{eq:pack-cover}, the packing numbers satisfy the same bounds up to rescaling $\varepsilon$.
\end{lemma}
\begin{proof}[Proof sketch]
By \cite[Prop.~1.6]{HaahKothariODonnellTang2023}, for $U,V\in U(d)$ the diamond distance
$\|\mathcal U_U-\mathcal U_V\|_\diamond$ is equivalent up to constants to the phase-optimized operator norm
$d_{\mathrm{ph}}(U,V)\coloneqq \inf_{\theta\in\mathbb R}\|U-e^{i\theta}V\|_\infty$.
Since $\mathcal U_U=\mathcal U_{e^{i\theta}U}$, this identifies $(\mathsf{U}_d,d_\diamond)$ (up to constants) with the quotient
space $U(d)/(U(1)I_d)$ equipped with the so-called quotient metric induced by $\|\cdot\|_\infty$, which coincides precisely with $d_{\mathrm{ph}}$.
Applying the covering-number bounds for such quotients \cite[Thm.~11]{NETHOMszarek1997metricentropyhomogeneousspaces}
yields exponent $\dim_{\mathbb R}(U(d)/(U(1)I_d))=d^2-1$, which gives \eqref{eq:unitary-channel-cover}.
\end{proof}
\begin{corollary}[Unitary-channel tomography lower bound]
\label{cor:unitary-tomo-lb}
Let $d\ge 2$ and $\varepsilon\in(0,1)$.
Any protocol that (even allowing fully adaptive strategies and dropping any definite causal-order constraint), given $Q$ uses of an unknown unitary channel
$\mathcal U_U(\rho)=U\rho U^\dagger$ with $U\in U(d)$, outputs an estimate
$\widehat{\mathcal U}$ satisfying $\|\widehat{\mathcal U}-\mathcal U_U\|_\diamond\le \varepsilon$
with probability at least $2/3$ must use
\begin{align}
Q\ =\ \Omega\left(\frac{d^2}{\varepsilon}\right),
\end{align}
where the $\Omega(\cdot)$ notation hides only constants.
This matches the upper bound $Q=\Theta(d^2/\varepsilon)$ of \cite{HaahKothariODonnellTang2023}.
\end{corollary}

\begin{proof}
By Lemma~\ref{lem:unitary-channel-cover} and \eqref{eq:pack-cover}, there exists a $2\varepsilon$-separated family of unitary
channels $\{\mathcal U_i\}_{i=1}^M\subset\mathsf U_d$ in diamond norm with
$M\ge (c_p/(2\varepsilon))^{d^2-1}$ for a constant $c_p>0$.
Applying Lemma~\ref{lem:Q-lb-M} with $d_{\mathrm{in}}=d_{\mathrm{out}}=d$, $k=1$ (so $D=d^2$) yields
\begin{align}
Q \ge\
(d^2-1)\left(\frac{1}{e}\left(\frac{2M}{3}\right)^{\frac{1}{d^2-1}}-1\right) \ge\
(d^2-1)\left(\frac{2}{3e}\left(\frac{c_p}{2\varepsilon}\right)-1\right),
\end{align}
which implies $Q=\Omega(d^2/\varepsilon)$.
\end{proof}

\subsubsubsection{Lower bound for measurement tomography}
\label{subsubsec:meas-lb}

We now prove that the upper bound in Corollary~\ref{co:binary-povm-learning}
for learning binary POVMs in operator norm is optimal in its dimension
dependence at constant accuracy: any worst-case protocol must use
\(N=\Omega(d^2)\) queries. In the next section, we complement this with a
separate accuracy- and confidence-dependent lower bound,
\(N=\Omega(d\varepsilon^{-2}\log(d/\delta))\), obtained from a classical
coin-bias estimation reduction.
We use the following standard covering/packing estimate.

\begin{lemma}[Covering/packing of rank-$r$ complex projectors (operator norm)]
\label{le:proj-covering}
Let $d\ge 2$ and $r\le d/2$. Define
\begin{align}
\mathsf{Proj}_r
\;\coloneqq\;
\bigl\{\Pi\in\mathcal L(\mathbb C^d):\ \Pi^2=\Pi=\Pi^\dagger,\ \Tr(\Pi)=r\bigr\},
\qquad
d_\infty(\Pi,\Pi')\coloneqq\|\Pi-\Pi'\|_\infty .
\end{align}
Then there exist constants $c,C>0$ such that for all $\eta\in(0,1)$,
\begin{align}
\Bigl(\frac{c}{\eta}\Bigr)^{2r(d-r)}
\ \le\
N\bigl(\mathsf{Proj}_r,d_\infty,\eta\bigr)
\ \le\
\Bigl(\frac{C}{\eta}\Bigr)^{2r(d-r)}.
\end{align}
The same bounds hold for the packing number (up to changing $c,C$).
\end{lemma}

\begin{proof}[Proof sketch]
A rank-$r$ projector is uniquely determined by its image, so $\mathsf{Proj}_r$ identifies with the so-called complex Grassmannian $\mathrm{Gr}(r,\mathbb C^d)
=
U(d)/(U(r)\times U(d-r)).$
(Intuitively: an $r$-dimensional subspace can be specified by choosing an orthonormal basis of $r$ vectors, i.e., by choosing the first $r$
columns of a unitary matrix. This identifies with the Stiefel manifold $U(d)/U(d-r)$.
Changing the orthonormal basis within the same subspace corresponds to right-multiplication by $U(r)$, hence such space is the
quotient $(U(d)/U(d-r))/U(r)=U(d)/(U(r)\times U(d-r))$.)
The covering bounds \cite[Eq.~(5.20)]{AubrunSzarek2017AliceBob} together with the parameters in \cite[Table~5.1]{AubrunSzarek2017AliceBob}
give exponent $2r(d-r)$ for the so-called quotient metric induced by the standard (called there ``extrinsic'') Schatten-$p$ metric.
By \cite[Sec.~5.1.5]{AubrunSzarek2017AliceBob}, this quotient metric is equivalent up to constants to the corresponding
Schatten-$p$ distance between projectors; taking $p=\infty$ yields the stated bounds for $d_\infty$.
\end{proof}

\begin{proposition}[Lower bound for binary POVM learning from projector packing]
\label{pr:binary-povm-lower-bound}
There exist constants $c>0$ and $\varepsilon_0\in(0,1/4)$ such that the following holds.
Fix $d\ge 2$ and $\varepsilon\in(0,\varepsilon_0]$. Let $\{E,\mathbb I-E\}$ be an unknown binary POVM on $\mathbb C^d$ and let
\begin{align}
\Lambda_E(\rho)
\;\coloneqq\;
\tr(E\rho)\,\ketbra{0}{0}
+
\tr\bigl((\mathbb I-E)\rho\bigr)\,\ketbra{1}{1}
\end{align}
be its measurement channel. Any (possibly adaptive) procedure that, given $Q$ uses of $\Lambda_E$, outputs $\widehat E$ such that
$\|\widehat E-E\|_\infty\le \varepsilon$ with probability at least $2/3$ must use
\begin{align}
Q\ \ge\ c\,d^2\,\varepsilon^{-\beta_d},
\qquad
\beta_d\ \coloneqq\ \frac{d^2-1}{4d^2-2}.
\end{align}
In particular, for constant accuracy $\varepsilon=\Theta(1)$ one has $Q=\Omega(d^2)$, which matches the $O(d^2)$ upper bound in Corollary~\ref{co:binary-povm-learning}.
\end{proposition}

\begin{proof}
It suffices to prove the claim for a subclass of binary POVMs. We therefore restrict to the case $E=\Pi$ where $\Pi$ ranges over projectors of
a rank $r$ that maximizes $r(d-r)$; equivalently, $r(d-r)=d^2/4$ when $d$ is even and $r(d-r)=(d^2-1)/4$ when $d$ is odd, so in
all such cases
\begin{align}
2r(d-r)\ \ge\ \frac{d^2-1}{2}.
\label{eq:best-r-bound}
\end{align}

Let $\mathsf{Proj}_r$ be the set of rank-$r$ projectors. By Lemma~\ref{le:proj-covering} and \eqref{eq:pack-cover}, there exists a
$(2\varepsilon)$-separated family $\{\Pi_i\}_{i=1}^M\subset\mathsf{Proj}_r$ in $\|\cdot\|_\infty$ with
$M\ge (c_p/\varepsilon)^{2r(d-r)}$ for a constant $c_p>0$.
By Lemma~\ref{le:binary-povm-diamond}, the corresponding channels satisfy
$\|\Lambda_{\Pi_i}-\Lambda_{\Pi_j}\|_\diamond=2\|\Pi_i-\Pi_j\|_\infty>4\varepsilon$ for $i\neq j$.
Moreover, $\Lambda_{\Pi_i}$ has minimal Kraus rank $d$.
Indeed its Choi operator is block-diagonal,
\begin{align}
J(\Lambda_{\Pi_i})
=
\Pi_i^{\mathsf T}\otimes \ketbra{0}{0}
+
(\mathbb I-\Pi_i)^{\mathsf T}\otimes \ketbra{1}{1},
\end{align}
so $\rank J(\Lambda_{\Pi_i})=\rank(\Pi_i)+\rank(\mathbb I-\Pi_i)=r+(d-r)=d$,
and the minimal Kraus rank equals $\rank J(\Lambda_{\Pi_i})$.
Thus in Lemma~\ref{lem:Q-lb-M} we may take $d_{\mathrm{in}}=d$, $d_{\mathrm{out}}=2$, and $k=d$, hence $D=2d^2$.
 Thus in Lemma~\ref{lem:Q-lb-M} we have $d_{\mathrm{in}}=d$, $d_{\mathrm{out}}=2$, and $D=k\,d_{\mathrm{in}}d_{\mathrm{out}}=2d^2.$
Applying Lemma~\ref{lem:Q-lb-M} with accuracy parameter $2\varepsilon$ gives $Q\ \ge\ \Omega\left(d^2\,\varepsilon^{-\frac{2r(d-r)}{2d^2-1}}\right).$
Using \eqref{eq:best-r-bound} yields $\frac{2r(d-r)}{2d^2-1}\ge \frac{d^2-1}{4d^2-2}=\beta_d$, and therefore $Q\ \ge\ \Omega\left(d^2\,\varepsilon^{-\beta_d}\right).$
\end{proof}

\subsection{A classical-coins reduction for the accuracy lower bound 
$\Omega(1/\varepsilon^2)$}\label{subsec:delta-lb-coins}

Throughout this subsection we assume \(d_{\mathrm{out}}\ge2\), and we fix two
orthogonal output states \(|0\rangle,|1\rangle\in\mathbb C^{d_{\mathrm{out}}}\).
For \(d_{\mathrm{out}}>2\), the states \(\rho_p\) below are understood as
supported on their span.

Here we show a lower bound for learning quantum channels in diamond distance whenever $k\ge 2d_{\mathrm{in}}$, and in the next subsection we generalize to any $k$. Precisely, here
we prove that
\begin{align}
  N \;=\; \Omega\!\left(\frac{d_{\mathrm{in}}}{\varepsilon^{2}}
  \log\frac{d_{\mathrm{in}}}{\delta}\right)
  \label{eq:lb-delta-statement}
\end{align}
queries are necessary (in the worst case) to learn a quantum channel to diamond-distance accuracy
$\varepsilon$ with failure probability at most $\delta$.

Combined with the lower bound from the previous section (Theorem~\ref{thm:lower-boundPROOF}), this yields the
following failure-probability dependent statement in the regime $k\ge 2d_{\mathrm{in}}$: any learning procedure that succeeds with probability at
least $1-\delta$ must use
\begin{align}
  N \;=\; \Omega\!\left(
    \frac{d_{\mathrm{in}} d_{\mathrm{out}} k}{\varepsilon^{\beta}}
    \;+\;
    \frac{d_{\mathrm{in}}}{\varepsilon^{2}}\log\!\frac{d_{\mathrm{in}}}{\delta}
  \right),
  \label{eq:combined-delta-lb}
\end{align}
where $\beta$ is the exponent appearing in Theorem~\ref{thm:lower-boundPROOF}. This matches the $\delta$-dependence of our upper bound in the corresponding regime (see
Theorem~\ref{th:th1Mainapp}), which states that%
\footnote{In Theorem~\ref{th:th1Mainapp} the confidence term is written as
$O\bigl(\frac{d_{\mathrm{in}}}{\varepsilon^{2}}\log(1/\delta)\bigr)$. Moreover,
\begin{align}
  \frac{d_{\mathrm{in}}}{\varepsilon^{2}}\log\!\frac{1}{\delta}
  \;=\;
  \frac{d_{\mathrm{in}}}{\varepsilon^{2}}\log\!\frac{d_{\mathrm{in}}}{\delta}
  \;-\;
  \frac{d_{\mathrm{in}}}{\varepsilon^{2}}\log d_{\mathrm{in}},
\end{align}
so the difference between the two logarithmic forms is the $\delta$-independent term
$\frac{d_{\mathrm{in}}}{\varepsilon^{2}}\log d_{\mathrm{in}}$, which can be absorbed into
$O\bigl(\frac{d_{\mathrm{in}} d_{\mathrm{out}} k}{\varepsilon^{2}}\bigr)$.}
\begin{align}
  N \;=\; O\left(
    \frac{d_{\mathrm{in}} d_{\mathrm{out}} k}{\varepsilon^{2}}
    \;+\;
    \frac{d_{\mathrm{in}}}{\varepsilon^{2}}\log\!\frac{d_{\mathrm{in}}}{\delta}
  \right).
\end{align}

To establish \eqref{eq:lb-delta-statement}, we restrict attention to a fully classical subfamily of channels.
Fix an orthonormal basis $\{|j\rangle\}_{j=1}^{d_{\mathrm{in}}}$ of the input space and consider the
$\,d_{\mathrm{in}}$-parameter family\ of CPTP maps
\begin{align}
  \Lambda_{\mathbf p}(\rho)
  \coloneqq
  \sum_{j=1}^{d_{\mathrm{in}}} \langle j|\rho|j\rangle\,\rho_{p_j},
  \label{eq:def-Lambda-p-final2}
\end{align}
where $\mathbf p=(p_1,\dots,p_{d_{\mathrm{in}}})\in (0,1)^{d_{\mathrm{in}}}$ and $\rho_{p}$ is the diagonal qubit
state encoding a biased coin,
\begin{align}
  \rho_p \coloneqq (1-p)\,|0\rangle\!\langle 0| + p\,|1\rangle\!\langle 1|.
  \label{eq:def-rho-p-final2}
\end{align}
Moreover, $\Lambda_{\mathbf p}$ has Kraus (Choi) rank $2d_{\mathrm{in}}$ because its Choi operator
$\sum_{j=1}^{d_{\mathrm{in}}}|j\rangle\!\langle j|\otimes \rho_{p_j}$ is block-diagonal and each block
$\rho_{p_j}$ has rank $2$ for $p_j\in(0,1)$.
Thus this family is contained in the class of Kraus-rank-\(\le k\) channels
whenever \(k\ge2d_{\mathrm{in}}\).

The following identity makes explicit that, within the multi-coin family,
diamond distance is exactly \(\ell_\infty\) distance between the corresponding
coin biases, with our convention
\(d_\diamond(\Lambda,\Gamma)=\frac12\|\Lambda-\Gamma\|_\diamond\).
\begin{lemma}[Diamond norm for the multi-coin channel family]
\label{lem:diamond-multicoin-final2}
For any $\mathbf p,\mathbf q\in[0,1]^{d_{\mathrm{in}}}$,
\begin{align}
  \|\Lambda_{\mathbf p}-\Lambda_{\mathbf q}\|_\diamond
  =
  2\max_{j\in[d_{\mathrm{in}}]} |p_j-q_j|.
  \label{eq:diamond-multicoin-final2}
\end{align}
\end{lemma}
\begin{proof}
Let $\Delta:=\Lambda_{\mathbf p}-\Lambda_{\mathbf q}$ and let $R$ be an arbitrary reference system.
For any state $\rho_{AR}$ on $A\otimes R$ with $A\simeq\mathbb C^{d_{\mathrm{in}}}$, set
$\rho_R^{(j)}\coloneqq \langle j|\rho_{AR}|j\rangle\succeq 0$, so that $\sum_j \mathrm{tr}(\rho_R^{(j)})=1$.
Using \eqref{eq:def-Lambda-p-final2}, only the diagonal blocks contribute and
\begin{align}
  (\Delta\otimes \mathrm{id}_R)(\rho_{AR})
  \;=\;
  \sum_{j=1}^{d_{\mathrm{in}}}\bigl(\rho_{p_j}-\rho_{q_j}\bigr)\otimes \rho_R^{(j)}.
\end{align}
By the triangle inequality and $\|X\otimes Y\|_1=\|X\|_1\|Y\|_1$ for $Y\succeq 0$,
\begin{align}
  \|(\Delta\otimes \mathrm{id}_R)(\rho_{AR})\|_1
  &\le
  \sum_{j=1}^{d_{\mathrm{in}}}\|\rho_{p_j}-\rho_{q_j}\|_1\,\|\rho_R^{(j)}\|_1
  \;=\;
  \sum_{j=1}^{d_{\mathrm{in}}} 2|p_j-q_j|\,\mathrm{tr}(\rho_R^{(j)}) \nonumber\\
  &\le
  2\max_j |p_j-q_j|.
\end{align}
Taking the supremum over $R$ and $\rho_{AR}$ yields the upper bound.
For the matching lower bound, choose the product input
$|j^\star\rangle\!\langle j^\star|\otimes \rho_R$ with
$j^\star\in\arg\max_j|p_j-q_j|$, which achieves value $2|p_{j^\star}-q_{j^\star}|$.
\end{proof}

\begin{lemma}[Classicalisation of $N$-query strategies]
\label{lem:coin-classicalization}
Fix $\mathbf p$. Any $N$-use quantum strategy interacting with $\Lambda_{\mathbf p}$ and producing a classical
output can be simulated by a classical adaptive procedure that, at each round $t$, selects an index
$J_t\in[d_{\mathrm{in}}]$ and observes a bit
\begin{align}
  X_t \mid (J_t=j)\sim \mathrm{Bernoulli}(p_j),
\end{align}
such that the induced distribution of the final output is identical under the quantum and classical
procedures for every fixed~$\mathbf p$.
\end{lemma}

\begin{proof}
Let $\Delta_A$ be the complete dephasing map in the $\{|j\rangle\}$ basis, $ \Delta_A(X)\coloneqq \sum_{j=1}^{d_{\mathrm{in}}}|j\rangle\!\langle j|X|j\rangle\!\langle j|.$
Since $\Lambda_{\mathbf p}$ depends only on the diagonal of its input in this basis, we have $\Lambda_{\mathbf p}=\Lambda_{\mathbf p}\circ \Delta_A.$
Therefore, inserting $\Delta_A$ on the channel input immediately before each use leaves the overall strategy
unchanged. In particular, right before a call, the joint state on the channel input register $A$ and the
strategy's auxiliary memory register $R$ may be assumed block-diagonal on $A$, i.e.
\begin{align}
  \rho_{AR}=\sum_{j=1}^{d_{\mathrm{in}}}|j\rangle\!\langle j|\otimes \sigma_R^{(j)},
  \qquad
  \sigma_R^{(j)}\ge 0,
  \qquad
  \sum_{j=1}^{d_{\mathrm{in}}}\mathrm{tr}(\sigma_R^{(j)})=1.
\end{align}
Applying $\Lambda_{\mathbf p}$ gives
\begin{align}
  (\Lambda_{\mathbf p}\otimes \mathrm{id}_R)(\rho_{AR})
  =
  \sum_{j=1}^{d_{\mathrm{in}}}\rho_{p_j}\otimes \sigma_R^{(j)}.
\end{align}

Each $\rho_{p_j}$ is diagonal in the basis $\{|0\rangle,|1\rangle\}$. Hence we may, without changing the
resulting outcome distribution, insert a measurement of the channel output in this basis immediately after
each use. This produces a classical bit $X\in\{0,1\}$ with $\Pr[X=1\mid J=j]=p_j$, where the label $J$ is
distributed according to $\Pr[J=j]=\mathrm{tr}(\sigma_R^{(j)})$.

Doing this at every channel use shows that the entire interaction can be viewed as producing a classical
transcript $(J_1,X_1,\dots,J_N,X_N)$ generated by an adaptive sampling procedure, where at round $t$ the
index $J_t$ is chosen (possibly at random) as a function of the previously observed pairs
$(J_1,X_1),\dots,(J_{t-1},X_{t-1})$, and then $X_t\sim\mathrm{Bernoulli}(p_{J_t})$.
Finally, any final measurement of the quantum strategy is a fixed processing of the
information contained in this transcript, and can therefore be absorbed into a classical (possibly randomised)
map applied to $(J_1,X_1,\dots,J_N,X_N)$.

This gives an adaptive classical coin-sampling
procedure with the same final output distribution as the original quantum
strategy.
\end{proof}

We now explain how any successful channel estimator yields an estimator of the
coin biases. Given an arbitrary output channel \(\widehat\Lambda\), define
\begin{align}
  \widehat p_j
  \coloneqq
  \langle 1|
  \widehat\Lambda(|j\rangle\!\langle j|)
  |1\rangle .
\end{align}
If the learner satisfies $\frac12\|\widehat\Lambda-\Lambda_{\mathbf p}\|_\diamond\le\varepsilon,$
then, for every \(j\), contractivity of trace distance under the two-outcome
measurement \(\{|1\rangle\!\langle1|,I-|1\rangle\!\langle1|\}\) gives
\begin{align}
  |\widehat p_j-p_j|
  &\le
  \frac12
  \left\|
    \widehat\Lambda(|j\rangle\!\langle j|)
    -
    \Lambda_{\mathbf p}(|j\rangle\!\langle j|)
  \right\|_1 \le
  \frac12\|\widehat\Lambda-\Lambda_{\mathbf p}\|_\diamond
  \le
  \varepsilon .
\end{align}
Hence channel tomography in diamond distance implies estimating
\(\mathbf p\) in \(\ell_\infty\) error at most \(\varepsilon\).

By Lemma~\ref{lem:coin-classicalization}, any $N$-query quantum learning algorithm for $\Lambda_{\mathbf p}$
can be simulated by a classical adaptive algorithm that samples the $d_{\mathrm{in}}$ coins $N$ times.

A standard lower bound~\cite{Tsybakov2008IntroductionTN} for adaptive
estimation of all \(d_{\mathrm{in}}\) Bernoulli biases to \(\ell_\infty\)
accuracy \(\varepsilon\) with failure probability \(\delta\) yields that any
estimator must use
\begin{align}
  N
  =
  \Omega\!\left(
    \frac{d_{\mathrm{in}}}{\varepsilon^{2}}
    \log\frac{d_{\mathrm{in}}}{\delta}
  \right).
  \label{eq:classical-coin-lb-import}
\end{align}

The same construction immediately gives also a lower bound for
operator-norm tomography of binary POVMs.

\begin{corollary}[Lower bound for binary POVM tomography from classical coins]
\label{cor:binary-povm-delta-lb}
There exists a constant $\varepsilon_0>0$ such that the following holds.
Let $d\ge1$, $0<\varepsilon\le\varepsilon_0$, and $0<\delta<1/4$. Any $N$-query protocol
that learns every binary POVM $\{I-E,E\}$ on $\mathbb C^d$ to operator-norm
accuracy $\varepsilon$, with success probability at least $1-\delta$, must satisfy
\[
  N
  =
  \Omega\!\left(\frac{d}{\varepsilon^2}\log\frac{d}{\delta}\right).
\]
\end{corollary}

\begin{proof}
It suffices to restrict to the diagonal subfamily
\begin{align}
  E_{\mathbf p}
  \coloneqq
  \sum_{j=1}^d p_j |j\rangle\!\langle j|,
  \qquad
  \mathbf p\in(0,1)^d .
\end{align}
The measurement channel induced by $\{I-E_{\mathbf p},E_{\mathbf p}\}$ is
\begin{align}
  \rho
  \mapsto
  \tr\bigl((I-E_{\mathbf p})\rho\bigr)|0\rangle\!\langle0|
  +
  \tr\bigl(E_{\mathbf p}\rho\bigr)|1\rangle\!\langle1|,
\end{align}
which is exactly the multi-coin family $\Lambda_{\mathbf p}$ in
\eqref{eq:def-Lambda-p-final2}, with $d_{\mathrm{in}}=d$. Hence
Lemma~\ref{lem:coin-classicalization} and the classical lower bound
\eqref{eq:classical-coin-lb-import} apply. Finally, if a POVM learner outputs
$\widehat E$ with $\|\widehat E-E_{\mathbf p}\|_\infty\le\varepsilon$, then
$\widehat p_j\coloneqq \langle j|\widehat E|j\rangle$ satisfies, for every $j$,
\begin{align}
  |\widehat p_j-p_j|
  =
  \bigl|
    \langle j|(\widehat E-E_{\mathbf p})|j\rangle
  \bigr|
  \le
  \|\widehat E-E_{\mathbf p}\|_\infty
  \le
  \varepsilon .
\end{align}
Thus POVM tomography on this subfamily implies
$\ell_\infty$ estimation of $d$ Bernoulli biases to accuracy $\varepsilon$, which
requires
$N
  =
  \Omega\!\left(
    \frac{d}{\varepsilon^2}\log\frac{d}{\delta}
  \right).$
\end{proof}

\subsubsection{Embedding independent coins under a Kraus-rank constraint}
\label{subsubsec:relax-krausrank-delta}

The family \eqref{eq:def-Lambda-p-final2} encodes \(d_{\mathrm{in}}\)
independent coin biases and has Kraus rank \(2d_{\mathrm{in}}\). We now obtain
an analogous lower bound under a general Kraus-rank budget \(k\) by encoding
only as many independent coins as fit within that budget. Throughout this
subsection we assume \(d_{\mathrm{out}}\ge2\), and we fix two orthogonal output
states \(|0\rangle,|1\rangle\in\mathbb C^{d_{\mathrm{out}}}\). For
\(d_{\mathrm{out}}>2\), the states \(\rho_p\) below are understood as supported
on the span of \(|0\rangle,|1\rangle\).

Assume
\(k\ge \lceil d_{\mathrm{in}}/d_{\mathrm{out}}\rceil\), since otherwise the
class of Kraus-rank-\(\le k\) channels is empty. Define
\begin{align}
  m_{\mathrm{coin}}
  \coloneqq
  \max\left\{
    r\in\{0,\ldots,d_{\mathrm{in}}\}:
    2r+
    \left\lceil\frac{d_{\mathrm{in}}-r}{d_{\mathrm{out}}}\right\rceil
    \le k
  \right\},
  \qquad
  s\coloneqq d_{\mathrm{in}}-m_{\mathrm{coin}},
  \qquad
  \ell\coloneqq \left\lceil\frac{s}{d_{\mathrm{out}}}\right\rceil .
  \label{eq:def-m-s-ell-relax}
\end{align}
Here \(m_{\mathrm{coin}}\) is the largest number of two-outcome coins that can
be embedded: \(2r\) Kraus operators are used for the \(r\) coin coordinates,
while \(\lceil(d_{\mathrm{in}}-r)/d_{\mathrm{out}}\rceil\) Kraus operators
suffice to pad the remaining input directions. In particular,
\(m_{\mathrm{coin}}=d_{\mathrm{in}}\) whenever \(k\ge2d_{\mathrm{in}}\). If
\(s=0\), the padding part below is understood to be empty.

Fix the input basis \(\{|j\rangle\}_{j=1}^{d_{\mathrm{in}}}\) used above, and
an orthonormal basis \(\{|i\rangle\}_{i=0}^{d_{\mathrm{out}}-1}\) of the output
space, with \(|0\rangle,|1\rangle\) hosting the coin states. For each block
index \(b\in\{0,1,\ldots,\ell-1\}\), define the partial isometry
\begin{align}
  \Pi_b
  \coloneqq
  \sum_{i=1}^{r_b} |i-1\rangle\!\langle bd_{\mathrm{out}}+i|,
  \qquad
  r_b \coloneqq \min\{d_{\mathrm{out}},\,s-bd_{\mathrm{out}}\}.
  \label{eq:def-Pi-b-relax}
\end{align}
These maps pad the first \(s\) input basis states, and satisfy
\begin{align}
  \sum_{b=0}^{\ell-1}\Pi_b^\dagger \Pi_b
  =
  \sum_{j=1}^{s}|j\rangle\!\langle j|
  \eqqcolon P_{\mathrm{pad}} .
  \label{eq:Pi-completeness-relax}
\end{align}

For
\(\mathbf p=(p_1,\ldots,p_{m_{\mathrm{coin}}})\in(0,1)^{m_{\mathrm{coin}}}\),
define the coin component acting on the last \(m_{\mathrm{coin}}\) input basis
states by
\begin{align}
  \Lambda_{\mathbf p}^{(m_{\mathrm{coin}})}(\rho)
  \coloneqq
  \sum_{t=1}^{m_{\mathrm{coin}}}
  \langle s+t|\rho|s+t\rangle\,\rho_{p_t},
  \qquad
  \rho_p \coloneqq
  (1-p)|0\rangle\!\langle0|+p|1\rangle\!\langle1|.
  \label{eq:def-Lambda-m-relax}
\end{align}
We then define the padded channel family
\begin{align}
  \widetilde{\Lambda}_{\mathbf p}(\rho)
  \coloneqq
  \sum_{b=0}^{\ell-1}\Pi_b\rho\Pi_b^\dagger
  +
  \Lambda_{\mathbf p}^{(m_{\mathrm{coin}})}(\rho).
  \label{eq:def-tildeLambda-relax}
\end{align}

\begin{lemma}[Validity and Kraus rank of the padded family]
\label{lem:padded-family-cptp-rank}
For every \(\mathbf p\in(0,1)^{m_{\mathrm{coin}}}\), the map
\(\widetilde{\Lambda}_{\mathbf p}\) in \eqref{eq:def-tildeLambda-relax} is
CPTP and satisfies
\begin{align}
  \mathrm{rank}_{\mathrm{Kraus}}(\widetilde{\Lambda}_{\mathbf p})
  \le
  \ell+2m_{\mathrm{coin}}
  \le
  k.
\end{align}
\end{lemma}

\begin{proof}
The padding map
\(\rho\mapsto\sum_{b=0}^{\ell-1}\Pi_b\rho\Pi_b^\dagger\) is completely
positive and has Kraus rank at most \(\ell\). Moreover,
\(\Lambda_{\mathbf p}^{(m_{\mathrm{coin}})}\) is completely positive and admits
the Kraus operators
\begin{align}
  K_{t,0}\coloneqq \sqrt{1-p_t}\,|0\rangle\!\langle s+t|,
  \qquad
  K_{t,1}\coloneqq \sqrt{p_t}\,|1\rangle\!\langle s+t|,
  \qquad
  t\in[m_{\mathrm{coin}}].
\end{align}
Thus its Kraus rank is at most \(2m_{\mathrm{coin}}\), and
\begin{align}
  \sum_{t=1}^{m_{\mathrm{coin}}}\sum_{a\in\{0,1\}}
  K_{t,a}^\dagger K_{t,a}
  =
  \sum_{t=1}^{m_{\mathrm{coin}}}|s+t\rangle\!\langle s+t|
  \eqqcolon P_{\mathrm{coin}} .
\end{align}
Combining this with \eqref{eq:Pi-completeness-relax} gives
\(P_{\mathrm{pad}}+P_{\mathrm{coin}}=\mathbb I\). Hence the Kraus set
\(\{\Pi_b\}_b\cup\{K_{t,a}\}_{t,a}\) defines a trace-preserving completely
positive map, so \(\widetilde{\Lambda}_{\mathbf p}\) is CPTP. The Kraus-rank
bound \(\le \ell+2m_{\mathrm{coin}}\) follows from this representation, and
\(\ell+2m_{\mathrm{coin}}\le k\) holds by the definition of
\(m_{\mathrm{coin}}\) in \eqref{eq:def-m-s-ell-relax}.
\end{proof}

For any
\(\mathbf p,\mathbf q\in[0,1]^{m_{\mathrm{coin}}}\), the padding branch in
\eqref{eq:def-tildeLambda-relax} cancels, so
\(\widetilde{\Lambda}_{\mathbf p}-\widetilde{\Lambda}_{\mathbf q}
=\Lambda_{\mathbf p}^{(m_{\mathrm{coin}})}
-\Lambda_{\mathbf q}^{(m_{\mathrm{coin}})}\). The proof of
Lemma~\ref{lem:diamond-multicoin-final2} therefore applies on the
\(m_{\mathrm{coin}}\)-dimensional coin subspace, yielding
\begin{align}
  \|\widetilde{\Lambda}_{\mathbf p}
  -\widetilde{\Lambda}_{\mathbf q}\|_\diamond
  =
  2\max_{t\in[m_{\mathrm{coin}}]}|p_t-q_t|
  =
  2\|\mathbf p-\mathbf q\|_\infty .
  \label{eq:diamond-tildeLambda-relax}
\end{align}

We now reduce learning \(\widetilde{\Lambda}_{\mathbf p}\) to estimating
\(m_{\mathrm{coin}}\) Bernoulli biases, using the same classicalisation idea as
in Lemma~\ref{lem:coin-classicalization}.

\begin{lemma}[Classical reduction to adaptive coin sampling]
\label{lem:padded-family-classicalization}
Fix
\(\mathbf p=(p_1,\ldots,p_{m_{\mathrm{coin}}})\in(0,1)^{m_{\mathrm{coin}}}\).
Any \(N\)-use quantum strategy interacting with
\(\widetilde{\Lambda}_{\mathbf p}\) and producing a classical output induces an
adaptive procedure that obtains at most \(N\) samples from the
\(m_{\mathrm{coin}}\) Bernoulli variables with biases \(\mathbf p\), together
with arbitrary side information independent of \(\mathbf p\).
\end{lemma}

\begin{proof}
Let
\(P_{\mathrm{coin}}\coloneqq
\sum_{t=1}^{m_{\mathrm{coin}}}|s+t\rangle\!\langle s+t|\) and
\(P_{\mathrm{pad}}\coloneqq\mathbb I-P_{\mathrm{coin}}\). The padding Kraus
operators have input support contained in \(P_{\mathrm{pad}}\), while the coin
Kraus operators have input support contained in \(P_{\mathrm{coin}}\). Hence all
cross terms between these two input subspaces are annihilated. Equivalently,
\begin{align}
  \widetilde{\Lambda}_{\mathbf p}
  =
  \widetilde{\Lambda}_{\mathbf p}\circ\mathcal P,
  \qquad
  \mathcal P(\rho)
  \coloneqq
  P_{\mathrm{pad}}\rho P_{\mathrm{pad}}
  +
  P_{\mathrm{coin}}\rho P_{\mathrm{coin}}.
\end{align}
Thus, immediately before each channel use, we may insert the projective
measurement \(\{P_{\mathrm{pad}},P_{\mathrm{coin}}\}\) on the input without
affecting the final output distribution.

Conditioned on the event \(P_{\mathrm{coin}}\), the channel action is given by
the coin component \eqref{eq:def-Lambda-m-relax}. This component ignores
off-diagonal terms in the basis
\(\{|s+t\rangle\}_{t=1}^{m_{\mathrm{coin}}}\) and outputs states diagonal in
\(\{|0\rangle,|1\rangle\}\). Therefore, without changing the final output
distribution, we may further dephase the input in this basis and measure the
channel output in the basis \(\{|0\rangle,|1\rangle\}\). Each such use is
equivalent to choosing an index \(T_r\in[m_{\mathrm{coin}}]\) and observing a
bit
\[
  X_r\sim \mathrm{Bernoulli}(p_{T_r}).
\]

Conditioned on the event \(P_{\mathrm{pad}}\), the channel output is generated
solely by the padding branch
\(\rho\mapsto\sum_b \Pi_b\rho\Pi_b^\dagger\), which is independent of
\(\mathbf p\). Padding outcomes can therefore be treated as
\(\mathbf p\)-independent side information that may influence the adaptive rule
for choosing future coin indices, but carries no direct information about
\(\mathbf p\). Consequently, over \(N\) channel uses, the strategy can generate
at most \(N\) genuine Bernoulli samples of the form
\(X_r\sim \mathrm{Bernoulli}(p_{T_r})\), and all remaining processing can be
absorbed into classical post-processing of the resulting transcript and
\(\mathbf p\)-independent side information.
\end{proof}

We now explain how any successful channel estimator yields an estimator of the
coin biases. Given an arbitrary output channel \(\widehat\Lambda\), define
\begin{align}
  \widehat p_t
  \coloneqq
  \langle 1|
  \widehat\Lambda(|s+t\rangle\!\langle s+t|)
  |1\rangle,
  \qquad
  t\in[m_{\mathrm{coin}}].
\end{align}
If the learner satisfies
\[
  \frac12
  \|\widehat\Lambda-\widetilde{\Lambda}_{\mathbf p}\|_\diamond
  \le
  \varepsilon,
\]
then, for every \(t\), contractivity of trace distance under the two-outcome
measurement \(\{|1\rangle\!\langle1|,I-|1\rangle\!\langle1|\}\) gives
\begin{align}
  |\widehat p_t-p_t|
  &\le
  \frac12
  \left\|
    \widehat\Lambda(|s+t\rangle\!\langle s+t|)
    -
    \widetilde{\Lambda}_{\mathbf p}(|s+t\rangle\!\langle s+t|)
  \right\|_1
  \le
  \varepsilon .
\end{align}
Thus any successful channel learner gives an estimator of \(\mathbf p\) in
\(\ell_\infty\) error at most \(\varepsilon\).

By Lemma~\ref{lem:padded-family-classicalization}, any \(N\)-query quantum
learning algorithm for \(\widetilde{\Lambda}_{\mathbf p}\) can be simulated by
a classical adaptive procedure obtaining at most \(N\) samples from the
\(m_{\mathrm{coin}}\) Bernoulli variables. The standard adaptive multi-coin
lower bound~\cite{Tsybakov2008IntroductionTN} therefore gives, for
\(m_{\mathrm{coin}}\ge1\),
\begin{align}
  N
  =
  \Omega\!\left(
    \frac{m_{\mathrm{coin}}}{\varepsilon^{2}}
    \log\frac{m_{\mathrm{coin}}}{\delta}
  \right),
  \label{eq:delta-lb-relaxed-k}
\end{align}
with the term understood to be absent when \(m_{\mathrm{coin}}=0\). In
particular, when \(k\ge2d_{\mathrm{in}}\), we have
\(m_{\mathrm{coin}}=d_{\mathrm{in}}\), and
\eqref{eq:delta-lb-relaxed-k} reduces to \eqref{eq:lb-delta-statement}.

\begin{corollary}[Coin lower bound under a Kraus-rank constraint]
\label{co:coin-lb-delta-general-k}
Let \(m_{\mathrm{coin}}\) be defined by \eqref{eq:def-m-s-ell-relax}. Any
\(N\)-query protocol that learns all Kraus-rank-\(\le k\) channels to
diamond-distance accuracy \(\varepsilon\), with failure probability at most
\(\delta\), must use
\begin{align}
  N
  =
  \Omega\!\left(
    \frac{m_{\mathrm{coin}}}{\varepsilon^{2}}
    \log\frac{m_{\mathrm{coin}}}{\delta}
  \right),
\end{align}
where the term is understood to be absent when \(m_{\mathrm{coin}}=0\).
\end{corollary}

\subsubsection{Kraus rank just above the minimal one}
\label{subsubsec:borderline-k-plus-one}

The independent-coin construction of Subsection~\ref{subsubsec:relax-krausrank-delta} gives a lower bound proportional to \(m_{\mathrm{coin}}\), but \(m_{\mathrm{coin}}\) can still vanish for Kraus ranks just above the minimal value. We now show that, as soon as the allowed Kraus rank is larger than the minimal Kraus rank allowed, one can already embed a one-parameter pure-state discrimination problem which forces $ N
  =
  \Omega\!\left(
    \varepsilon^{-2}\log\frac{1}{\delta}
  \right)$.

Let
\(k_{\min}\coloneqq \left\lceil d_{\mathrm{in}}/d_{\mathrm{out}}\right\rceil\).
At the minimal Kraus-rank boundary \(k=k_{\min}\), a uniform \(\Omega(\varepsilon^{-2})\) lower bound cannot hold in general: in the square case \(d_{\mathrm{in}}=d_{\mathrm{out}}\), this class contains unitary channels, for which Heisenberg scaling \(O(1/\varepsilon)\) is achievable. The next proposition shows that once \(k>k_{\min}\), an \(\varepsilon^{-2}\) lower-bound contribution is already unavoidable.

\begin{proposition}
\label{prop:delta-lb-borderline-k-plus-one}
Assume \(d_{\mathrm{out}}\ge2\) and \(k\ge k_{\min}+1\). There exist constants \(\varepsilon_0,\delta_0>0\) such that, for every \(0<\varepsilon\le\varepsilon_0\) and \(0<\delta\le\delta_0\), there is a family of channels \(\{\Gamma_\theta\}_{\theta\in(0,\pi/4)}\) with
\(\mathrm{rank}_{\mathrm{Kraus}}(\Gamma_\theta)\le k\) such that any \(N\)-query protocol that outputs an estimator \(\widehat{\Lambda}\) satisfying
\[
  \frac12\|\widehat{\Lambda}-\Gamma_\theta\|_\diamond\le \varepsilon
\]
with probability at least \(1-\delta\), uniformly over \(\theta\), must use
\begin{align}
  N
  =
  \Omega\!\left(
    \frac{1}{\varepsilon^{2}}\log\frac{1}{\delta}
  \right).
  \label{eq:delta-lb-eps2-borderline}
\end{align}
\end{proposition}

\begin{proof}
It suffices to construct a hard one-parameter family of channels with Kraus rank at most \(k_{\min}+1\). Since \(k\ge k_{\min}+1\), the same family is contained in the Kraus-rank-\(\le k\) class.

Let \(s\coloneqq d_{\mathrm{in}}-1\) and \(\ell_0\coloneqq \left\lceil s/d_{\mathrm{out}}\right\rceil\). Since \(s\le d_{\mathrm{in}}\), we have \(\ell_0\le k_{\min}\). Fix orthonormal bases \(\{|j\rangle\}_{j=1}^{d_{\mathrm{in}}}\) for the input space and \(\{|i\rangle\}_{i=0}^{d_{\mathrm{out}}-1}\) for the output space. For each \(b\in\{0,\dots,\ell_0-1\}\), define
\begin{align}
  \Pi_b
  \coloneqq
  \sum_{i=1}^{r_b} |i-1\rangle\!\langle b d_{\mathrm{out}}+i|,
  \qquad
  r_b\coloneqq \min\{d_{\mathrm{out}},\, s-b d_{\mathrm{out}}\}.
\end{align}
If \(s=0\), the padding sum is empty; otherwise,
\(\sum_{b=0}^{\ell_0-1}\Pi_b^\dagger\Pi_b=\sum_{j=1}^{s}|j\rangle\!\langle j|\).
For \(\theta\in(0,\pi/4)\), set
\(|\psi_\theta\rangle\coloneqq \cos\theta\,|0\rangle+\sin\theta\,|1\rangle\) and
\(K_\theta\coloneqq |\psi_\theta\rangle\!\langle d_{\mathrm{in}}|\), and define
\begin{align}
  \Gamma_\theta(\rho)
  \coloneqq
  \sum_{b=0}^{\ell_0-1}\Pi_b\,\rho\,\Pi_b^\dagger
  +
  K_\theta\,\rho\,K_\theta^\dagger .
  \label{eq:def-Gamma-borderline}
\end{align}
Then \(\Gamma_\theta\) is CPTP because
\[
  \sum_{b=0}^{\ell_0-1}\Pi_b^\dagger\Pi_b+K_\theta^\dagger K_\theta
  =
  \sum_{j=1}^{s}|j\rangle\!\langle j|
  +
  |d_{\mathrm{in}}\rangle\!\langle d_{\mathrm{in}}|
  =
  \mathbb I_{d_{\mathrm{in}}},
\]
and it has Kraus rank at most \(\ell_0+1\le k_{\min}+1\le k\).

Fix \(0<\varepsilon\le\varepsilon_0\) and set \(\Delta\coloneqq \arcsin(3\varepsilon)\). For \(\varepsilon_0\) small enough, \(\Delta\in(0,\pi/6)\). Choose
\(\theta_0\coloneqq \Delta/2\) and \(\theta_1\coloneqq 3\Delta/2\), so \(|\theta_0-\theta_1|=\Delta\) and both angles lie in \((0,\pi/4)\). On input
\(\rho^\star\coloneqq |d_{\mathrm{in}}\rangle\!\langle d_{\mathrm{in}}|\) we have
\(\Gamma_\theta(\rho^\star)=|\psi_\theta\rangle\!\langle\psi_\theta|\), hence
\begin{align}
  \frac12\|\Gamma_{\theta_0}-\Gamma_{\theta_1}\|_\diamond
  &\ge
  \frac12
  \bigl\|
    |\psi_{\theta_0}\rangle\!\langle\psi_{\theta_0}|
    -
    |\psi_{\theta_1}\rangle\!\langle\psi_{\theta_1}|
  \bigr\|_1 \nonumber\\
  &=
  \sqrt{1-|\langle\psi_{\theta_0}|\psi_{\theta_1}\rangle|^2}
  =
  \sin\Delta
  =
  3\varepsilon .
  \label{eq:diamond-sep-borderline}
\end{align}
Therefore, any learner that achieves diamond-distance error at most \(\varepsilon\) for both channels with success probability at least \(1-\delta\) yields a binary discriminator between \(\Gamma_{\theta_0}\) and \(\Gamma_{\theta_1}\) with error probability at most \(\delta\): decide \(i\) when \(\widehat{\Lambda}\) lies within diamond distance \(\varepsilon\) of \(\Gamma_{\theta_i}\). The two \(\varepsilon\)-balls are disjoint by \eqref{eq:diamond-sep-borderline}.

Now observe that the only \(\theta\)-dependence of \(\Gamma_\theta\) sits on the one-dimensional input subspace
\(P_{\mathrm{coin}}\coloneqq |d_{\mathrm{in}}\rangle\!\langle d_{\mathrm{in}}|\).
Writing \(P_{\mathrm{pad}}\coloneqq \mathbb I-P_{\mathrm{coin}}\) and
\[
  \mathcal P(\rho)\coloneqq
  P_{\mathrm{pad}}\rho P_{\mathrm{pad}}
  +
  P_{\mathrm{coin}}\rho P_{\mathrm{coin}},
\]
we have \(\Pi_b=\Pi_bP_{\mathrm{pad}}\) and \(K_\theta=K_\theta P_{\mathrm{coin}}\), hence \(\Gamma_\theta=\Gamma_\theta\circ\mathcal P\). Thus we may insert \(\mathcal P\) immediately before each use without changing the joint output state. That is, without loss of generality, before each query we may measure \(\{P_{\mathrm{pad}},P_{\mathrm{coin}}\}\) on the input without changing the distribution of any final classical outcome.

After this insertion, each query either lands in \(P_{\mathrm{coin}}\), in which case it produces one copy of the pure state \(|\psi_{\theta_i}\rangle\!\langle\psi_{\theta_i}|\), or lands in \(P_{\mathrm{pad}}\), in which case the output and any resulting side information are independent of \(\theta_i\). Consequently, any \(N\)-query discrimination strategy can be simulated by a measurement on at most \(N\) copies of \(|\psi_{\theta_i}\rangle\), together with a \(\theta\)-independent auxiliary system. In particular, its error probability is lower bounded by the optimal Helstrom error for discriminating
\[
  |\psi_{\theta_0}\rangle^{\otimes N}
  \quad\text{versus}\quad
  |\psi_{\theta_1}\rangle^{\otimes N}.
\]

Let \(c\coloneqq|\langle\psi_{\theta_0}|\psi_{\theta_1}\rangle|=\cos\Delta\), since
\(\langle\psi_{\theta_0}|\psi_{\theta_1}\rangle=\cos(\theta_0-\theta_1)\). Then the overlap of the \(N\)-copy pure states equals
\[
  |\langle\psi_{\theta_0}^{\otimes N}|\psi_{\theta_1}^{\otimes N}\rangle|
  =
  c^N .
\]
For binary discrimination between two equiprobable states \(\rho_0,\rho_1\), Helstrom's theorem gives
\[
  p_{\mathrm{err}}^\star(\rho_0,\rho_1)
  =
  \frac{1}{2}\left(1-\frac{1}{2}\|\rho_0-\rho_1\|_1\right).
\]
For pure states \(\rho_i=|\phi_i\rangle\!\langle\phi_i|\), one has
\[
  \bigl\|
    |\phi_0\rangle\!\langle\phi_0|
    -
    |\phi_1\rangle\!\langle\phi_1|
  \bigr\|_1
  =
  2\sqrt{1-|\langle\phi_0|\phi_1\rangle|^2}.
\]
Hence, applied to \(|\phi_i\rangle=|\psi_{\theta_i}\rangle^{\otimes N}\),
\begin{align}
  p_{\mathrm{err}}^\star
  =
  \frac{1}{2}\left(1-\sqrt{1-c^{2N}}\right).
\end{align}
Using \(\sqrt{1-x}\le 1-\tfrac{x}{2}\) for \(x\in[0,1]\), we get
\(p_{\mathrm{err}}^\star\ge \tfrac14 c^{2N}\). Thus any discriminator with error at most \(\delta\) must satisfy
\[
  \delta\ge \frac14\cos^{2N}\Delta .
\]
For \(\Delta\in(0,1)\), we have the elementary bounds
\(\cos\Delta\ge 1-\tfrac{\Delta^2}{2}\) and
\(1-y\ge e^{-2y}\) for \(y\in[0,1/2]\). Hence
\(\cos\Delta\ge 1-\tfrac{\Delta^2}{2}\ge e^{-\Delta^2}\), and therefore
\[
  \delta\ge \frac14 e^{-2N\Delta^2},
\]
which implies
\begin{align}
  N
  \ge
  \frac{1}{2\Delta^2}\log\frac{1}{4\delta}.
\end{align}
For \(\delta_0>0\) sufficiently small, \(\log(1/(4\delta))=\Omega(\log(1/\delta))\) for all \(0<\delta\le\delta_0\). Finally, since
\(\Delta=\arcsin(3\varepsilon)=\Theta(\varepsilon)\) for
\(0<\varepsilon\le\varepsilon_0\), we obtain
\[
  N
  =
  \Omega\!\left(
    \frac{1}{\varepsilon^2}\log\frac1\delta
  \right),
\]
as claimed.
\end{proof}

Combining the independent-coin construction with the one-parameter construction
above, define
\begin{align}
  m_{\mathrm{coin}}
  \coloneqq
  \max\left\{
    r\in\{0,\ldots,d_{\mathrm{in}}\}:
    2r+
    \left\lceil\frac{d_{\mathrm{in}}-r}{d_{\mathrm{out}}}\right\rceil
    \le k
  \right\},
  \qquad
  m
  \coloneqq
  \max\left\{
    m_{\mathrm{coin}},
    \mathbf 1_{\{k>k_{\min}\}}
  \right\}.
  \label{eq:def-effective-m-final}
\end{align}
Thus \(m=0\) at the minimal Kraus-rank boundary, \(m>0\) for every
non-minimal Kraus rank, and \(m=d_{\mathrm{in}}\) whenever
\(k\ge2d_{\mathrm{in}}\). The preceding two constructions therefore give the
accuracy-dependent contribution
\[
  \Omega\!\left(
    \frac{m}{\varepsilon^2}\log\frac{m}{\delta}
  \right),
\]
with the convention that this term is absent when \(m=0\). Combining this with
the dimension-dependent lower bound from Theorem~\ref{thm:lower-boundPROOF}
gives the final lower-bound statement.

\begin{theorem}[Combined lower bound with explicit failure probability]
\label{prop:combined-lb-delta-general-k}
Assume \(d_{\mathrm{out}}\ge2\), let
\(k_{\min}\coloneqq\left\lceil d_{\mathrm{in}}/d_{\mathrm{out}}\right\rceil\),
and assume \(k\ge k_{\min}\). There exist constants
\(\varepsilon_0,\delta_0>0\) such that, for every
\(0<\varepsilon\le\varepsilon_0\) and \(0<\delta\le\delta_0\), the following
holds. In the worst case over all channels
\[
  \Lambda:
  \mathcal L(\mathbb C^{d_{\mathrm{in}}})
  \to
  \mathcal L(\mathbb C^{d_{\mathrm{out}}})
\]
of Kraus rank at most \(k\), any \(N\)-query tomography protocol that outputs
an estimator \(\widehat\Lambda\) satisfying
\[
  \frac12\|\widehat\Lambda-\Lambda\|_\diamond\le\varepsilon
\]
with probability at least \(1-\delta\) must use
\begin{align}
  N
  =
  \Omega\!\left(
    \frac{d_{\mathrm{in}}d_{\mathrm{out}}k}{\varepsilon^{\beta}}
    +
    \frac{m}{\varepsilon^{2}}\log\frac{m}{\delta}
  \right),
  \label{eq:combined-delta-lb-general-k}
\end{align}
where \(m\) is defined in \eqref{eq:def-effective-m-final}, the second term is
understood to be absent when \(m=0\), and
\[
  \beta
  \coloneqq
  \frac{
    2d_{\mathrm{in}}d_{\mathrm{out}}k-d_{\mathrm{in}}^2-k^2
  }{
    2\bigl(d_{\mathrm{in}}d_{\mathrm{out}}k-1\bigr)
  } .
\]
\end{theorem}

\begin{proof}
The first term is the dimension-dependent lower bound from
Theorem~\ref{thm:lower-boundPROOF}. The second term follows from
\eqref{eq:delta-lb-relaxed-k} when \(m=m_{\mathrm{coin}}\), and from
Proposition~\ref{prop:delta-lb-borderline-k-plus-one} when
\(m=\mathbf 1_{\{k>k_{\min}\}}\). Taking the stronger of the two
accuracy-dependent lower bounds gives the definition of \(m\) in
\eqref{eq:def-effective-m-final}.
\end{proof}

\end{document}